\documentclass{iopart}
\usepackage{iopams}  
\usepackage{braket,graphicx,amsmath,cite}
\usepackage[colorlinks,linkcolor={blue},citecolor={blue}]{hyperref}
\begin{document}

\title{Frustration shapes multi-channel Kondo physics: a star graph perspective}
\author{Siddhartha Patra$^1$, Abhirup Mukherjee$^1$, Anirban Mukherjee$^1$, N. S. Vidhyadhiraja$^4$, A. Taraphder$^5$ and Siddhartha Lal$^1$}
\eads{\mailto{sp14ip022@iiserkol.ac.in}, \mailto{am18ip014@iiserkol.ac.in}, \mailto{mukherjee.anirban.anirban@gmail.com}, \mailto{raja@jncasr.ac.in}, \mailto{arghya@phy.iitkgp.ernet.in}, \mailto{slal@iiserkol.ac.in}}

\address{$^1$Department of Physical Sciences, Indian Institute of Science Education and Research-Kolkata, W.B. 741246, India}

\address{$^4$Theoretical Sciences Unit, Jawaharlal Nehru Center for Advanced Scientific Research, Jakkur, Bengaluru 560064, India}

\address{$^5$Department of Physics, Indian Institute of Technology Kharagpur, Kharagpur 721302, India}

\date{\today}

\begin{abstract}
	We study the overscreened multi-channel Kondo (MCK) model using the recently developed unitary renormalization group (URG) technique. Our results display the importance of ground state degeneracy in explaining various important properties like the breakdown of screening and the presence of local non-Fermi liquids. The impurity susceptibility of the intermediate coupling fixed point Hamiltonian in the zero-bandwidth (or star graph) limit shows a power-law divergence at low temperature. Despite the absence of inter-channel coupling in the MCK fixed point Hamiltonian, the study of mutual information between any two channels shows non-zero correlation between them. A spectral flow analysis of the star graph reveals that the degenerate ground state manifold possesses topological quantum numbers. Upon disentangling the impurity spin from its partners in the star graph, we find the presence of a local Mott liquid arising from inter-channel scattering processes. The low energy effective Hamiltonian obtained upon adding a finite non-zero conduction bath dispersion to the star graph Hamiltonian for both the two and three-channel cases displays the presence of local non-Fermi liquids arising from inter-channel quantum fluctuations. Specifically, we confirm the presence of a local marginal Fermi liquid in the two channel case, whose properties show logarithmic scaling at low temperature as expected.
Discontinuous behaviour is observed in several measures of ground state entanglement, signalling
the underlying orthogonality catastrophe associated with the degenerate ground state manifold. 
We extend our results to underscreened and perfectly screened MCK models through duality arguments.
A study of channel anisotropy under renormalisation flow reveals a series of quantum phase transitions due to the change in ground state degeneracy. Our work thus presents a template for the study of how a degenerate ground state manifold arising from symmetry and duality properties in a multichannel quantum impurity model can lead to novel multicritical phases at intermediate coupling.   

\end{abstract}

\ioptwocol

\section{Introduction}

A local antiferromagnetic exchange interaction between a spin-\(\frac{1}{2}\) impurity and its host metal gives rise to the well-understood phenomenon of the Kondo effect (\cite{kondo1964resistance}, see ~\cite{hewson1993} for an extensive discussion). Here, the impurity local moment is screened by the conduction electrons at temperatures below a certain scale called the Kondo temperature~\cite{anderson1969exact,anderson1970exact,anderson1970,wilson1975,andreiKondoreview,
tsvelickKondoreview,affleck1993exact,affleck1995conformal}, leading to a spin singlet ground state and local Fermi liquid excitations above that ground state~\cite{nozieres1974fermi,hewsonp}. The screening manifests as an initial increase at temperatures below the Kondo temperature, followed by a saturation of the resistivity of the metal~\cite{kondo1964resistance,Zhang2013}, and in the saturation of the impurity contribution to the magnetic susceptibility at low temperatures~\cite{wilson1975,andreiKondoreview,tsvelickKondoreview}.
Generalisations of this model are obtained by taking impurities of higher spin~\cite{hewson1993,Noz_blandin_1980,affleck1993exact}, adding interactions between them~\cite{affleck_1992,affleck_ludwig_jones_1992,georges_1999,zarand_chung_2006,
mitchell_logan_2011,mitchell_sela_2012,konig_coleman_2021}, by promoting the model to a lattice of Kondo impurities~\cite{coleman_1987,millis_lee_1987,auerbach_1986,rice_ueda_1986,stewart_1984} or by considering the case of a correlated host metal~\cite{granath_1998}. One can also construct multi-channel Kondo (MCK) models by allowing \(K\) conduction electron channels \( \left(K > 1\right) \) to interact with a spin-\(S_{d}\) impurity~\cite{Noz_blandin_1980,zawadowski_1980,cox_zawadowski_1998} via a common exchange coupling \(\mathcal{J}\). Using conformal field theory~\cite{affleck_1991_overscreen,affleck_ludwig_1991,affleck_pang_cox_1992,affleck1993exact,parcollet_olivier_large_N,affleck_2005}, bosonization~\cite{emery_kivelson,clarke_giamarchi_1993,zarand_2000,vondelft_prl_1998,schofield_1997}, numerical renormalization group~\cite{bullaNRGreview,affleck_pang_cox_1992,pang_cox_1991}, Bethe ansatz~\cite{andrei_destri_1984,Tsvelick1984,Tsvelick_1985,andrei_jerez_1995,zarand_costi_2002} and other methods~\cite{sengupta_1994,fabrizio_nozieres_1995,Coleman_tsvelik,fabrizio_gogolin_1995}, it has been shown that for the over-screened systems \(\left(K > 2S_d\right)\), the low energy physics is of the non-Fermi liquid type obtained at an intermediate coupling fixed point of the RG flows~\cite{Noz_blandin_1980}. Further, the system displays anomalous behaviour in thermodynamic properties near \(T=0\) and a fractional zero temperature entropy in the thermodynamic limit (upon ensuring that temperature is taken to zero after the system size is taken to infinity~\cite{rozhkov_1998,vondelft_prl_1998}).
This anomalous divergent behaviour is a signature of the fact that the MCK system with a single channel-symmetric exchange coupling is quantum critical: the ground state is susceptible to perturbations that introduce channel anisotropy in the exchange couplings~\cite{Noz_blandin_1980,andrei_jerez_1995,affleck_pang_cox_1992,zarand_2000,zheng_2021}. This makes the experimental realisation of such states quite challenging.
Nevertheless, several features of the two-channel Kondo model have been reproduced using structural two-level systems~\cite{zawadowski_1980,vladar_1983} interacting with a conduction bath~\cite{cichorek_2005,ralph_buhrman_1992,ralph_ludwig_1994,Iftikhar2015,Zhu2016}.
More recently, it has become possible to tune a two-channel quantum dot system across the quantum phase transition~\cite{Potok2007,Keller2015}, revealing the fractionalisation of the low-energy degrees of freedom~\cite{emery1995,Coleman_tsvelik,mebrahtu_2013}.

The importance of ground state degeneracy and quantum-mechanical \textit{frustration} in the MCK problem has not received sufficient attention, even though they play crucial roles in determining the quantum criticality and associated non-Fermi liquid physics of the system. This is, at least partially, due to the lack of an intermediate-coupling effective Hamiltonian that describes the low-energy physics of the MCK (i.e., reveals the nature of the excitations). Frustration in this context refers to the inability of the impurity spin to bind with the conduction channels, and be screened completely by forming a maximally-entangled singlet ground state. The absence of such frustration in the single-channel spin-1/2 Kondo model (and more generally, the exact screening variant of the MCK) means that imperfect screening \((K \neq 2S_d)\) leads to dramatic differences in the multi-channel models. Indeed, we demonstrate that there exists an orthogonality catastrophe in the imperfectly screened MCK models that leads to the non-Fermi liquid behaviour of its low-energy excitations, and that this is directly related to the existence of a degenerate ground state manifold. The effects of these non-Fermi liquid low energy excitations are expected to appear in measures of many-particle entanglement; the lack of an intermediate coupling Hamiltonian has, however, prevented the study of the renormalisation group (RG) evolution of such measures~\cite{alkurtass_affleck_2016,kim_shim_2021}. 
Importantly, our approach enables a comparison of, for instance, the inter-channel mutual information along the RG flow and as functions of the excitation energies, providing insight into the nature of the fixed point theory.

To obtain the RG flow of the MCK model, we have employed the unitary renormalization group (URG) technique developed recently by  some of us~\cite{anirbanurg1,anirbanurg2}. The method has been applied on several fermionic and spin problems like the single-channel Kondo model~\cite{kondo_urg}, the kagome antiferromagnet~\cite{santanukagome}, the 1D~\cite{1dhubjhep} and 2D Hubbard models~\cite{anirbanmott1,anirbanmott2,mukherjeeMERG2022}, the reduced-BCS model~\cite{siddharthacpi} as well as other generalised models of interacting electrons with and without translation invariance~\cite{anirbanurg2}.
The URG proceeds by resolving the number fluctuations in the electronic Fock states at high energies by the iterative application of many-particle unitary transformations, leading to their successive decoupling from the states lying at lower energies. Applying this URG method to the channel-isotropic MCK model leads to a low-energy fixed point Hamiltonian at the intermediate (Kondo) coupling fixed point of Ref.\cite{Noz_blandin_1980}. In further analysing the effective Hamiltonian, we focus initially on the zero bandwidth limit of this Hamiltonian (which corresponds to a frustrated quantum spin model on a star graph). We show thereby that several important properties of the MCK problem, such as the ground state degeneracy and breakdown of screening, can be understood from the zero bandwidth problem.
We then proceed to studying the low energy effective Hamiltonian for the non-Fermi liquid excitations lying above the degenerate ground state manifold. This enables the computation of a plethora of quantities, such as various thermodynamic measures (e.g., specific heat and susceptibility), many-particle entanglement and the self-energy of the propagating degrees of freedom. We also explore various signatures of criticality, and various duality transformations of the MCK Hamiltonian. Finally, we investigate the evolution of ground state degeneracy under RG of the channel anisotropic MCK model.

\subsection*{Summary of Main Results}
We start in section \ref{rg_section} by describing the URG flows for the MCK, the intermediate coupling URG fixed point Hamiltonian and the features of the zero bandwidth (star graph) limit of the effective Hamiltonian.
In section \ref{sec:props_star}, we explore some properties of the zero bandwidth model: namely, the degree of compensation, impurity magnetisation and impurity susceptibility.
The degree of compensation is simply the average correlation between the impurity spin and the conduction channel local spins; it is found to be maximum at exact screening, and decreases for both over- and under-screening.
The magnetisation and susceptibility of the star graph model show a discontinuity and a \(T^{-1}\) divergence respectively.
We also explore the topological properties of the degenerate ground state manifold of the star graph model using non-local twist and translation operators.
On decoupling the impurity spin from the conduction spin degrees of freedom within the star graph, we obtain an all-to-all Hamiltonian that represents a local Mott liquid phase.
In section \ref{sec:excitations}, we obtain the low energy effective Hamiltonian for the low-lying excitations by including a dispersion into the conduction bath.
Notable features of this effective Hamiltonian include (i) the complete absence of local Fermi liquid terms, and (ii) the emergence of local non-Fermi liquid terms.
Both are shown to result from the ground state degeneracy of the underlying star graph model.
The excitations of the two-channel Kondo problem are found to pertain to a local marginal Fermi liquid, signalling an orthogonality catastrophe. 

In section \ref{sec:ent_prop}, we study various entanglement signatures (e.g., von-Neumann entanglement entropy, mutual information, multipartite information, Bures distance) of the low energy fixed-point ground state of the MCK problem, with and without dispersion into the bath. The discontinuous behaviour observed in these measures for the MCK problem reveal clear distinctions from the single-channel case. In \ref{sec:duality}, we discuss two duality transformations that exist for the MCK model - a strong-weak duality, and an over-screened-under-screened duality. It is found that the transformations are constrained by the fixed point structure, and impose relations between the under-screened and over-screened models.
In \ref{anisotropic_rg}, we show the robustness of the ground state degeneracy of the star graph model against the impurity-channel coupling anisotropy, and correlate it with a URG treatment of the channel anisotropic MCK.
The isotropic fixed point is found to be unstable to asymmetry, leading to impurity phase transitions that lead to dramatic changes in the ground state degeneracy.
For the sake of brevity, some supporting evidence and details of certain lengthy calculations are presented in Supplementary Materials~\cite{SM}. Finally, we conclude in section \ref{conclusions}.

\section{Fixed point theory of over-screened MCK model}
\label{rg_section}
\subsection{RG flows towards intermediate coupling}
\label{rg_flow_section}
We start with the \(K\)- conduction channel Kondo model Hamiltonian with isotropic couplings \cite{Noz_blandin_1980}:
\begin{eqnarray}
	\label{mc_ham}
	\hspace*{-5pt}H = \sum_{\substack{l,k\\\alpha=\uparrow,\downarrow}}\epsilon_{k,l} \hat n_{k\alpha,l} + \frac{\mathcal{J}}{2}\sum_{\substack{l,kk^\prime\\\alpha,\alpha^\prime= \uparrow,\downarrow}} \vec{S_d}\cdot\vec{\sigma}_{\alpha\alpha^\prime}c_{k\alpha,l}^\dagger c_{k^\prime\alpha^\prime, l}~,
\end{eqnarray}
where \(\mathcal{J}\) is the Kondo spin-exchange coupling, \(c_{k\alpha,l}\) is the fermionic field operator at momentum \(k\), spin \(\alpha\) and channel \(l\), \(\epsilon_{k,l}\) represents the dispersion of the \(l^\mathrm{th}\) conduction channel, \(\vec \sigma\) is the vector of Pauli matrices and \(\vec S_d = \frac{1}{2}\vec \sigma_d\) is the impurity spin operator. Here, \(l\) sums over the \(K\) channels of the conduction bath, \(k,k^\prime\) sum over all the momentum states of the bath and \(\alpha,\alpha^\prime\) sum over the two spin indices of a single electron.

We now perform a renormalisation group analysis of the MCK Hamiltonian using the recently developed URG method \cite{anirbanmott1,anirbanmott2,anirbanurg1,anirbanurg2,siddharthacpi,santanukagome,1dhubjhep}.
The RG proceeds {by applying unitary transformations in order to block-diagonalize the Hamiltonian by removing number fluctuations of the high energy degrees of freedom}.
At the $j^{th}$ RG step, we focus on decoupling the highest energy electronic state denoted by $\ket{j}$.
The Hamiltonian will in general not conserve the number of particles in this state: \(\left[H_{(j)}, \hat n_{j}\right] \neq 0\), and the unitary transformation \(U_{(j)}\) causes this number fluctuation to vanish
~\cite{anirbanurg1,anirbanurg2}:
\begin{eqnarray}
	H_{(j-1)} = U_{(j)} H_{(j)} U^\dagger_{(j)}~, ~\left[H_{(j-1)}, \hat n_{j}\right] =0~.
\end{eqnarray}
The unitary transformations are given in terms of a fermionic generator \(\eta_{(j)}\)~\cite{anirbanurg1,anirbanurg2}:
\begin{eqnarray}
	\label{unitary}
	U_{(j)} = \frac{1}{\sqrt 2}\left(1 + \eta_{(j)} - \eta_{(j)}^\dagger\right)~,~ \quad\left\{ \eta_{(j)},\eta_{(j)}^\dagger \right\} = 1~,
\end{eqnarray}
where \(\left\{\cdot\right\}\) is the anticommutator. Further, $\left[ \eta_{(j)},\eta_{(j)}^\dagger \right] = 2\hat n_{j}-1$. The unitary operator \(U_{(j)}\) mentioned in Eq.~\eqref{unitary} can be thought of as a special case of the most general form \(U = e^\mathcal{S}\) of a unitary operator defined by an anti-Hermitian generator \(\mathcal{S}\). For our case, the generator turns out to be \(\mathcal{S} = \frac{\pi}{4}\left( \eta^\dagger_{(j)} - \eta_{(j)} \right) \), such that the unitary operator is \(U_{(j)} = e^{\frac{\pi}{4}\left(\eta^\dagger_{(j)} - \eta_{(j)}\right) }\)~\cite{anirbanurg1}. If one then uses the anti-commutation and commutation properties of \(\eta_{(j)},\eta^\dagger_{(j)}\)~\cite{anirbanurg1,anirbanurg2}, the exponential can be expanded in its Taylor series and then resummed into the form in Eq.~\eqref{unitary}.

The generator itself is given by ~\cite{anirbanurg1,anirbanurg2}
\begin{eqnarray}
	\label{eta_dag}
	\eta^\dagger_{(j)} = \frac{1}{\hat \omega_{(j)} - \mathrm{Tr}\left(H_{(j)} \hat n_{j}\right) } c^\dagger_{j} \mathrm{Tr}\left(H_{(j)}c_{j}\right)~.
\end{eqnarray}
The operators \(\eta_{(j)},\eta^\dagger_{(j)}\) can be thought of as the many-particle analogue of the single-particle field operators \(c_j,c^\dagger_j\) that change the occupation number of the single-particle Fock space, \(\ket{n_j}\). The presence of the operators in $\eta_{(j)}$ other than $c_j$ are required to ensure that the decoupling of the specific Fock state under consideration is carried out through a unitary modification of the spectrum of the remnant degrees of freedom. Since the \(j^\mathrm{th}\) degree of freedom is coupled with many others, a rotation of the basis of \(j\) also involves a rotation of the many-particle basis of the remaining degrees of freedom in the Hamiltonian \(H_{(j)}\); this is encoded within the \(\mathrm{Tr}\left(H_{(j)} c_j\right) \) term.
The operator \(\hat \omega_{(j)}\) encodes the quantum fluctuation scales arising from the interplay of the kinetic energy terms and the interaction terms in the Hamiltonian:
\begin{eqnarray}
	\hat \omega_{(j)} = H_{(j-1)} - H^i_{(j)}~.
\end{eqnarray}
\(H^i_{(j)}\) is that part of \(H_{(j)}\) that commutes with \(\hat n_j\) but does not commute with at least one \(\hat n_l\) for \(l < j\). The RG flow continues till a fixed point is reached at an energy \(D^*\). The Greens function in front of $c_j$ (in Eq.~\ref{eta_dag}) accounts for the dynamical cost of this rotation in terms of an operator for quantum fluctuations (\(\hat \omega\)), as well as the kinetic and self-energies \(\mathrm{Tr}\left( H_{(j)}\hat n_j \right) \).

There exist other renormalisation group schemes that depend on canonical transformations, such as Wegner's continuous unitary transformations (CUT) RG~\cite{wegner1994flow} and the similarity RG method by Glazek and Wilson~\cite{glazek1993renormalization}. There are at least two specific distinctions between the CUT RG and the URG method adopted in the present work.
\begin{itemize}
	\item While most RG methods that apply successive unitary transformations rely on a gradual decrease in the off-diagonal content of the Hamiltonian, the URG proceeds by decoupling each discrete Fock space node at a time. In other words, the URG achieves a block-diagonalisation of the Hamiltonian at each RG step, while the CUT RG involves a gradual decay of the off-diagonal terms of the Hamiltonian in order to make it more band diagonal.
	\item The CUT RG involves choosing a truncation scheme that allows closing the RG equations. The URG, on the other hand, is able to resum the coupling expansion series through the denominator structure and the quantum fluctuation operator.  
\end{itemize}
Detailed comparisons of the URG with other methods (e.g., the functional RG, spectrum bifurcation RG etc.) can be found in Refs.~\cite{anirbanmott1,anirbanurg1}.

The derivation of the RG equation for the over-screened regime \((2S < K)\) of the spin-\(S\)-impurity \(K\)-channel Kondo problem is shown in detail in the Supplementary Materials~\cite{SM}.
On decoupling circular isoenergetic shells at energies \(D_{(j)}\), the change in the Kondo coupling at the \(j^\mathrm{th}\) RG step, \(\Delta {\mathcal{J}}_{(j)}\), is given by
\begin{eqnarray}
	\Delta {\mathcal{J}}_{(j)} = -\frac{{\mathcal{J}}_{(j)}^2 \mathcal{N}_{(j)}}{\omega_{(j)} - \frac{D_{(j)}}{2} + \frac{{\mathcal{J}}_{(j)}}{4}}\left( 1 - \frac{1}{2}\rho {\mathcal{J}}_{(j)} K \right)~, \label{mckRG}
\end{eqnarray}
where \(\mathcal{N}_{(j)}\) is the number of electronic states at the energy shell \(D_{(j)}\). We work in the low quantum fluctuation regime \(\omega_{(j)} < \frac{D_{(j)}}{2}\). There are three fixed points of the RG equation. One arises from the vanishing of the denominator, and was present in the single-channel Kondo RG equation as well \cite{kondo_urg}. As shown there, this fixed-point goes to \({\mathcal{J}}^* = \infty\) as the bare bandwidth of the conduction electrons is made large. The other fixed point is the trivial one at \({\mathcal{J}}^* = 0\). The third fixed point is reached when the numerator vanishes: \({\mathcal{J}}^* = \frac{2}{K \rho}\) \cite{Gan_mchannel_1994,Kogan_2018,Kuramoto1998,Noz_blandin_1980}. Only the intermediate fixed point is found to be stable. This is consistent with results from Bethe ansatz calculations~\cite{Tsvelick_Weigmann_mchannel_1984,andrei_destri_1984,zarand_costi_2002,andrei_jerez_1995,Tsvelick_1985,Tsvelick1984}, CFT calculations~\cite{affleck_1991_overscreen,affleck1993exact,affleck_ludwig_1991}, bosonization treatments~\cite{emery_kivelson,vondelft_prl_1998} and NRG analysis~\cite{pang_cox_1991,mitchell_bulla_2014}.
\begin{figure}[htpb]
	\centering
	\includegraphics[width=0.45\textwidth]{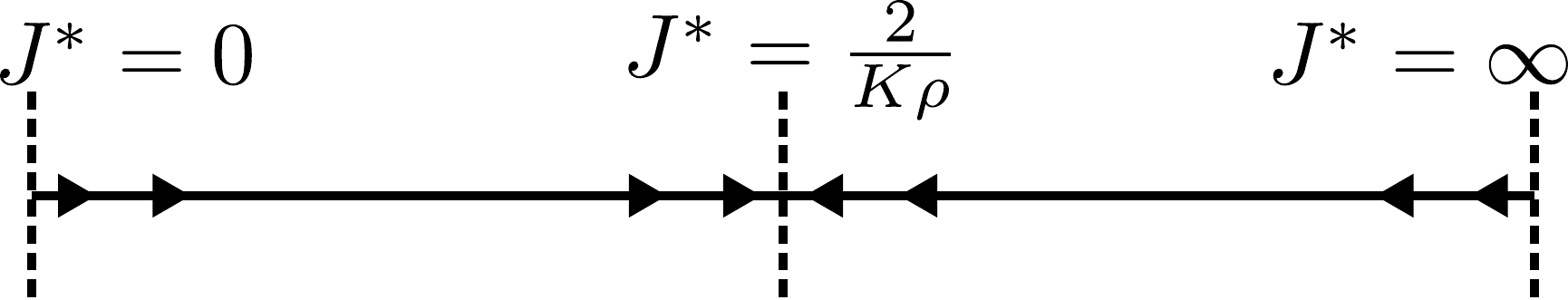}
	\caption{The three fixed points of the over-screened RG equation. Only the intermediate one is stable.}
	\label{rg_flow}
\end{figure}

The RG equation reduces to the perturbative form \(\Delta {\mathcal{J}}_{(j)} \simeq \frac{{\mathcal{J}}_{(j)}^2 \mathcal{N}_{(j)}}{D_{(j)}}\left( 1 - \frac{1}{2}\rho {\mathcal{J}}_{(j)} K \right)\)~\cite{Kogan_2018,Kuramoto1998,Noz_blandin_1980,tripathi2018landau} when one replaces \(\omega_{(j)}\) with the ground state energy \(-\frac{D_{(j)}}{2}\) and assumes \({\mathcal{J}} \ll D_{(j)}\).

\subsection{The star graph as the zero-bandwidth limit of the fixed point Hamiltonian}
\label{sec:star graph}
The fixed point Hamiltonian takes the form
\begin{eqnarray}
	H^* = \sum_l\left[ \sum^*_{k}\epsilon_{k,l} \hat n_{k\alpha,l} + {\mathcal{J}}\sum_{kk^\prime\atop{\alpha,\alpha^\prime}}^* \vec{S_d}\cdot\frac{1}{2}\vec{\sigma}_{\alpha\alpha^\prime}c_{k\alpha,l}^\dagger c_{k^\prime\alpha^\prime, l}\right].~ ~ ~ ~
\end{eqnarray}
We have not explicitly written the decoupled degrees of freedom \(D_{(j)} > D^*\) in the Hamiltonian. The \(*\) over the summations indicate that only the momenta inside the window \(D^*\) enter the summation.

\begin{figure}[htpb]
	\centering
	\includegraphics[width=0.35\textwidth]{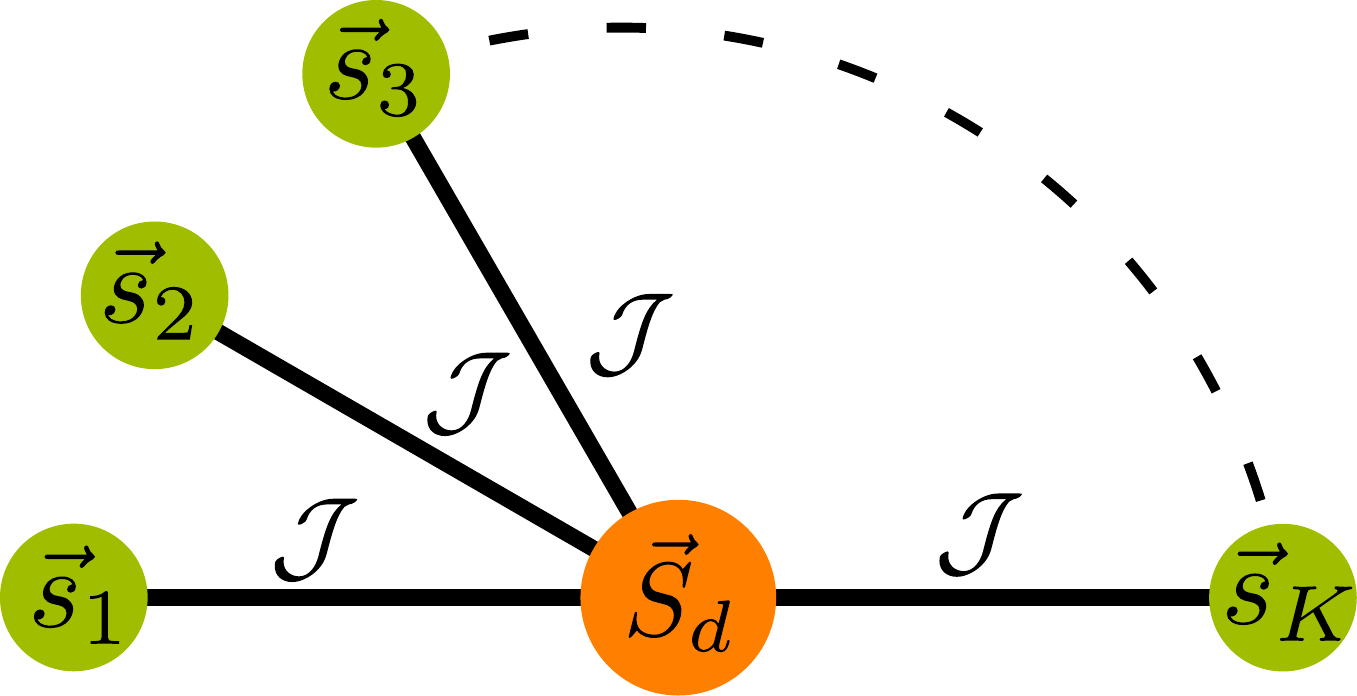}
	\caption{Zero bandwidth limit of the fixed point Hamiltonian. The central yellow node is the impurity; it is interacting with the \(K\) green outer nodes that represent the local spins of the channels.}
	\label{fig:star graph}
\end{figure}

We will first study the zero bandwidth limit of the fixed point Hamiltonian, obtained by compressing the sum over the momentum states to a single state at the Fermi surface for each conduction channel (see Fig.\ref{fig:star graph}). Upon setting the chemical potential equal to the Fermi energy, the kinetic energy part vanishes and the zero bandwidth model becomes a Heisenberg spin-exchange Hamiltonian
\begin{eqnarray}
	\label{star graph}
	H^* = {\mathcal{J}}\sum_l\sum_{kk^\prime,\atop{\alpha,\alpha^\prime}}^* \vec{S_d}\cdot\frac{1}{2}\vec{\sigma}_{\alpha\alpha^\prime}c_{k\alpha,l}^\dagger c_{k^\prime\alpha^\prime, l} = {\mathcal{J}}\vec{S_d}\cdot\vec S~.
\end{eqnarray}
At the last step, we defined the total bath local spin operator \(S = \sum_l \vec{S}_l = \frac{1}{2}\sigma_l = \frac{1}{2}\sum_{kk^\prime}^*\sum_{\alpha\alpha^\prime}\vec{\sigma}_{\alpha\alpha^\prime}c_{k\alpha,l}^\dagger c_{k^\prime\alpha^\prime, l}\).
The star graph commutes with several operators, including the total spin operator \(J^z = S_d^z + S^z\) along \(z\), the square of the total bath local spin operator (\(S^2\)) and the string operators 
\begin{eqnarray}
\pi^{x,y,z} = \sigma_d^{x,y,z} \otimes_{l=1}^K \sigma_l^{x,y,z}~.
\end{eqnarray}
If we define the global spin operator \(\vec J = \vec S + \vec S_d\), the star graph Hamiltonian can be written as \(\frac{1}{2}\mathcal{J}\left[J^2 - S_d^2 - S^2\right] \).
\begin{itemize}
	\item The ground state is achieved for the maximal value of \(S\), \(S=\frac{K}{2}\), and the corresponding minimal value of \(J\), \(J = |\frac{K}{2} - S_d|\). The ground state energy is therefore 
\begin{eqnarray}
\label{ground-state-energy}
E_g = \frac{1}{2}\mathcal{J}\left[J(J+1) - S_d(S_d+1) - S(S+1)\right] \\
= \begin{cases}
	-\mathcal{J}S_d\left(\frac{K}{2} + 1\right) & \mathrm{ when } K \geq 2S_d, ~\mathrm{ and}\\
	-\mathcal{J}\frac{K}{2}\left(S_d + 1\right) & \mathrm{ otherwise}~.
\end{cases}
\end{eqnarray}
	\item The value of \(J\) corresponds to a multiplicity of \(2J+1 = |K - 2S_d|+1\) in \(J^z\), and since the Hamiltonian does not depend on \(J^z\), these orthogonal states \(\ket{J^z}\) constitute a degeneracy of \(g^{S_d}_K = |K - 2S_d|+1\).
\end{itemize}
The \(\pi^z\) acts on the eigenstates \(\ket{J^z}\) and reveals the odd-even parity of the eigenvalue \(J^z\), and is hence a parity operator. Interestingly, {the string operator \(\pi^z\) is a Wilson loop operator~\cite{fradkin2013field} that wraps around all the nodes of the star graph}:
\begin{eqnarray}
	\label{w_loop}
	\pi^z = \exp\left[i \frac{\pi}{2} \left(\sigma_d^z + \sum_{l=1}^K \sigma^z_l - K\right)\right] = e^{i \pi \left(J^z - \frac{1}{2}K\right)}~.\qquad
\end{eqnarray}
\(\pi^x\) and \(\pi^y\) are 't Hooft operators~\cite{fradkin2013field} and mix states of opposite parity. For example, it can be shown that \(\pi^x \ket{J^z} = -\ket{J^z}\).

There are several reasons for working with the star graph in particular and zero mode Hamiltonians in general.
In the single-channel Kondo model, the star graph is just the two spin Heisenberg problem, and it reveals the stabilization of the Kondo model ground state, as well as certain thermodynamic properties (e.g., the impurity contribution to the susceptibility)~\cite{varma_yafet_1976,yosida_1966,wilson1975renormalization,moca_zarand_2021,
varma_yafet_1976,kondo_urg}.
Similarly, in the MCK model, the star graph captures accurately the nature of the RG flows. At weak coupling \({\mathcal{J}} \to 0^+\), the central spin is weakly coupled to the outer spins and  prone to screening because of the \(S^\pm\) terms in the star graph; at strong coupling \({\mathcal{J}} \to \infty^-\), the outer spin-half objects tightly bind with the central spin-half object to form a single spin object that interacts with the remaining states through an exchange coupling which is RG relevant.
This renders both the terminal fixed points unstable.
The true stable fixed point must then lie somewhere in between, and we recover the schematic phase diagram of fig.~\ref{rg_flow}. 

Moreover, the RG flows of the MCK model have been show to {preserve the degeneracy of the ground state}~\cite{pang_cox_1991,kroha_kolf_2007,zitko_fabrizio_2017}, and the star graph captures this degeneracy in its entirety.
This is important, because it will be shown in a later section that the lowest excitations of the intermediate fixed point are described by a non-Fermi liquid phase, and it can be argued that this non-Fermi liquid physics arises solely from the ground state degeneracy of the underlying zero mode Hamiltonian. As mentioned previously, the ground state degeneracy of the more general star graph with a spin-\({S_d}\) impurity and \(K\) channels is given by \(g^{S_d}_K = |K - 2{S_d}|+1\).
The cases of \(K=2{S_d}\), \(K<2{S_d}\) and \(K>2{S_d}\) correspond to exactly screened, under-screened and over-screened regimes respectively. The latter two cases correspond to a multiply-degenerate manifold with \(g^S_K > 1\), and simultaneously have non-Fermi liquid phases~\cite{Noz_blandin_1980,Gan_Andrei_Coleman_1993,emery_kivelson,Gan_mchannel_1994,
Tsvelick_Weigmann_mchannel_1984,Tsvelick_weigmann_mchannel_1985,parcollet_olivier_large_N,
kimura_taro_Su_N_kondo,PhysRevB.73.224445,cox_jarrell_two_channel_rev,affleck_1991_overscreen,
Coleman_tsvelik,affleck1993exact,coleman_pepin_2003,roch_nicolas_costi_2009,schiller_avraham_2008,
Durganandini_2011}, while the first regime has a unique ground state with \(g^S_K = 1\) and is 
described by a local Fermi liquid (LFL) phase ~\cite{wilson1975,nozieres1974fermi,Noz_blandin_1980,andreiKondoreview,tsvelickKondoreview}, thereby substantiating the claim that a degeneracy greater than unity is closely tied to non-Fermi liquid physics.

We now comment on the validity of our study of the zero-bandwidth  model at the intermediate coupling fixed point (ICFP) of the MCK  problem. Nozieres and Blandin indeed present an argument for why the  zero-bandwidth model at the strong coupling fixed point ($J =\infty$)  is unstable due to the existence of non-trivial RG relevant quantum  fluctuations~\cite{Noz_blandin_1980}. The resulting RG flow of the Kondo coupling $J$ to  the ICFP resolves these quantum fluctuations. The RG stability of the  Kondo coupling $J$ at the ICFP then allows for a renormalised perturbation  theory approach to studying the fixed point Hamiltonian. Specifically, the  zero-bandwidth model can be safely extracted (in the RG sense) as the zeroth  approximation of a renormalised perturbation theory in the ratio of  the conduction bath hopping amplitude ($t$) to $J$. A study of the  zero-bandwidth model at the ICFP offers, in turn, new perspectives  into the problem, e.g., the ground state degeneracy and it's relation  to the breakdown of exact screening etc., as described in the two preceding paragraphs. It also informs us on the ground state manifold about which to carry out a systematic renormalised perturbative expansion to obtain the low-lying excitations (see Sec.\ref{sec:excitations} below, and Sec.III of the supplementary materials). Moreover, in the following sections, we will show how the inherent quantum-mechanical frustration of singlet order that is present in the Hamiltonian leads to the non-trivial physics of the fixed points in terms of non-Fermi liquid phase, diverging thermodynamic quantities, quantum criticality as well as emergent gauge theories.
\section{Important properties of the star graph}
\label{sec:props_star}

\subsection{Degree of compensation: a measure of the frustration}
\label{sec:deg_of_comp}
\begin{figure}[htpb]
	\centering
	\includegraphics[width=0.45\textwidth]{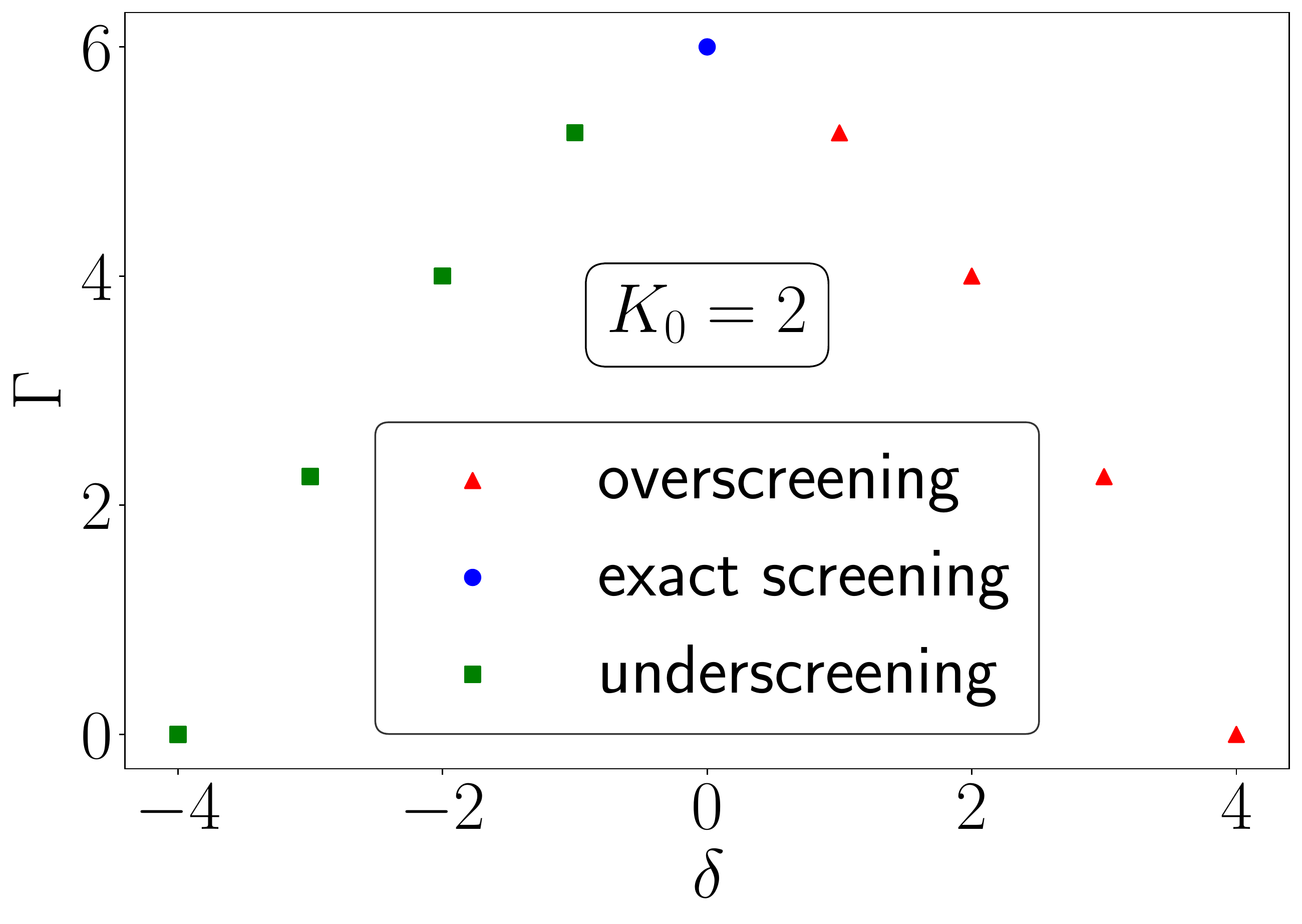}
	\caption{Variation of the degree of compensation $\Gamma$ upon tuning from under-screening to over-screening by varying the deviation from exact screening $\delta=K_{0}/2-S$. The maximum spin compensation occurs at exact-screening \(\delta=0\).}
	\label{compen}
\end{figure}

One can quantify the screening of the local moment at the impurity site by defining a degree of compensation \(\Gamma\). Such a quantity also measures the inherent singlet frustration in the problem: the higher the degree of compensation, the better the spin can be screened into a singlet and lower is the frustration. It is given by the antiferromagnetic correlation existing between the impurity spin and conduction electron channels: $\Gamma \equiv - \left< \vec{S_d}\cdot \vec{S}\right>$.
The expectation value is calculated in the ground state. Since the inner product is simply the ground state energy of a spin-\(S_{d}\) impurity \(K-\)channel MCK model in units of the exchange coupling \(J\), we have \(\Gamma = \frac{1}{2} \left[ l_\mathrm{imp}^2 + l_c^2 - g^S_K\left( g^S_K - 1 \right)\right]\), where \(l_\mathrm{imp}^2 = S_d(S_d+1)\) is the length-squared of the impurity spin. Similarly, \(l_c^2 = \frac{K}{2}\left(\frac{K}{2} + 1\right) \) is the length-squared of the total conduction bath spin. \(g^S_K = |K - S_d| + 1\) is the ground state degeneracy. We will explore the three regimes of screening by defining \(K = K_0 + \delta, S_{d} = \frac{K_0}{2} - \delta\). \(\delta=0\) represents the exactly-screened case of \(K = 2S_{d} = K_0\). Non-zero \(\delta\) represents either over- or under-screening. In terms of \(K_0\) and \(\delta\), the degree of compensation becomes
\begin{eqnarray}
	\label{gamma}
	\Gamma = \frac{1}{4}\left[\left( K_0 + 1 \right) ^2 - \left(|\delta| + 1 \right) ^2\right]~.
\end{eqnarray}
{For a given \(K_0\), the degree of compensation $\Gamma$ is maximised for exact screening \(\delta=0\), and is reduced for \(\delta \neq 0\) (see Fig.\ref{compen}). This shows the inability of the system to form a unique singlet ground state and reveals the quantum-mechanical frustration inherent in the zero mode Hamiltonian and therefore in the entire problem.} The degree of compensation is symmetric under the Hamiltonian transformation \(\delta \to -\delta\), and this represents a duality transformation between over-screened and under-screened MCK models. This topic will be discussed in more detail later.

\subsection{Impurity magnetization and susceptibility}
\label{sec:imp_suscept}
The impurity magnetic susceptibility can be obtained by diagonalising the zero-mode Hamiltonian \(H(h)\) in the presence of a local magnetic field \(h\) on the impurity:
\begin{eqnarray}
	\label{star graph_field_hamiltonian}
	H(h) = {\mathcal{J}^*} \vec{S_d}\cdot\vec{S} + h S_d^z~.
\end{eqnarray}
The Hamiltonian commutes with \(\vec S\), so it is already block-diagonal in terms of the eigenvalues \(M\) of \(S\). \(M\) takes values in the range \(\left[M_\mathrm{min}, M_\mathrm{max}\right]\), where \(M_\mathrm{max} = K/2\) for a \(K-\)channel Kondo model, and \(M_\mathrm{min} = 0\)  if \(K\) is even, otherwise \(\frac{1}{2}\). The diagonalisation of the Hamiltonian is shown in the Supplementary Materials~\cite{SM}. Defining \(\alpha = \frac{1}{2}\left({\mathcal{J}}m + h\right) + \frac{{\mathcal{J}}}{4}\) and \(x^M_m = M(M+1) - m(m+1)\), the partition function can be written as
\begin{eqnarray}
	Z(h) =\sum_{M=M_\mathrm{min}}^{M_\mathrm{max}}r^K_M\bigg[&\sum_{m=-M, \atop{m\in \mathbb{Z}}}^{M-1}2e^{\beta \frac{{\mathcal{J}}}{4}}\cosh \beta\sqrt{{\mathcal{J}}^2x^M_m/4 + \alpha^2} \nonumber\\
& + 2e^{-\beta {\mathcal{J}}M/2}\cosh \beta h/2\bigg]~.
\end{eqnarray}
Here, \(\beta = \frac{1}{k_B T}\), \(M\) sums over the eigenvalues of \(S\) while \(m\) sums over \({\mathcal{J}}^z - \frac{1}{2}\) and the additional degeneracy factor \(r^K_M= {}^{K-1}C_{K/2 - M}\) arises from the possibility that there are multiple subspaces defined by \(S=M\). 
To calculate the impurity magnetic susceptibility, we will use the expression
\begin{eqnarray}
	\chi = \frac{1}{\beta}\lim_{h \to 0}\left[\frac{Z^{\prime\prime}(h)}{Z(h)} - \left(\frac{Z^{\prime}(h)}{Z(h)}\right)^2 \right] ~,
\end{eqnarray}
where the \(\prime\) indicates derivative with respect to \(h\). At low temperatures, the susceptibility takes the form
\begin{eqnarray}
	\chi \to \frac{\beta\Sigma_\mathrm{max}}{2M_\mathrm{max}\left(2M_\mathrm{max}+1\right)^2} = \frac{\beta(K-1)}{12(K+1)} \sim \frac{1}{T}~.
\end{eqnarray}

\begin{figure}[!htpb]
\centering
\includegraphics[width=0.49\textwidth]{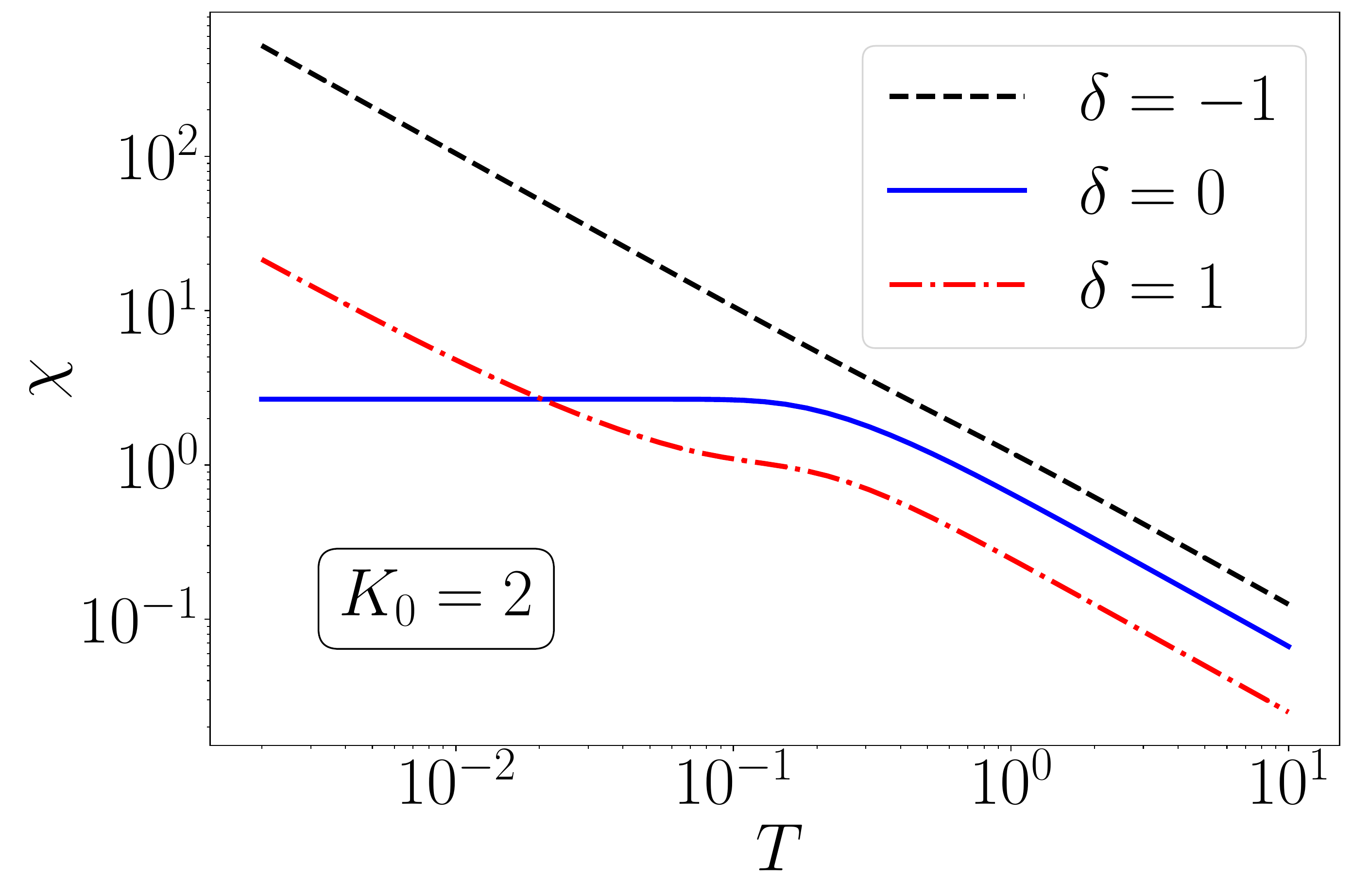}
\caption{Variation of impurity susceptibility against temperature. The exactly screened case ($\delta=0$) saturates to a constant value at low temperatures, indicating complete screening. The cases of inexact screening show a divergence of the susceptibility, which means there is a remnant local spin at the impurity. Since the axes are in log scale, the behaviour is $\log T \sim -\log \chi$ which translates to $\chi \sim 1/T$.  }
\label{fig:suseptibility_impurity}
\end{figure}

The \(\chi\) is seen to diverge as \(T^{-1}\) at low temperatures.
Such a non-analyticity in a response function is in contrast to the behaviour in the exactly-screened fixed point where the ground state is unique.
There, the susceptibility becomes constant at low temperatures: \(\chi(T\to 0) = \frac{W}{4 T_K}\), \(T_K\) being the single-channel Kondo temperature and \(W\) the Wilson number \cite{wilson1975renormalization,nozieres1974fermi,bullaNRGreview,kondo_urg}.
In Fig.~\ref{fig:suseptibility_impurity}, we have checked the case of general spin-\(S_{d}\) impurity by diagonalising the star graph Hamiltonian numerically, and the general conclusion is that all exactly-screened models show a constant impurity susceptibility at \(T \to 0\), while the over-screened and under-screened cases show a diverging impurity susceptibility in the same limit.
A similar divergence is also seen in the susceptibility of the outer spins, calculated by inserting a magnetic field purely on the outer spins. Some additional results regarding susceptibility are presented in the Supplementary Materials~\cite{SM}.

Departure from single-channel behaviour is most apparent in the presence of the highly polarised states \(\ket{J^z=\pm \frac{K-1}{2}}\) in the ground-state spectrum,
\begin{eqnarray}
	\ket{J^z=\sigma \frac{K-1}{2}} =& \frac{1}{\sqrt{1 + K}} \ket{\sigma}_d\otimes\ket{S^z=\sigma(\frac{K}{2}-1)} \nonumber\\
					&- \frac{\sqrt K}{\sqrt{1 + K}}\ket{\bar\sigma}_d\otimes\ket{S^z=\sigma\frac{K}{2}},
\end{eqnarray}
where \(\sigma=\pm 1\) and \(\ket{\sigma}_d\) represents the up and down configurations of the impurity spin and \(\ket{S^z}\) represents the configuration of the bath spins. These states lead to large unquenched magnetisation (\(m(h=0^\pm)\)) at zero temperatures
that displays discontinuous behaviour in the limit of a vanishing field $h\to 0\pm$. This can be seen as follows. In the presence of a field, the ground-state becomes unique: \(\ket{\psi}_\mathrm{gs}(h) = \theta(h)\ket{J^z=\frac{K-1}{2}} + \theta(-h)\ket{J^z=-\frac{K-1}{2}}\), and the impurity magnetisation can be calculated from the state:
\begin{eqnarray}
	m(h=0^\pm) &= \theta(h)\bra{J^z=\frac{K-1}{2}} S_d^z \ket{J^z=\frac{K-1}{2}} \nonumber\\
		   &+ \theta(-h)\bra{J^z=-\frac{K-1}{2}}S_d^z\ket{J^z=-\frac{K-1}{2}}\nonumber\\
	&= \pm  \frac{1}{2} \frac{1-K}{1 + K}\label{imp-magnetisation}~.
\end{eqnarray}

\begin{figure}[!htpb]
	\centering
	\includegraphics[width=0.49\textwidth]{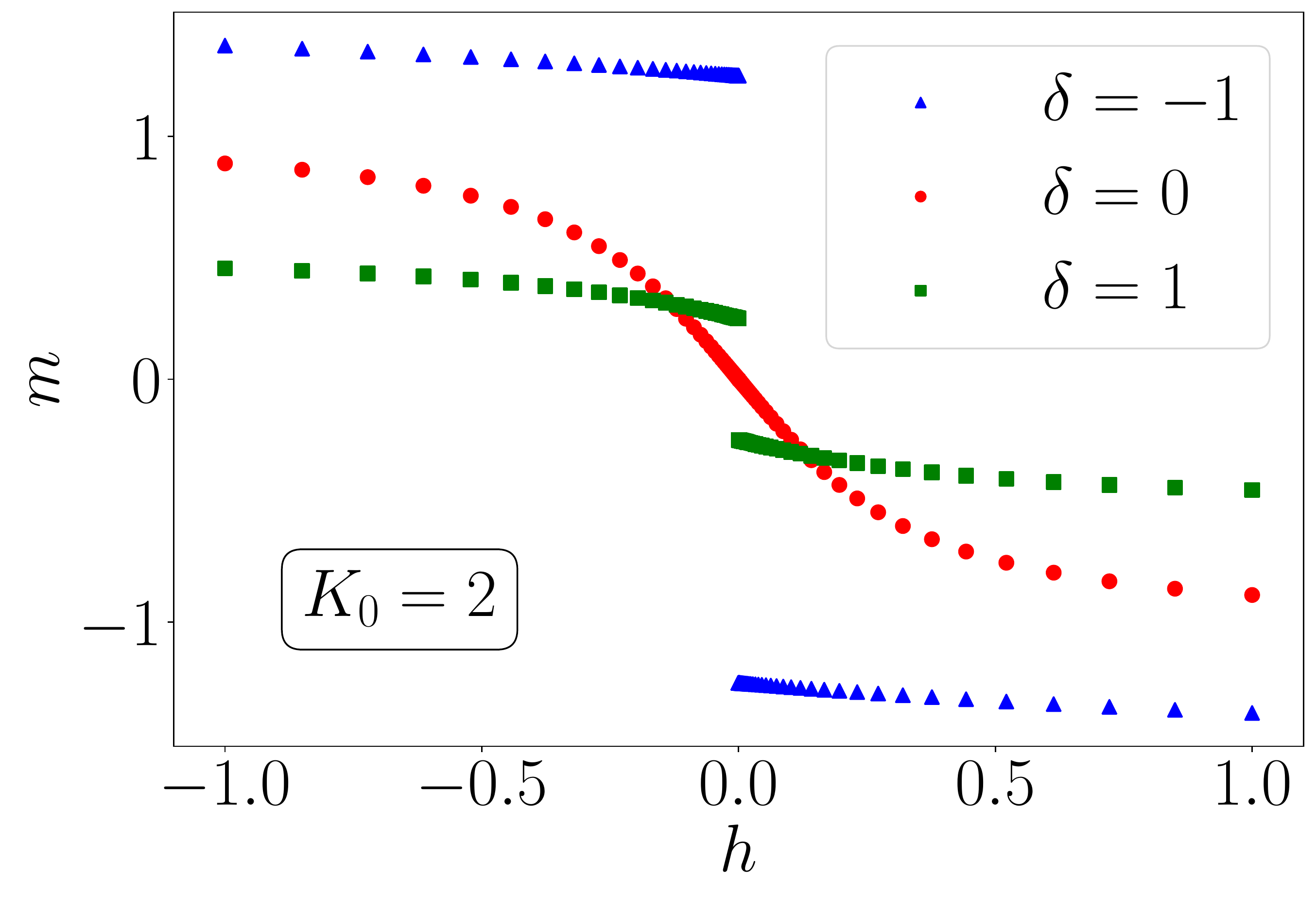}
	\caption{Behaviour of the impurity magnetization for three values of \(\left(K, 2S\right) = \left(2, 4\right), \left( 3,3 \right), \left(4, 2\right)  \). Only the case of \(K=2S=3 \left(\delta=0\right)\) is analytic near zero. The non-analyticity of the other cases arises because of the frustration brought about by the degeneracy of the star graph ground state.}
	\label{mag_crit}
\end{figure}

The magnetization is therefore discontinuous as \(h\to 0\pm\) for \(K>1\). The non-analyticity occurs because the magnetic field is able to flip the states with \(J^z= (K-1)/2\) into the state of the opposite magnetisation and vice versa. The available space for scattering is simply the frustration that we discussed earlier. Indeed, we have checked (see Fig.~\ref{mag_crit}) by diagonalising numerically the star graph Hamiltonian that the non-analyticity exists for all \(\delta \neq 0\), where \(\delta = \frac{K}{2} - S_{d}\) is the deviation from exact screening. 

These polarised states can also be used to obtain the scattering phase shift suffered by the conduction channels at the Fermi surface. The gapless spin-flip fluctuations between the impurity and the bath lead to the delocalisation of the impurity degree of freedom into the conduction bath, specifically at the Fermi surface. This results in an "excess charge" being contributed by the impurity to the momentum states. As shown in the Supplementary Materials~\cite{SM}, within the polarised states, the impurity contributes an excess charge of \(n_\mathrm{exc}^{(l)} = 1/K\) to each conduction channel. Following Friedel's sum rule, this leads to a total scattering phase shift \((\Delta^{(l)})\) of \(\pi/K\) for each channel:
\begin{eqnarray}\label{phase-stargraph}
	\Delta^{(l)} = \pi n_\mathrm{exc}^{(l)} = \frac{\pi}{K}, ~ ~ l \in \left[1, K\right] ~.
\end{eqnarray}
For \(K > 1\), each channel therefore experiences an unrenormalised phase shift of value between \(0\) and \(\pi\) that, as we will show in a later section, is responsible for the non-Fermi liquid behaviour observed at the ICFP. For example, we will show that these phase shifts are linked to the critical non-Fermi liquid exponents of various thermodynamic quantities at the fixed point. That this phase shift is tied to the frustration that leads to incomplete Kondo screening at the ICFP can be evidenced 
through its relation to the impurity magnetisation \(m (h=0\pm)\) obtained in eq.~\ref{imp-magnetisation}:
\begin{eqnarray}\label{magnetisation-phase}
	m(h=0^\pm) = \pm \frac{1}{2} \frac{\Delta^{(l)} - \pi}{\Delta^{(l)} + \pi}~.
\end{eqnarray}
When the phase shift \(\Delta^{(l)}\) acquires a fractional value (in units of \(\pi\)) at the ICFP, the impurity magnetisation also becomes fractional (in units of \(\frac{1}{2}\)). We will see shortly that this unrenormalised phase shift value of \(0 < \delta^{(l)} < \pi\)~\cite{anderson_1990} is also at the heart of 
an orthogonality catastrophe, i.e., a dramatic change in the ground-state wavefunction of the conduction bath degrees of freedom as the impurity is introduced among them~\cite{anderson1967infrared,varma2002singular}.

\subsection{Topological properties of ground state manifold}
\label{sec:topo_ground_state}
We now present the non-local twist and translation operators which can be used to explore the degenerate ground state manifold of the star graph model.
We begin by defining two operators $\hat{T}$ and $\hat{O}$, which we will call the translation and twist operators respectively: 
\begin{eqnarray}
\hat{T} = e^{i\frac{2\pi}{K} \hat{\Sigma}} ~,~\hat{O} = e^{i\hat{\phi}}~,~~\hat{\Sigma}=[\hat{J}^z-(K-1)/2]~.
\end{eqnarray}
One can see that the generators of thee above operators commute with the Hamiltonian: $[H,J^x]=[H,J^z]=[H,J^z]=0$. In the large \(K\) limit, we can perform a semi-classical approximation. 
Thus, we define
\begin{eqnarray}
e^{i\hat{\phi}} = \frac{J_{x}}{\sqrt{J^{2} - J_{z}^{2}}} + i\frac{J_{y}}{\sqrt{J^{2} - J_{z}^{2}}}~,
\end{eqnarray}
such $[e^{i\hat{\phi}},\hat{H}]=0$. We can label the ground states by the eigenvalues of the translation operator $\hat{T}$. We can also use the ground states labeled by the eigenvalues of $J^z$ (say $M$). The ground states are now written as 
\begin{eqnarray}
|M_1\rangle,|M_2\rangle,|M_2\rangle,\cdots ,|M_K\rangle~.
\end{eqnarray}
The operations of the translation operators on these states are then given by
\begin{eqnarray}
\hat{T}|M_i\rangle = e^{i\frac{2\pi}{K} [M_i-(K-1)/2]} |M_i\rangle = e^{i2\pi\frac{p_i}{K} } |M_i\rangle~,~p_i\in[K]~.
\end{eqnarray}
The braiding rule between the twist and the translation operators is
\begin{eqnarray}
\hat{T}\hat{O}\hat{T}^{\dagger}\hat{O}^{\dagger} = e^{\frac{2\pi }{K}[i\hat{\Sigma},i\hat{\phi}]}=e^{i\frac{2\pi }{K}} ~,\nonumber\\
\hat{T}\hat{O}^m\hat{T}^{\dagger}\hat{O}^{\dagger m} = e^{\frac{2\pi }{K}[i\hat{\Sigma},im\hat{\phi}]}=e^{i2\pi \frac{m}{K}}~.
\end{eqnarray}
The states $\hat{O}^m |M_i\rangle$ (with distinct $m$) are labelled by different eigenvalues of the translation operations, and are thus orthogonal to each other
\begin{eqnarray}
\hat{T} \hat{O}^m |M_i\rangle = \hat{O}^m \hat{T} e^{i2\pi \frac{m}{K}} |M_i\rangle =\hat{O}^m e^{i \frac{2\pi(m+p_i)}{K} } |M_i\rangle \\
\hat{T} \bigg(\hat{O}^m |M_i\rangle \bigg) = e^{i \frac{2\pi(m+p_i)}{K} } \bigg( \hat{O}^m |M_i\rangle \bigg)~.
\end{eqnarray}
As the twist operator $\hat{O}$ commutes with the Hamiltonian, we can write
\begin{eqnarray}
\langle M_i| \hat{O}^{m \dagger}  H \hat{O}^m |M_i\rangle =\langle M_i|  H |M_i\rangle ~.
\end{eqnarray}
Thus, the energy eigenvalues of all the orthogonal states are equal, and they form the $K$ fold degenerate ground state subspace. By the application of the twist operator $\hat{O}$, we can go from one degenerate ground state to another. We note that a similar exploration of the degenerate ground state manifold of spin-1/2 Heisenberg antiferromagnets on geometrically frustrated lattices via twist and translation operators has been conducted in Refs.\cite{pal2019magnetization,pal2020topological}. 

\subsection{Local Mott liquid}
\label{sec:loc_mott_liquid}
In order to obtain the effective theory for the K zero modes, we will now use the URG method to decouple the impurity site from the star graph. The starting point is the star graph Hamiltonian
\begin{eqnarray}
H = {\mathcal{J}} \vec{S}_d.\displaystyle\sum_i \vec{S}_i ={\mathcal{J}} S_d^zS^z + \frac{{\mathcal{J}}}{2} (S_d^+S^-+ S_d^-S^+) ~,\nonumber\\
H_D = {\mathcal{J}} S^z_d S^z~,~~ H_X = \frac{{\mathcal{J}}}{2} (S_d^+S^-+ S_d^-S^+)~.
\end{eqnarray}
In order to remove the quantum fluctuations between the impurity spin and the rest, we perform one step of URG:
\begin{eqnarray}
\Delta H = H_X ({\hat{\omega}-H_D})^{-1} H_X ~.
\end{eqnarray}
The total bath spin operator \(S^z\) is not a good quantum number for the zero mode ground state, and because there is no net $S^z$ field, the ground state manifold has a vanishing expectation value of $S^z$, $\langle S^z \rangle=0$. We use this expectation value to replace the denominator of the above RG equation.
\begin{eqnarray}
\beta_{\uparrow} ({\mathcal{J}},\omega_{\uparrow}) = ({\mathcal{J}}^2 \Gamma_{\uparrow})/2 ~,~~\Gamma_{\uparrow}=(\omega_{\uparrow}-{\mathcal{J}}(S_d^z-1))^{-1}~.\qquad
\end{eqnarray}
The effective Hamiltonian is therefore
\begin{eqnarray}
H_{eff} 
 =\frac{\beta_{\uparrow}({\mathcal{J}},\omega_{\uparrow})}{4} (S^+S^-+S^-S^+)   ~.
\end{eqnarray}
In terms of the electronics degree of freedom, this can be written as
\begin{eqnarray}
	\frac{ \beta_{\uparrow}({\mathcal{J}},\omega_{\uparrow}) }{4}  \sum_{i\neq j}\sum_{\alpha_i,\beta_i}\sum_{\alpha_j,\beta_j}&\left(\vec{\sigma}_{\alpha_i\beta_i}\vec{\sigma}_{\alpha_j\beta_j}  c_{0\alpha_i}^{(i)\dagger}  c_{0\beta_i}^{(i)} c_{0\alpha_j}^{(j)\dagger}  c_{0\beta_j}^{(j)} \right.\nonumber\\
																   &\left. +\textrm{h.c.}\right) ~.
\label{eq:all-to-all_1}
\end{eqnarray}
The case $\frac{\beta_{\uparrow}({\mathcal{J}},\omega_{\uparrow})}{2} <0$ leads to a ferromagnetic effective Hamiltonian. In this case, the ground state is realised for $S$ being maximum and $S^z$ being minimum. The effective Hamiltonian can be rewritten in this case as
\begin{eqnarray}
H_{eff}   =-|\beta_{\uparrow}({\mathcal{J}},\omega_{\uparrow})| \times(S^+S^{-}+ S^-S^{+}) /4   ~.
\end{eqnarray}
The complete set of commuting observables for this Hamiltonian contains $H,S,S^z$. In the ground state, $S$ is maximum, and  therefore $S=K/2$, and $S^z$ can take $2S+1=K+1$ values. Defining the dual operator \(\hat \phi\) with the algebra \([\hat{\phi,S^z}]=i\), we get the twist operator $\hat{Q}=\exp(i\hat{\phi}\Phi/\Phi_0)$. Applying this operator, we get the twisted Hamiltonian 
\begin{eqnarray}
	H(\Phi) = -\frac{|\beta_{\uparrow}({\mathcal{J}},\omega_{\uparrow})|}{2} \left[S^2 - \left(S^{z}-\frac{\Phi}{\Phi_0} \right)^2\right] ~.
\label{eq:flux_spectral}
\end{eqnarray}
One can thus explore different $S^z$ ground states via this spectral flow (flux insertion) argument.

For a $K$ channel star graph problem the ground state is $K$ fold degenerate associated with different $J^z$ values $\{-(K-1)/2,-(K-3)/2,\cdots  ,(K-1)/2 \}$. After removing the quantum fluctuations between the impurity spin and the outer spins, we get an all-to-all model (eq.\eqref{eq:flux_spectral}) as the effective Hamiltonian. The eigenstates of this all-to-all Hamiltonian can be labeled by the eigenvalues of $S^z$. There are $K+1$ such states made out of only the outer spins. The total state including the impurity spin can be written as $|J^z\rangle = |S_d^z\rangle \otimes |S^z\rangle$ labeled by $J^z=S_d^z+S^z$. As the all-to-all model has $Z_2$ symmetry in impurity sector, there are $2(K+1)$ total states. For $S_d^z=1/2$, $ \{J^z\}=\{ -(K-1)/2, \cdots, (K-1)/2, (K+1)/2 \}$ and for $S_d^z=-1/2$, $\{J^z\}=\{-(K+1)/2, -(K-1)/2, \cdots,  (K-1)/2  \}$. We can see that in both the cases, all $K$ states of the star graph ground state manifold are present in the spectrum of the all-to-all model, one of them being the ground state. Using the twist operator, we can cycle between the various states. Some additional topological features of the Mott liquid are presented in the Supplementary Materials~\cite{SM}.

\section{Local non-Fermi liquid excitations of the 2CK model}
\label{sec:excitations}

\subsection{Effective Hamiltonian}
\label{sec:eff_ham_2ch}

We will now proceed to extract the nature of the low-energy excitations of the two-channel Kondo problem. This will be done by treating real-space hopping as a perturbation to the zero-bandwidth Hamiltonian, as it accounts for the lowest-energy electronic quantum fluctuations in the conduction bath~\cite{nozieres1974fermi}. The zero-bandwidth Hamiltonian obtained from the URG of the two-channel Kondo problem then acts as the zeroth-level Hamiltonian
\begin{eqnarray}
H^{(0)}_{2CK}= {\mathcal{J}} \vec{S}_d.(\vec{S}_1+\vec{S}_2)~~,~~\vec{S}_i =  \frac{1}{2}~ c_{i,\alpha}^{\dagger}~ \vec{\sigma}_{\alpha,\beta}~ c_{i,\beta}~.
\label{eq:channel-2-spin}
\end{eqnarray}
Here, $\vec{S}_i=1,2$ represents the spin degree of freedom present at the origin of the $i^{th}$ channel. This zeroth-level Hamiltonian has two degenerate ground states labeled by $\ket{J^z=\pm 1/2}$ with energy $-\mathcal{J}$ and $6$ excited states~\cite{kunj_slal_1999}.
The perturbation Hamiltonian is the real-space hopping:
\begin{eqnarray}
H_{X} = -t \displaystyle\sum_{<1,l_1>}\sum_{<2,l_2>} (c^{\dagger}_{1,\sigma}c_{l_1,\sigma}+c^{\dagger}_{2,\sigma}c_{l_2,\sigma}+ ~\textrm{h.c.}),
\end{eqnarray}
and its effects will now be accounted for using degenerate perturbation theory. Here $l_i$ represents the nearest site to the origin of the $i^{th}$ channel.

Since the perturbation is a single-particle hopping that transfers electrons, we need to rewrite the spin operators in the zeroth Hamiltonian in terms of fermionic operators:
\begin{eqnarray}
\mathcal{H} = \frac{{\mathcal{J}}\hbar}{2}~ \vec{S}_d. \displaystyle\sum_{i=\{1,2\}} \displaystyle\sum_{\alpha,\beta\in\{\uparrow,\downarrow\}}c_{i\alpha}^{\dagger} \vec{\sigma}_{\alpha\beta} c_{i\beta} + H_X~.
\label{eq:excitation_hamiltonian}
\end{eqnarray}
In this expanded basis, there are $32$ degenerate ground states:
\begin{eqnarray}
\label{gstate-set}
\left\{ \ket{\alpha_i} \right\} = \left\{ \ket{J^z}\otimes\ket{n_{l_1,\uparrow},n_{l_1,\downarrow},n_{l_2,\uparrow},n_{l_2,\downarrow}} \right\}~.
\end{eqnarray}
The \(32-\)fold degeneracy arises from the two-fold degeneracy of \(J^z\) (\(J^z = \pm 1/2\)) multiplying the \(16-\)fold degeneracy of the rest of the sites \(n_{l_i,\sigma}\) (each of the four \(n_{l_i,\sigma}\) can be 0 or 1, leading to a \(2^4\)-fold degeneracy).

The first and the second order corrections to the low energy effective Hamiltonian are
\begin{eqnarray}
H^{(1)} = \sum_{ij} |\alpha_i\rangle \langle \alpha_i  | V| \alpha_j \rangle \langle \alpha_j |~,\nonumber\\
H^{(2)} = \sum_{ij} \sum_l |\alpha_i\rangle \frac{\langle \alpha_i  | V| \mu_l \rangle \langle \mu_l  | V| \alpha_j \rangle}{E_0-E_{l}}\langle \alpha_j |~.
\end{eqnarray}
Here, $|\alpha_i\rangle$ represents the ground states with energy $E_0$ and is the set of states defined in eq.~\ref{gstate-set}, and $\mu_l$ represents the excited states with energy $E_l$. It is easy to see that the diagonal contribution at any odd order is zero - the final state is never equal to the initial state. The off-diagonal part at first order is also zero. At second order we get both diagonal and off-diagonal contributions to the effective Hamiltonian. 

\subsubsection{Diagonal renormalisation}
We first calculate the diagonal renormalisation at second order coming from each of the states \(\ket{\tilde \alpha_0} = \ket{J^z = \frac{1}{2}}\otimes\ket{\psi_\mathrm{rest}}, \ket{\tilde \alpha_1} = \ket{J^z = -\frac{1}{2}}\otimes\ket{\psi_\mathrm{rest}}\), where \(\ket{\psi_\mathrm{rest}}\) refers to the configuration of the remaining sites apart from the two that form the zeroth Hamiltonian. These second order corrections are given by
\begin{eqnarray}
H^{(2)}_\mathrm{diag}\left(\tilde \alpha_0\right)  = \frac{2t^2}{E_0} \hat{I} - \Omega_l, ~ ~ H^{(2)}_\mathrm{diag}\left(\tilde \alpha_1\right) = \frac{2t^2}{E_0} \hat{I} + \Omega_l ~,
\end{eqnarray}
where $S_{l_i}^z=(n_{l_i\uparrow}-n_{l_i\downarrow})/2$ and $\Omega_l=\frac{2t^2}{3E_0} ( S_{l_1}^z + S_{l_2}^z)$.  The total diagonal second-order contribution to the effective Hamiltonian is obtained by adding these two contributions, and the result is a trivial shift: $H^{(2)}_\mathrm{diag} = H^{(2)}_\mathrm{diag}\left(\tilde \alpha_0\right) + H^{(2)}_\mathrm{diag}\left(\tilde \alpha_1\right) = -4t^2 \hat{I}$. The fact there is no non-trivial diagonal renormalisation to the low-energy excitations shows the absence of (local) Fermi liquid terms.

\subsubsection{Off-diagonal renormalisation}
The off-diagonal contribution to the second order effective Hamiltonian is obtained purely from \(H^{(2)}_{od}\):
\begin{eqnarray}
	H_{NFL} &= \sum_{i\neq j} \sum_l |\alpha_i\rangle \frac{\langle \alpha_i  | V| \mu_l \rangle \langle \mu_l  | V| \alpha_j \rangle}{E_0-E_{l}}\langle \alpha_j | \nonumber\\
			       &= -\frac{8t^2}{3} [ (S_1^z)^2 c_{2\uparrow}^{\dagger}c_{2\downarrow}  (  c_{l_1\uparrow}c_{l_1\downarrow}^{\dagger} +  c_{l_2\uparrow}c_{l_2\downarrow}^{\dagger}  ) \nonumber\\
&+ (S_2^z)^2 c_{1\uparrow}^{\dagger}c_{1\downarrow}  (  c_{l_1\uparrow}c_{l_1\downarrow}^{\dagger} +  c_{l_2\uparrow}c_{l_2\downarrow}^{\dagger}  ) ] + \textrm{h.c.} ~.
\label{eq:hamiltonian_NFL}
\end{eqnarray}
This effective Hamiltonian is purely of the non-Fermi liquid kind, arising due to the degeneracy of the ground state manifold. Using this low energy effective Hamiltonian, we have calculated different thermodynamic quantities like susceptibility, specific heat and the Wilson ratio. The first two measures show logarithmic behavior at low temperatures, in agreement with known results in the literature~\cite{Noz_blandin_1980,Gan_Andrei_Coleman_1993,emery_kivelson,Gan_mchannel_1994,
Tsvelick_Weigmann_mchannel_1984,Tsvelick_weigmann_mchannel_1985,parcollet_olivier_large_N,
kimura_taro_Su_N_kondo,PhysRevB.73.224445,cox_jarrell_two_channel_rev,affleck_1991_overscreen,
Coleman_tsvelik,affleck1993exact,coleman_pepin_2003,roch_nicolas_costi_2009,schiller_avraham_2008,
Durganandini_2011}. These results are shown in the Supplementary Material~\cite{SM}. 

Following this method, we have also calculated the effective Hamiltonian for the three channel Kondo model. We again find the absence of all Fermi liquid terms up to $2^{nd}$ order, and the presence of non-Fermi liquid terms in the off-diagonal part of the effective Hamiltonian, and these are also shown in the Supplementary Material~\cite{SM}.

\subsection{Local marginal Fermi liquid and orthogonality catastrophe}
\label{sec:MFL}

The real space local low energy Hamiltonian that takes into account the excitations above the ground state is given by eq.~\ref{eq:hamiltonian_NFL} and can be written as
\begin{eqnarray}
	\label{nfl_terms}
	V_\mathrm{eff} =& \frac{2t^2}{{\mathcal{J}^*}}\left[\left(\sigma^z_{0,1}\right)^2 s^+_{0,2} + \left(\sigma^z_{0,2}\right)^2 s^+_{0,1}\right] \left(s^-_{1,1} + s^-_{1,2}\right) \nonumber\\
			&+ \mathrm{h.c.};
\end{eqnarray}
here, \(\sigma^z_{0,l} = \hat n_{0\uparrow,l} - \hat n_{0\downarrow,l}, s^+ = c^\dagger_{0 \uparrow,l}c_{0 \downarrow,l}\) and \(s^- = \left(s^+\right)^\dagger\). The notation \(0\sigma,l\) has the site index \(i=0,1,2,\ldots\) as the first label, the spin index \(\sigma=\uparrow,\downarrow\) as the second label and the channel index \(l=1,2\) as the third label.
Such non-Fermi liquid (NFL) terms in the effective Hamiltonian and the absence of any Fermi-liquid term at the same order should be contrasted with the local Fermi liquid excitations induced by the singlet ground state of the single-channel Kondo model~\cite{nozieres1974fermi,wilson1975renormalization,hewson1993}.
We now take the MCK Hamiltonian to strong-coupling, and perform a perturbative treatment of the hopping.
At \(J \to \infty\), the perturbative coupling \(t^2/J\) is arbitrarily small and we again obtain Eq.~\ref{nfl_terms}. Such a change from the strong coupling model with parameter \(J\) to a weak coupling model with parameter \(t^2/J\) amounts to a duality transformation~\cite{kroha_kolf_2007,zitko_fabrizio_2017}.
It can be shown that the duality transformation leads to an identical MCK model~\cite{kroha_kolf_2007} (self-duality), which implies we can have identical RG flows, and our transformation simply extracts the NFL piece from the dual model.
The self-duality also ensures that the critical intermediate-coupling fixed point is unique and can be reached from either of the models.

The diagonal part of eq.~\ref{nfl_terms} is
\begin{eqnarray}
	V_\mathrm{eff} = \frac{2t^2}{J}\sum_{l=1,2}\left(\sum_\sigma \hat n_{0\sigma,l}\right) s^+_{0,\bar l}s^-_{1,\bar l} + \mathrm{h.c.}~,
\end{eqnarray}
where \(\bar l = 3 - l\) is the channel index complementary to \(l\). We will Fourier transform this effective Hamiltonian into \(k-\)space. The NFL part becomes
\begin{eqnarray}
	\label{k_space_od}
	\frac{2t^2}{J}\sum_{\left\{k_i,k_i^\prime\right\}}\sum_{\sigma, l} &\left(e^{i\left(k_1 - k_1^\prime\right)a}c^\dagger_{k\sigma,l}c_{k^\prime\sigma,l}c^\dagger_{k_2 \uparrow, \bar l}c_{k_2^\prime \downarrow,\bar l}c^\dagger_{k_1 \downarrow,\bar l}c_{k_1^\prime \uparrow, \bar l}\right.\nonumber\\
									   &\left. + \mathrm{h.c.}\right) ~.\qquad
\end{eqnarray}
Such a three particle interaction term was also obtained for the NFL phase of the 2D Hubbard model from a URG treatment (see Appendix B of Ref.\cite{anirbanmott1}). The channel indices in Eq.~\ref{k_space_od} can be mapped to the normal directions in~\cite{anirbanmott1}. The 2 particle-1 hole interaction in Eq.~\ref{k_space_od} has a diagonal component which can be obtained by setting \(k=k^\prime, k_1 = k_2^\prime\) and \(k_2 = k_1^\prime\):
\begin{eqnarray}
	H_\mathrm{MFL} &= \sum_{k, k_1, \atop{k_2,\sigma,  l}} \frac{2t^2e^{i\left(k_1 - k_2\right)a}}{J} \hat n_{k\sigma,l} \hat n_{k_2 \uparrow, \bar l}\left(1 - \hat n_{k_1 \downarrow,\bar l}\right) + \mathrm{h.c.}\nonumber\\
	&= \sum_{k, k_1, \atop{k_2,\sigma,  l}} \frac{4t^2}{J} \cos a\left(k_1 - k_2\right)  \hat n_{k\sigma,l} \hat n_{k_2 \uparrow, \bar l}\left(1 - \hat n_{k_1 \downarrow,\bar l}\right).\qquad 
\end{eqnarray}
The most dominant contribution comes from \(k_1 = k_2 = k^\prime\), revealing the non-Fermi liquid metal~\cite{cox_jarrell_two_channel_rev,andrei_jerez_1995}:
\begin{eqnarray}
	\label{mfl_large}
	H^*_\mathrm{MFL} = \frac{4t^2}{J} \sum_{\sigma, k, k^\prime, l} \hat n_{k\sigma,l} \hat n_{k^\prime \uparrow, \bar l}\left(1 - \hat n_{k^\prime \downarrow,\bar l}\right)~.
\end{eqnarray}
A non-local version of this effective Hamiltonian was found to describe the normal phase of the Mott insulator of the 2D Hubbard model, as seen from a URG analysis~\cite{anirbanmott1,anirbanmott2}. Following~\cite{anirbanmott1}, one can track the RG evolution of the dual coupling \(R_j = \frac{4t^2}{J_j}\) at the \(j^\mathrm{th}\) RG step, in the form of the URG equation
\begin{eqnarray}
	\Delta R_j =- \frac{R_j^2}{\omega - \epsilon_{j}/2 - R_j/8}~.
\end{eqnarray}
In the RG equation, \(\epsilon_{j}\) represents the energy of the \(j^\mathrm{th}\) isoenergetic shell. It is seen from 
the RG equation that \(R\) is relevant in the range of \(\omega < \frac{1}{2}\epsilon_j\) that has been used throughout, leading to a fixed-point at \(R^*/8 = \omega - \frac{1}{2}\epsilon^*)\). The relevance of \(R\) is expected because the strong coupling \(J\) is irrelevant and \(R \sim 1/J\).

The renormalisation in \(R\) leads to a renormalisation in the single-particle self-energy~\cite{anirbanmott1}. The \(k-\)space-averaged self-energy renormalisation is
\begin{eqnarray}
	\Delta \Sigma(\omega) = \rho {R^*}^2\int_0^{\epsilon^*} \frac{d\epsilon_j}{\omega - \epsilon_j/2 + R_j/8}~.
\end{eqnarray}
The density of states can be approximated to be \(N^*/R^*\), where \(N^*\) is the total number of states over the interval \(R^*\). As suggested by the fixed point value of \(R_j\), we can approximate its behaviour near the fixed point by a linear dependence on the dispersion \(\epsilon_j\). The two limits of the integration are the starting and ending points of the RG. We start the RG very close to the Fermi surface and move towards the fixed point \(\epsilon^*\). Near the starting point, we substitute \(\epsilon = 0\) and \(R = \omega\), following the fixed point condition. From the fixed point condition, we also substitute \(R^*/8 = \omega - \frac{1}{2}\epsilon^*\). On defining \(\bar \omega = N^* \left(\omega - \frac{1}{2}\epsilon^*\right)\), we can write
\begin{eqnarray}
	\label{self_energy}
	\Delta \Sigma(\omega) \sim  \bar \omega \ln \frac{N^* \omega}{\bar \omega}~.
\end{eqnarray}
The self-energy also provides the quasiparticle residue for each channel\cite{anirbanmott1}:
\begin{eqnarray}
	Z(\bar\omega) = \left(2 - \ln \frac{2\bar\omega}{N^* \omega}\right) ^{-1}~.
\end{eqnarray}
As \(\omega \to 0\), the \(Z\) vanishes, implying that the ground state is {\it not adiabatically connected} to the Fermi gas in the presence of the NFL terms.
This is the orthogonality catastrophe~\cite{varma2002singular,anderson_infraredcat,yamada_catastrophe,yamada1979orthogonality} in the two-channel Kondo problem, and it is brought about by the presence of the channel-non diagonal terms in Eq.~\ref{mfl_large}.
Such terms were absent in the single-channel Kondo model, because there was no multiply-degenerate ground state manifold that allowed scattering.
This line of argument shows that the extra degeneracy of the ground state subspace and the frustration of the singlet order that comes about when one upgrades from the single-channel Kondo model to the MCK models is at the heart of the NFL behaviour, and the orthogonality catastrophe is expected to be a general feature of all such frustrated MCK models.
A local NFL term and a similar self-energy was also obtained by Coleman, et al.~\cite{Coleman_tsvelik} in terms of Majorana fermions at the strong-coupling fixed point of the \(\sigma-\tau\) model, which they claimed was equivalent to the intermediate-coupling fixed point of the two-channel Kondo model.
Indeed, along with Ref.\cite{schofield_1997}, this demonstration shows the universality between the two-channel Kondo and the \(\sigma-\tau\) models. 

Further, we argue below that the orthogonality catastrophe for the 2-channel problem can also be understood as a drastic change in the ground state and lowest-lying excited state wavefunctions and related quantum numbers. The detailed calculation has been shown in the Supplementary Materials~\cite{SM}. Upon taking account of the hopping between the star graph and the momentum states of the conduction bath, the full ground-state \(\ket{\psi_-^{(l)}}\) and low-lying excited states \(\ket{\psi_+^{(l)}}\)  in the \(l^\text{th}\) channel  take the form:
\begin{equation}\begin{aligned}
\ket{\psi_\pm^{(l)}} = \frac{1}{\sqrt 2}\left(\ket{\chi_{-\frac{1}{2}}}\otimes\ket{e_\uparrow^{(l)}} \pm \ket{\chi_{\frac{1}{2}}}\otimes\ket{e_\downarrow^{(l)}}\right)~,
\end{aligned}\end{equation}
where \(\ket{\chi_{\frac{\sigma}{2}}} = \frac{1}{\sqrt 6}\left( \ket{\sigma,\sigma,\bar\sigma} + \ket{\sigma,\bar\sigma,\sigma} -2\ket{\bar\sigma,\sigma,\sigma} \right)\) are the ground-states of the two-channel star graph, and \(\sigma\) can take values \(\pm 1\). The states \(\ket{e_\sigma^{(l)}} \equiv e^\dagger_{\sigma,(l)}\ket{\phi} = \sum_{k\in\text{FS}}c^\dagger_{k\sigma,(l)}\ket{\phi}\) are gapless excitations close to the Fermi surface (FS) \(\ket{\phi}\). The residue for scattering processes of the form \(e^\dagger_{\sigma,(l)} e_{\bar\sigma,(l)}\) between the new ground and excited states is then found to vanish:
\begin{equation}
	Z = |\braket{\psi_+^{(l)}| e^\dagger_{\uparrow,(l)} e_{\downarrow,(l)} | \psi_-^{(l)}}|^2 = \left|\bra{\psi_+^{(l)}}\left(\ket{\chi_{\frac{1}{2}}}\ket{e_\uparrow^{(l)}}\right)\right|^2 = 0~.
\end{equation}
This shows that single-particle excitations (a signature of the local Fermi liquid associated with the single channel Kondo fixed point) are no longer long-lived, signalling the orthogonality catastrophe.

The orthogonality catastrophe in the infrared wavefunction arises from an unrenormalised scattering phase shift in the conduction electrons at the fixed point. We have already seen from the zero mode approximation of the fixed point Hamiltonian that each channel suffers a phase shift of \(\Delta^{(l)} = \pi/K\) (eq.~\ref{phase-stargraph}), which is non-zero but less than the limit of unitarity. We will now see that this phase shift guides the renormalisation group flow of the Kondo coupling \(J\). Following refs.~\cite{si_kotliar_1993,giamarchi_varma_1993,giamarchi_book}, the URG equation for \(J\) (eq.\ref{mckRG}) can be recast in the form
\begin{equation}\begin{aligned}
	\Delta {\mathcal{J}}_{(j)} = -\frac{{\mathcal{J}}_{(j)}^2 \mathcal{N}_{(j)}}{\omega_{(j)} - \frac{D_{(j)}}{2} + \frac{{\mathcal{J}}_{(j)}}{4}}\left( 1 - \frac{\rho {\mathcal{J}}_{(j)}}{2\Delta^{(l)}/\pi} \right)~.
\end{aligned}\label{scattphaseRG}\end{equation}
The Kondo coupling $J$ continues to flow under RG until it obtains a value dictated by the phase shift of the star graph: \(\rho \mathcal{J}^* = 2\Delta^{(l)}/\pi = 2/K\). Following eq.~\ref{magnetisation-phase}, this also leads to the fractional value of \(1/4\) for the impurity magnetisation \(m(0^+)\) in the spin-half two-channel Kondo model. Indeed, the values of both \(\Delta^{(l)}\) and \(m(0^+)\) point to the frustrated nature of the MCK problem and the absence of complete screening of the impurity moment at the ICFP. Importantly, the phase shifts \(\Delta^{(l)}\) also affect the critical exponents of well-known algebraic behaviour of several thermodynamic properties at the ICFP. The impurity contributions to the specific heat linear coefficient \(\gamma = C_v/T\), the magnetic susceptibility \(\chi\) and the thermal entropy \(S\) take the forms~\cite{andrei_destri_1984,Tsvelick_Weigmann_mchannel_1984,affleck1993exact}
\begin{eqnarray}
	\gamma &\sim  \left( T/T_K \right)^\frac{1 - 2\Delta^{(l)}/\pi}{1 + 2\Delta^{(l)}/\pi} = \left( T/T_K \right)^\frac{K-2}{K+2}\sim \chi~,\nonumber\\ 
	S &\sim  \ln \left[2|\cos \left(\frac{\Delta^{(l)}}{1 + 2\Delta^{(l)}/\pi}\right)|\right] \nonumber \\
	  &= \ln \left[2|\cos \left(\frac{\pi}{K + 2}\right)|\right]~. \label{powerlawquants}
\end{eqnarray}

\section{Entanglement properties}
\label{sec:ent_prop}

\subsection{Entanglement properties of the star graph}
\label{sec:EE_star graph}

We will now present the results of our study of various entanglement measures in each of the $K$ degenerate ground states $|J^z\rangle$ of a $K$ channel star graph, labeled by the eigenvalues of $J^z$.
\subsubsection{Entanglement entropy between impurity and the rest}
In a 1-channel Kondo model, the ground state is unique and a singlet~\cite{wilson1975renormalization}: \(\ket{J,J_z=0,0} = \frac{1}{\sqrt{2}} \left(\ket{\uparrow_{d},\downarrow_0} - \ket{\downarrow_d, \uparrow_0}\right)\), and the impurity entanglement entropy ($EE_d$) is at the  maximum possible value of $\log 2$.
We will now calculate the same quantity for the ground states 
\begin{eqnarray}
\ket{J^z} \equiv \ket{S_d=1/2,S=K/2;J=(S-1/2),J^z}~,
\end{eqnarray}
of the $K$ channel model.

In order to compute \(EE_d\) for a particular state \(\ket{J^z}\), we calculate the von-Neumann entropy of the reduced density matrix \(\rho_d\) obtained by partially tracing the density matrix $\rho=|J^z\rangle\langle J^z|$ associated with the state \(\ket{J^z}\) over the impurity states \(\ket{S_d^z = \pm \frac{1}{2}}\): 
\begin{eqnarray}
\rho_{d}= \textrm{Tr}_{d} \rho=\sum_{S_d^z} \langle S_d^z| \rho | S^z_d\rangle ~,~~EE_d = -\rho_{d} \log \rho_{d}~.
\end{eqnarray}

\begin{figure}[!htpb]
\centering
\includegraphics[width=0.49\textwidth]{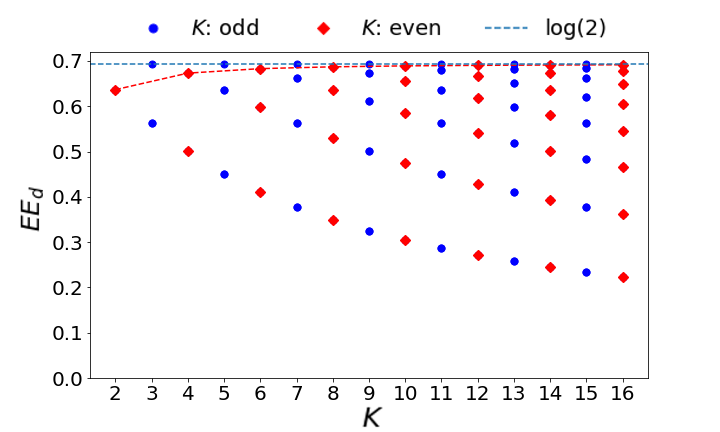}
\caption{Variation of impurity entanglement entropy ($EE_d$) for multiple values of \(K\). The maximum \(EE_d\) at large \(K\) is $\log 2$.}
\label{fig:EE_d}
\end{figure}
The \(EE_d\) for states of various \(J^z\) are shown as functions of the number of channels \(K\) in fig.~\ref{fig:EE_d}. At any given value of \(K\) on the \(x-\)axis, all the points directly above it represent values of \(EE_d\) for various values of \(J^z\) in the ground state manifold of the MCK problem defined by that particular value of \(K\). The minimum entanglement entropy occurs in the maximum \(J^z\) state $\left(|J^z=J\rangle\right)$, and this minimum value decreases with increase in \(K\). On the other hand, the maximum entanglement entropy is attained in the state with minimum \(J^z\). This maximum value of the entanglement entropy is always \(\ln 2\) for odd values of \(K\); for even values of \(K\), the value asymptotically reaches $\log 2$ at large \(K\). 

\begin{figure}[!htpb]
\centering
\includegraphics[width=0.49\textwidth]{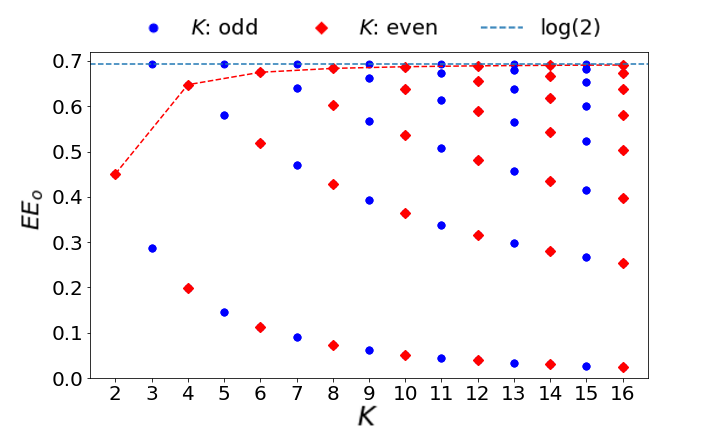}
\caption{Variation of the entanglement entropy of an outer spin ($EE_o$) with the rest for different channels $(K)$. The maximum entanglement entropy in the large $K$ limit is $\log 2$.}
\label{fig:EE_outer}
\end{figure}

A very similar computation gives the entanglement entropy between one outer spin and the rest of the spins, and it is shown in Fig.~\ref{fig:EE_outer}.
We again find that the minimum entanglement entropy is associated with the state $|J^z=J\rangle$ and it falls with increasing \(K\), approaching zero asymptotically; the maximum entanglement entropy is, as before, associated with the state $|\left(J^z\right)_\mathrm{min}\rangle$ and it rises to $\log 2$ in the limit $K\gg 1$.

Two interesting features emerge from the study:
\begin{itemize}
	\item For the odd \(K\) models, the impurity is maximally entangled with the other spins in the minimum \(J^z\), $|J^z=0\rangle$, member of the ground state manifold.
	At large \(K\), the even \(K\) models also acquire this property, the minimum \(J^z\) there being \(\pm \frac{1}{2}\).
	\item 
	Both \(EE_d\) and \(EE_0\) are individually equal for the states $|\pm J^z\rangle$, which shows that both types of entanglement entropy are invariant under the transformation \(\ket{J^z} \to \pi \ket{J^z}\). This reflects a parity symmetry in the state space in terms of entanglement.
\end{itemize}

\subsubsection{Mutual Information}

\begin{figure}[!htpb]
\includegraphics[width=0.49\textwidth]{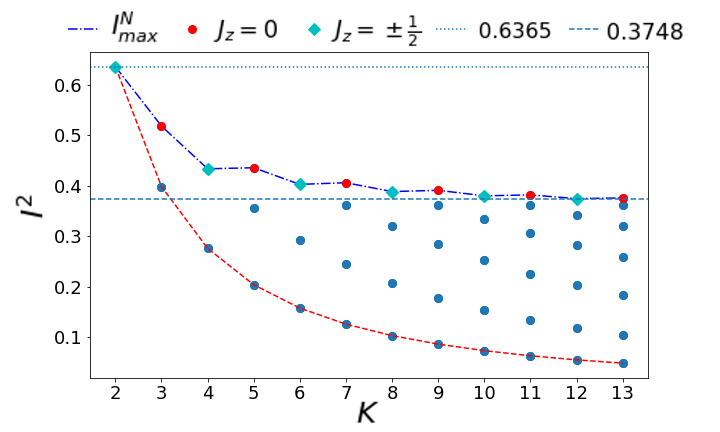}
\caption{Variation of mutual information $I^2(d:o)$ of the impurity with the rest, for different values of $K$. The maximum $I^2(d:o)$ is $0.37$.}
\label{fig:MI_d_o}
\end{figure}

\begin{figure}[!htpb]
\includegraphics[width=0.49\textwidth]{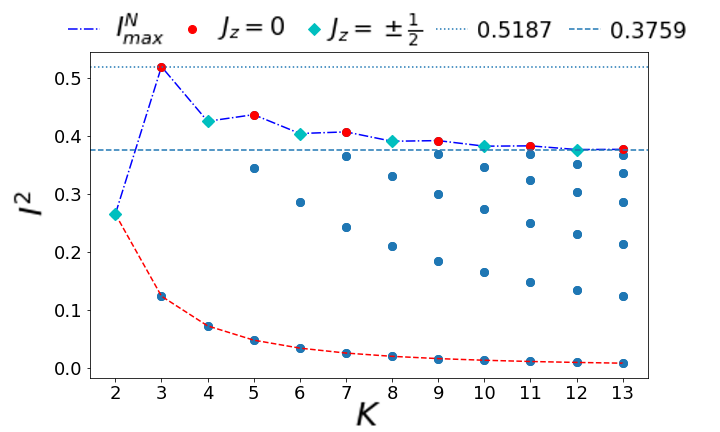}
\caption{Variation of mutual information $I^2(o:o)$ between 2 outer spins, for multiple $K$. $I^2(o:o)$ vanishes for large $K$.}
\label{fig:MI_o_o}
\end{figure}

Mutual information is a measure which captures the correlations present between two subsystems (A,B), in a particular state. It is defined as
\begin{eqnarray}
I^2(A:B)=S_A+S_B-S_{A\cup B}~,
\end{eqnarray}
where $S_{A(B)}$ is the entanglement entropy of the subsystem $A(B)$ with the rest, and \(S_{A\cup B}\) is the entanglement entropy of \(A\) and \(B\) with the rest.
We are interested in two types of mutual information: firstly, the mutual information $I^2(d:o)$ between the impurity and one of the other spins, and secondly, the mutual information $I^2(o:o)$ between any two of the outer spins.
Both of these have been computed for various channel numbers \(K\), and plotted in Figs.~\ref{fig:MI_d_o} and ~\ref{fig:MI_o_o}.

In both cases we find that the maximum and minimum mutual information are associated with the $|\left(J^z\right)_\mathrm{min}\rangle$ and $|J^z=J\rangle$ states respectively.
We also find that the  mutual information is the same in the states $|J^z\rangle$ and $\pi^x|J^z\rangle$, indicating the parity symmetry in the mutual information measure.
In the large channel limit ($K\gg 1$), the maximum \(I^2(d:o)\) and \(I^2(o:o)\) saturate to the common value of $0.375$.

\subsubsection{Multipartite information}
Similar to mutual information, one can calculate higher order multipartite information to study the nature of correlations present in different ground states. We define the tripartite information among three subsystems $A,B,C$ as $I^3_{ABC} = (S_A+S_B+S_C)-(S_{AB}+S_{BC}+S_{CA})+S_{ABC}~$. \(I^3\) shows behavior similar to the mutual information - the highest \(I^2\) value is associated with the $\left(J^z\right)_\mathrm{min}$ state and it saturates in the limit of $K\gg 1$. The $N-$partite information for a collection of subsystems (CSS) $\{\mathcal{A}_N\}\equiv\{A_1,A_2,\cdots,A_N\}$ can also be defined 
as~\cite{siddharthatee} 
\begin{eqnarray}
	I^{N}_{\{\mathcal{A}_N\}} = &\bigg[\displaystyle\sum_{m=1}^{N} (-1)^{m-1} \displaystyle\sum_{Q \in \mathcal{B}_m(\{\mathcal{A}\})} S_{V_{\cup}({Q})} \bigg] \nonumber\\
				    &- S_{V_{\cap}(\mathcal{A}_N)}~.
\label{eq:I_N_definition}
\end{eqnarray}
where  $\mathcal{B}_m(\{\mathcal{A}_N\})\equiv \{ Q~| ~Q\subset \mathcal{P}(\{\mathcal{A}_{N}\}), |Q|=m \}$, and $\mathcal{P}(\{\mathcal{A}_{N}\})$ is the powerset of $\{\mathcal{A}_N\}$. For simplicity, we also define the union and intersection of all the subsystems present in $Q$ as ${V}_{\cup}({Q})\equiv \bigcup_{A\in Q} A$ and ${V}_{\cap}({Q})\equiv \bigcap_{A\in Q} A$ respectively.
\begin{figure}
\includegraphics[width=0.49\textwidth]{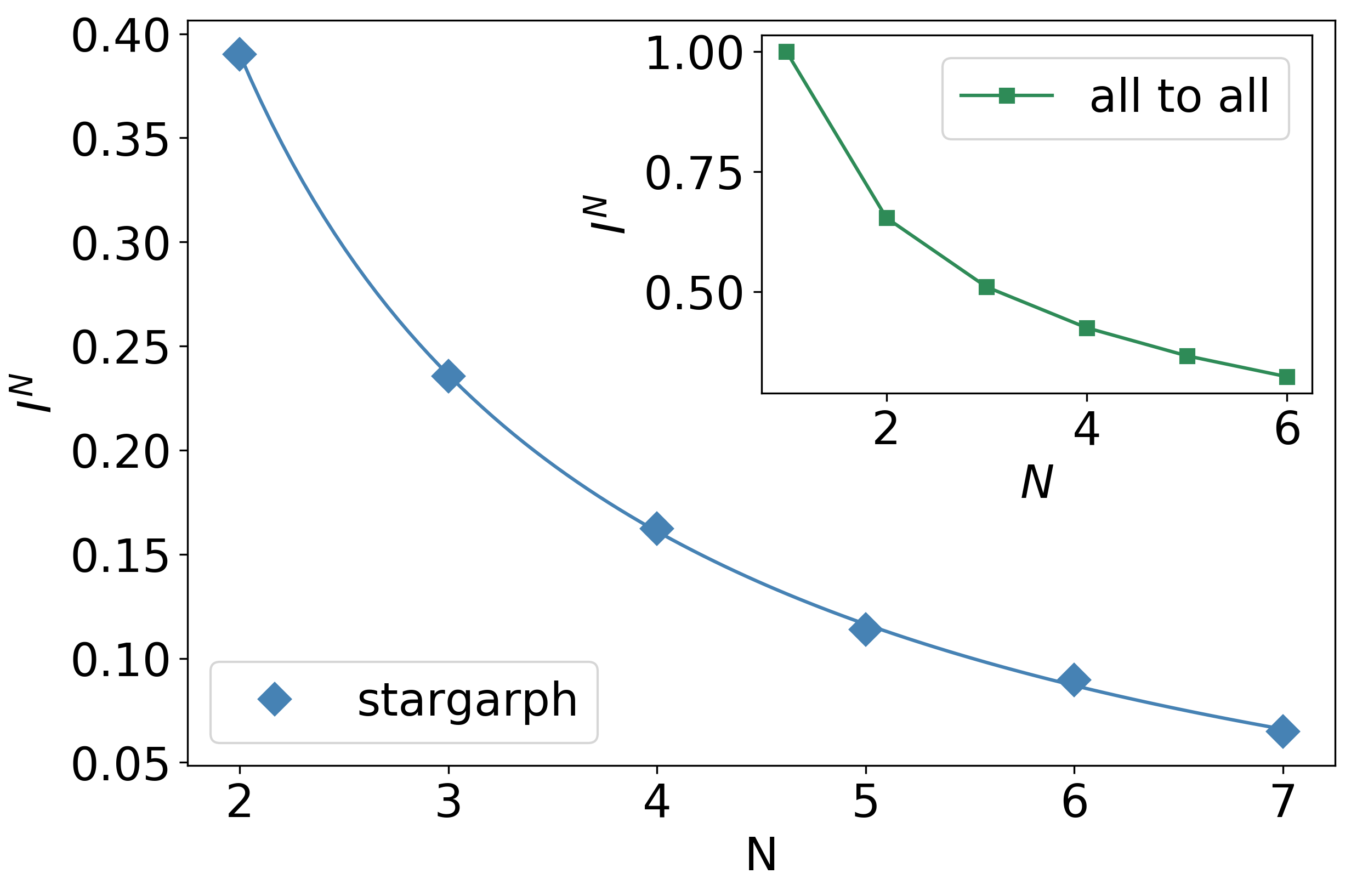}
\caption{Variation of multipartite information $I^N$ among \(N\) outer spins as a function of $N$, for $K=8$. The inset shows $I^N$ vs $N$ for an all-to-all model. Both show power law behavior.}
\label{fig:Im_vs_m}
\end{figure}

The study of such measures of multipartite information reveals the nature of the multi-party correlations present among the outer spins of the star graph. Our study, as presented in fig.~\eqref{fig:Im_vs_m}, shows that the higher order multi-partite correlations ($I^N$) decrease as you increase the order ($N$), and the behaviour follows a power law. In the inset of the same figure, we have shown the $I^N$ vs $N$ plot that is observed in the ground state of the all-to-all effective Hamiltonian (Eq.\ref{eq:flux_spectral}), and it shows a similar power-law behaviour. The similarities in the behavior suggest that one can capture entanglement properties either from the star graph or from the corresponding all-to-all model.

\subsection{Entanglement properties at the fixed point of the MCK model}
\label{sec:EE_excitation}

\begin{figure}[!htpb]
\includegraphics[width=0.49\textwidth]{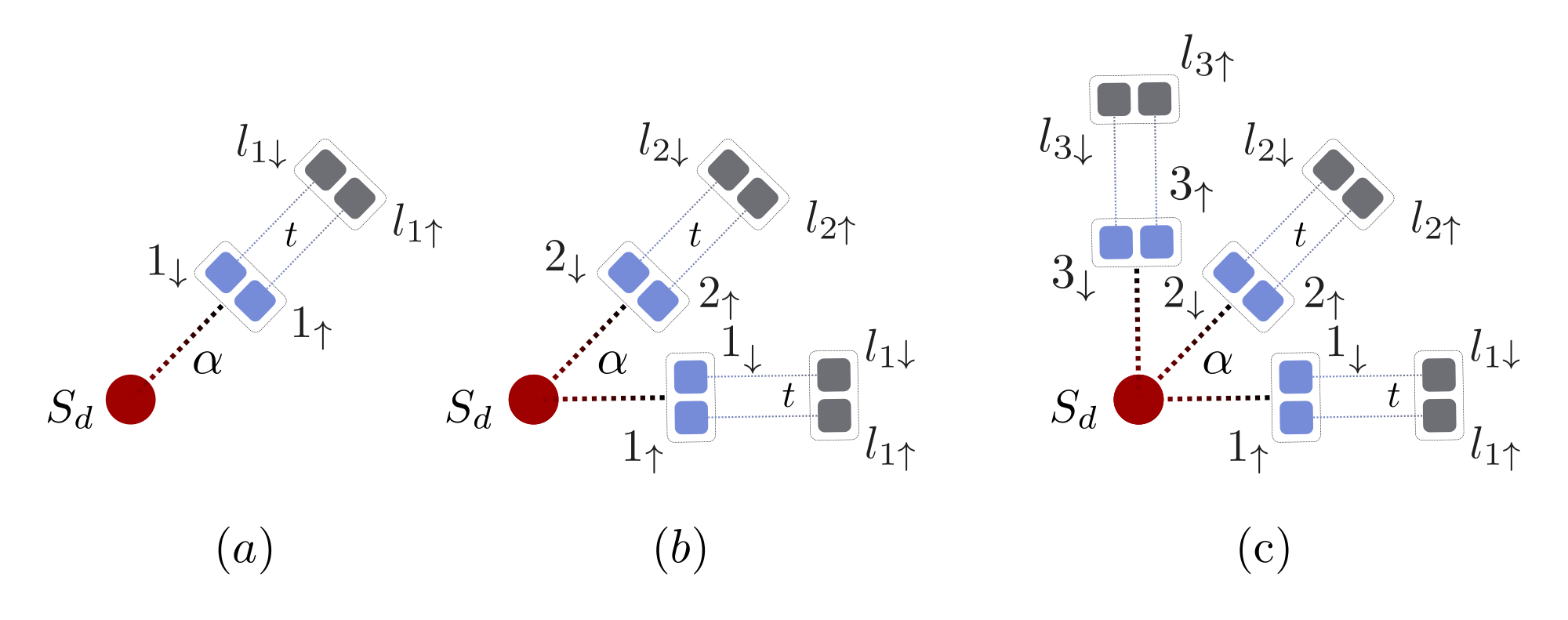}
\caption{This is a schematic diagram of (a) single channel and (b) two channel problem. $S_d$ is the impurity spin. For more details refer to the text.}
\label{fig:schematic_hopping}
\end{figure}

To study the nature of correlations at the MCK fixed point, we introduce excitations into the star graph ground state. We will work with the Hamiltonian Eq.\eqref{eq:excitation_hamiltonian} where we consider nearest-neighbor real space hopping \(t\) from the zeroth site into the lattice (Fig.~\ref{fig:schematic_hopping}). Setting $t=0$ recovers the zero bandwidth version of the MCK. In the \(K=1\) single channel case, the impurity ($S_d$) is coupled to the spin degree of freedom at the real space origin ($\{1_{\uparrow},1_{\downarrow}\}$) of the single conduction bath via a Heisenberg spin-exchange coupling. For $K=2$, the impurity is interacting with $2$ distinct local spins $\{1_{\uparrow},1_{\downarrow}\}$ and $\{2_{\uparrow},2_{\downarrow}\}$ belonging to zeroth sites of different conduction channels, and the real space hopping connects these zeroth sites to their nearest neighbor sites. As we increase the hopping strength $t$ from zero to non-zero, the lattice sites start interacting with each other.

\subsubsection{Impurity entanglement entropy}
\begin{figure}[!htpb]
\centering
\includegraphics[width=0.49\textwidth]{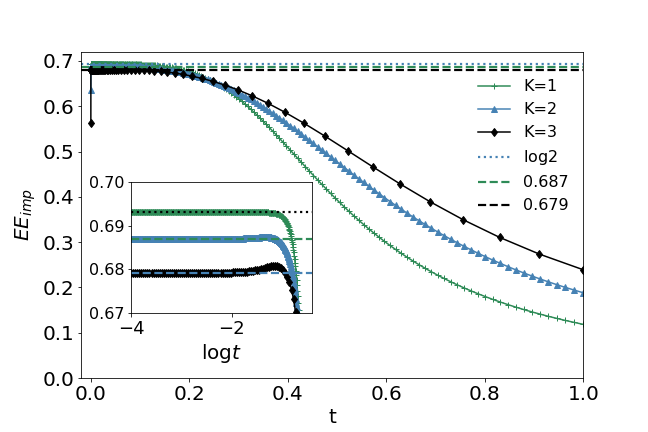}
\caption{Variation of impurity entanglement entropy $EE_{imp}$ against $t$, for $K=1,2,3$. }
\label{fig:EE_imp_vs_t_K}
\end{figure}

The impurity entanglement entropy $EE_d(t)$ in the ground state is shown as a function of $t$ in Fig.~\ref{fig:EE_imp_vs_t_K} for three values of channel number \(K\). At low hopping strength ($0<t\ll 1$), $EE_{d}$ is independent of $t$ and achieves a constant value. We also find that in this range of \(t\), \(EE_d\) decreases with increasing \(K\). This behaviour is reversed at high $t$, and \(EE_d\) is seen to increase with increasing \(K\). Though the values of the impurity entanglement entropy for various values of \(K\) are quite similar at low hopping strength, 
as the single channel \(EE_d\) varies smoothly at \(t \to 0^+\), whereas the \(EE_d\) for \(K > 1\) show a discontinuity at $t \to 0^+$.

\subsubsection{Intra-channel Mutual information}
Here, we will calculate the mutual information between two electronic states in the ground state wavefunction, as a function of the hopping strength $t$. 

\par
\begin{figure}[!htpb]
\includegraphics[width=0.49\textwidth]{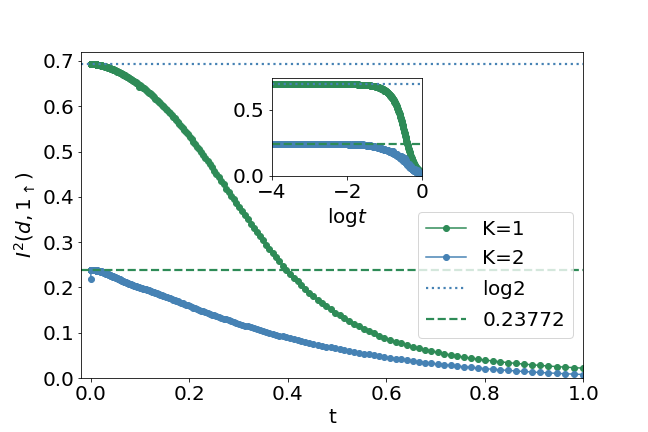}
\caption{This figure show the variation of mutual information between the impurity spin and the $1_{\uparrow}$ state against $t$. For more details please refer to the text. The error-bar $\sigma_2$ of the two channel data at low hopping strength $\sigma_2\approx 0.002$.}
\label{fig:MI_imp_1_vs_t_K}
\end{figure}

 \textit{Case 1:} We first study the mutual information $MI(d,1\uparrow)$ between the impurity spin $S_d$ and the state $1_\uparrow$ (fig.~\ref{fig:schematic_hopping}), where $1$ represents real-space origin of the $1^\mathrm{st}$ conduction channel. For a single channel model, this site is unique because there is just one channel, but in the presence of $K$ channels, there are $K$ possible choices corresponding to each of the channels. However, because of the symmetry of the Hamiltonian under the exchange of the channel indices, all such choices will show identical mutual information signatures. Similarly, due to the $SU(2)$ spin-rotation symmetry of the Hamiltonian, $MI(d,1\downarrow)$ will be identical to $MI(d,1\uparrow)$. We have numerically computed and plotted $MI(d,1\uparrow)$ as a function of $t$ in fig.~\ref{fig:MI_imp_1_vs_t_K}. The inset shows that at low hopping strength, the mutual information for the single channel and the two-channel models saturate to different values. In the single-channel case, the saturation value at $t=0$ is $\log 2$, and the mutual information changes smoothly as \(t\) is turned on. This is similar to the behaviour in the impurity entanglement entropy studied previously. The value of $\log 2$ shows the maximal entanglement between the zeroth site and the impurity, and the perfect screening of the impurity spin. For the two channel problem, we find that the mutual information at \(t=0\) is $MI(d,1_{\uparrow})=0.2401$. The reduction of the value from \(\ln 2\) shows the breakdown of screening. Also note that unlike the single channel case, there is a discontinuity in $MI(d,1_{\uparrow})$ as \(t\) is increased from 0.

\par
\begin{figure}[!htpb]
\includegraphics[width=0.49\textwidth]{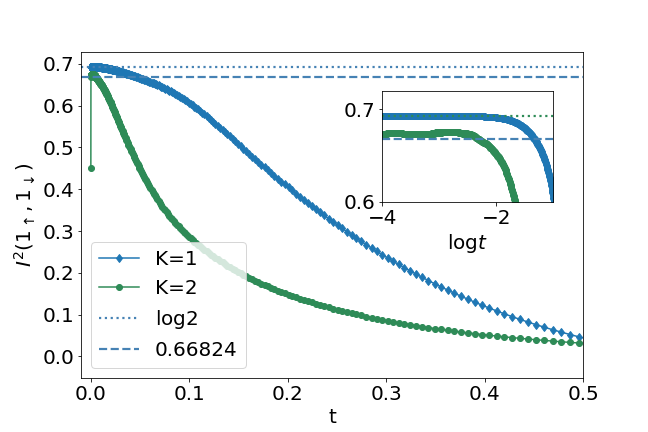}
\caption{Variation of mutual information ($i^2(1_{\uparrow},1_{\downarrow})$) among the two states present on the real-space origin ($1$) of a conduction channel, against $t$. The error-bar $\sigma_2$ for the two channel at low hopping strength is $\sigma_2\approx 0.013$.}
\label{fig:MI_1_2_vs_t_K}
\end{figure}

\textit{Case 2:} We have also computed the mutual information $MI(1_{\uparrow},1_{\downarrow})$  between the two electronic states on the zeroth site of the same conduction channel, and plotted them in fig.~\ref{fig:MI_1_2_vs_t_K}. We find that it is a smooth function of $t$ for \(K=1\), and for \(K=2\) there is a discontinuity at $t=0^{+}$. Note that for small values of \(t\), we find  $MI(1_{\uparrow},1_{\downarrow})>MI(d,1_{\uparrow})$ for the \(K=2\) model, showing the presence of strong intra-site correlations.

\par
\begin{figure}
\includegraphics[width=0.49\textwidth]{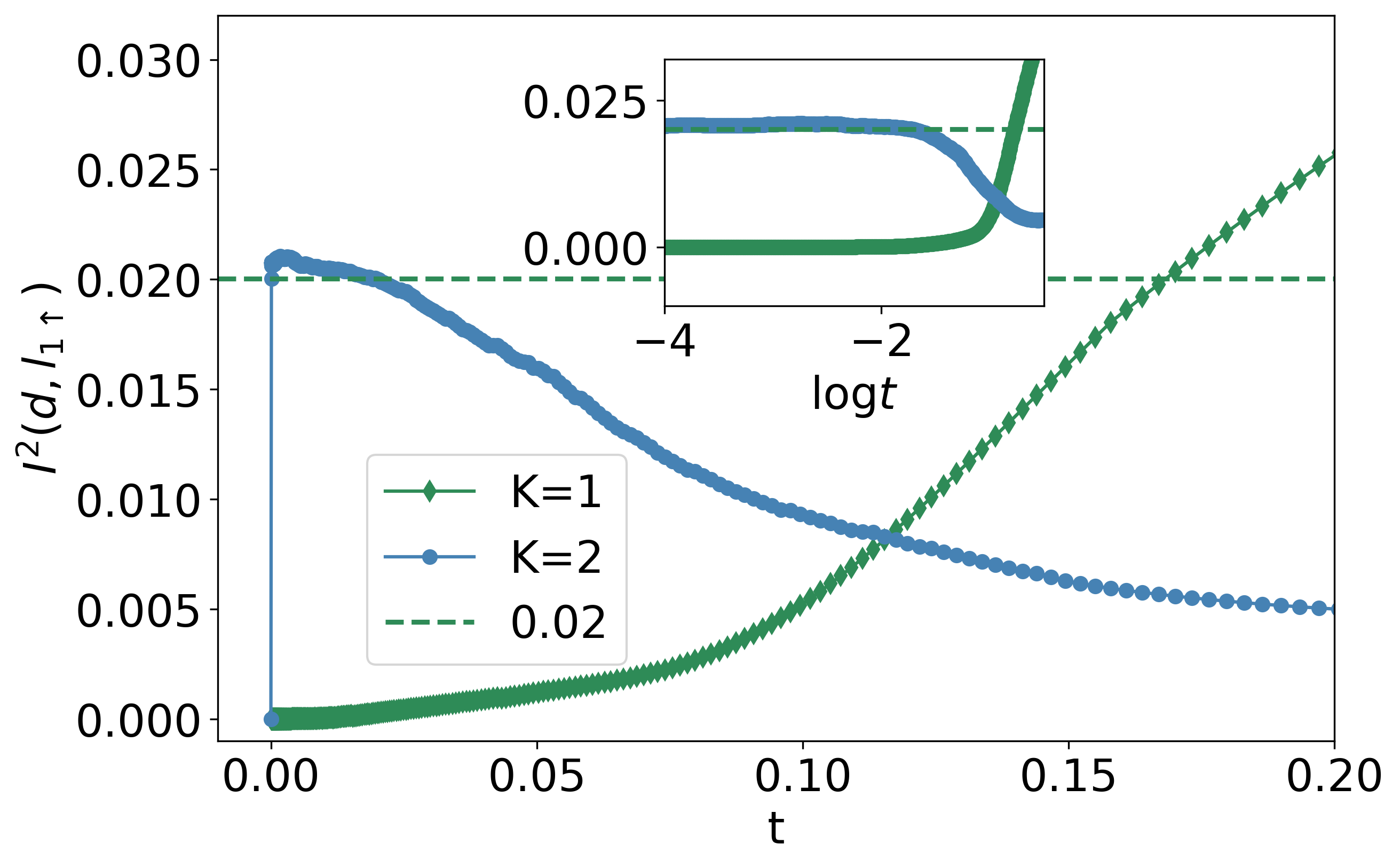}
\caption{Variation of the mutual information $MI(d,l_{1\uparrow})$ between the impurity spin and the $l_{1\uparrow}$ state, against $t$. The error-bar $\sigma_2$ for the two channel case is $\sigma_2\approx 0.001$.}
\label{fig:MI_d_l1}
\end{figure}

 \textit{Case 3:} Next, we measure the mutual information $MI(d,l_{1\uparrow})$ between the impurity site and an electronic state of the site that is nearest-neighbour to the real-space origin of one of the conduction channels (see fig.~\ref{fig:schematic_hopping}).  The results are shown in fig.~\ref{fig:MI_d_l1}. Vanishing mutual information for the single channel case shows the perfect screening of the impurity spin. The non-zero value of mutual information at small values of $t$ in the \(K=2\) model is again a result of the imperfect screening. As in the previous cases, we find a discontinuity in the mutual information for \(K=2\) as $t \to =0^+$.

\par
\begin{figure}[!htpb]
\includegraphics[width=0.49\textwidth]{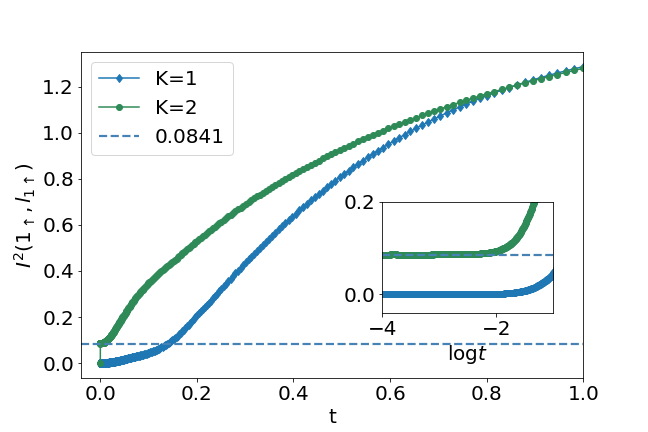}
\caption{Variation of the mutual information $MI(1_{\sigma},l_{1\sigma})$ between two nearest neighbor sites against \(t\). The error-bar for the two channel case is $\sigma_2\approx 0.005$.}
\label{fig:MI_l_l1}
\end{figure}

 \textit{Case 4:} A complementary study can be made by calculating the mutual information between the origin of a particular conduction channel and it's nearest neighbour site. They are plotted in fig.~\ref{fig:MI_l_l1}, and we find that at low hopping strength the single channel mutual information vanishes, showing the decoupling of those two states. On the other hand, in the two channel case, there is a non-zero mutual information which indicates the presence of scattering between the local Mott liquid and the local non-Fermi liquid states.

\begin{figure}[!htpb]
\centering
\includegraphics[width=0.48\textwidth]{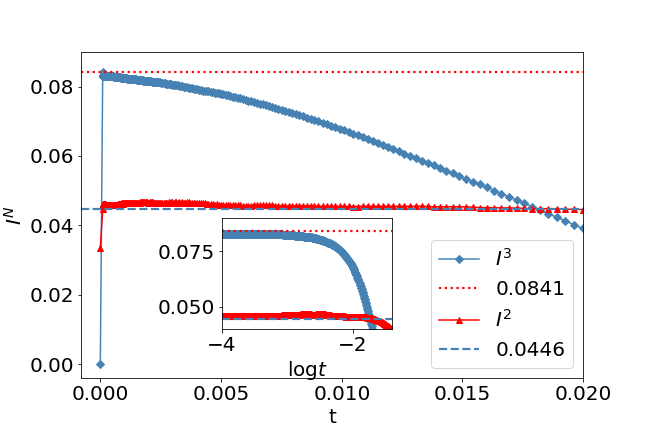}
\caption{Red: Mutual information $I^2(1_{\uparrow},2_{\uparrow})$ against \(t\). Blue: Tripartite information $I^3(1_{\uparrow},l_{1\uparrow},l_{1\downarrow})$ against \(t\). The error-bar ($\sigma$) for the $I^3$ and $I^2$ near the low hopping strength are $0.002$ and $0.003$ respectively.}
\label{fig:I3_I2_two_channel}
\end{figure}

Apart from these, we have also computed an intra-channel tripartite information $I^3(1_{\uparrow},l_{1\uparrow},l_{1\downarrow})$ of the two-channel ground state (blue curve in Fig.~\ref{fig:I3_I2_two_channel}) - it shows that there is a discontinuity in the tripartite information at $t=0^+$, and at low hopping strength the tripartite information is independent of $t$ with a value $0.084$. This non-zero value of the tripartite information is consistent with the presence of a non-Fermi liquid effective Hamiltonian having more than just two-particle interactions within it. We have also studied inter-channel mutual information $I^2(1_{\uparrow},2_{\uparrow})$ for \(K=2\) (red curve in fig.~\ref{fig:I3_I2_two_channel}). It reveals the correlation and quantum entanglement between these two channels, and demonstrates the all-to-all nature of the local Mott liquid.

\subsubsection{Bures distance and the orthogonality catastrophe}
\begin{figure}[!htpb]
\includegraphics[width=0.49\textwidth]{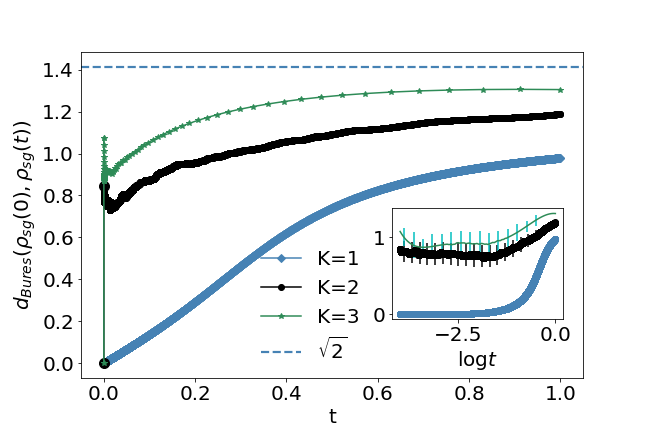}
\caption{Variation of Bures distance between the reduced density matrices of the star graph ground states at $t=0$ and $t\neq 0$, against $t$. The error-bars ($\sigma$) for \(K=2\) and \(K=3\) are $0.14$ and $0.08$ respectively, and near $t\approx 1$ the $\sigma$ are $0.04$ and $0.07$ respectively.}
\label{fig:bures_distance}
\end{figure}
In addition to these entanglement measures, we have also calculated the Bures distance between the states at \(t=0\) and \(t>0\). The Bures distance~\cite{uhlmann1976transition,bures1969extension,hubner1992explicit,hubner1993computation,dittmann1994some,marian2003bures} between two density matrices \(\rho_1,\rho_2\) is defined as 
\begin{eqnarray}
d_{Bures}(\rho_1,\rho_2) = \sqrt{2} \bigg[1- \textrm{Tr}\bigg\{ \bigg(\rho_1^{1/2}  \rho_2 \rho_{1}^{1/2}\bigg)^{1/2} \bigg\}\bigg]~,
\end{eqnarray}
and is a measure of the distance, in the Hilbert space, between the states forming the density matrices. One can see from the above definition that the maximum and minimum possible distances are $\sqrt{2}$ and $0$ respectively.
\par We get the ground state $|\psi(t)\rangle$ of the problem by solving the Hamiltonian in Eq.~\eqref{eq:excitation_hamiltonian}. This ground state $|\psi(t)\rangle(t)$ and the associated density matrix $\rho(t)= |\psi(t)\rangle \langle \psi(t)|$ are functions of the hopping strength \(t\). We are interested in the reduced density matrix $\rho_{sg}(t)$ of only the sites that form the star graph (impurity site and the zeroth sites of the conduction channels), so we trace out all the nearest neighbor sites from the full density matrix:, $\rho_{sg}(t)=\textrm{Tr}_{n-n}(\rho(t))$. The Bures distance $d_{Bures}(\rho_{sg}(0),\rho_{sg}(t))$ is then calculated between the reduced density matrices at \(t=0\) and \(t > 0\), for \(K=1\), \(K=2\) and \(K=3\).

We have shown the variation of the Bures distance as a function of the hopping strength \(t\) in Fig.~\ref{fig:bures_distance}. Although the Bures distance is \(0\) (by definition) at \(t=0\) for all \(K\), a difference arises as \(t\) is made slightly non-zero: there is a discontinuity in the Bures distance for the two and three channel models, while the single channel Bures distance is continuous. The smooth variation of the Bures distance in the single channel case shows the adiabatic continuity of the reduced star graph density matrix from the $t=0$ case to $t\neq 0$ case. On the other hand, the abrupt jump in the Bures distance for higher-channel cases at $t=0^+$ shows the breakdown of adiabatic continuity between the \(t=0\) and \(t = 0^+\) Hamiltonians and signals an {\it orthogonality catastrophe} between these two ground states. The discontinuity increases as \(K\) is increased, indicating that considering the case of more channels will lead to the maximum Bures distance of $d_{Bures}=\sqrt{2}$ and hence an exact orthogonality.

\section{Duality properties of the MCK model}
\label{sec:duality}

We start from a strong coupling \((J \to \infty)\) spin-\(S_{d}\) impurity MCK Hamiltonian in the over-screened regime \(\left( K > 2S_{d} \right) \),
\begin{eqnarray}
	\label{strong_ham}
	H(J) = \sum_{k,\sigma,l}\epsilon_{k,l} \hat n_{k\sigma,l} + J \vec{S_d}\cdot\vec{S}~.
\end{eqnarray}
Here, \(\vec S\) is the total spin \(\sum_l \sum_{kk^\prime \alpha\beta} \vec \sigma_{\alpha\beta}c^\dagger_{k\alpha,l}c_{k^\prime\beta,l}\) of all the zero modes. At strong coupling, the ground states of the star graph eq.\eqref{star graph} act as a good starting point for a perturbative expansion. As argued previously, there are \(K-2S_d+1\) ground states, labeled by the \(K\) values of the total spin angular momentum \(J^z = S_d^z + S^z = -\frac{K}{2} + S_d, -\frac{K}{2} + S_d + 1, \ldots, \frac{K}{2} - S_d\). Since a general spin-\(s\) object is simply a \(2s+1\) level system, the \(K-\)fold degenerate ground state manifold can be used to define a new impurity spin \(\mathbb{S}_d\) of multiplicity \(2\mathbb{S}_d + 1 = K-2S_d+1\) which implies that we need \(\mathbb{S}_d = \frac{K}{2} - S_d\). In other words, the spin-\(S_d\) impurity has a dual described by a spin-\((K-2S_d+1)\) impurity. The states of this new spin are defined by
\begin{eqnarray}
	\hat{\mathbb{S}_d^z} \ket{S^z} = S^z \ket{S^z},\nonumber\\
	\hat{\mathbb{S}_d^\pm} \ket{S^z} = \sqrt{\mathbb{S}_d\left( \mathbb{S}_d + 1 \right) - S^z\left( S^z \pm 1\right)} \ket{S^z \pm 1}~.
\end{eqnarray}
While the ground state subspace gives rise to the new central spin object, the excited states of the star graph can be used to define bosonic operators that mediate interactions between the central spin and the next-nearest neighbour lattice sites~\cite{kroha_kolf_2007}. In terms of the zero-bandwidth spectrum, the bosonic operators represent scattering between the ground state subspace and the excited subspaces. Through a Schrieffer-Wolff transformation in the small coupling \(\frac{t^2}{J}\), one can then remove this interaction and generate an exchange-coupling between the new impurity \(\vec {\mathbb{S}}_d\) and the new zero modes formed out of the remaining sites in the lattice \cite{kroha_kolf_2007} (by remaining, we mean those real space sites that have not been consumed into forming the new spin). The new Hamiltonian, characterized by the small super-exchange  coupling \(\mathbb{J}\) of the general form \(\gamma t^2/J\), has the form
\begin{eqnarray}
	H^\prime(\mathbb{J}) = \sum_{k,\sigma,l}\epsilon_{k,l} \hat n_{k\sigma,l} + \mathbb{J} \vec{\mathbb{S}_d}\cdot\vec{\mathbb{S}}~.
\end{eqnarray}
$\vec{\mathbb{S}}$ is the local bath spin formed by the new zero modes. This Hamiltonian is very similar to the one in eq.\eqref{strong_ham}, and that is the essence of the strong-weak duality: One can go from the over-screened strong coupling spin-\(S_{d}\) MCK model to another over-screened weak coupling spin-\((K-2S_{d}+1)\) MCK model. For the case of \(K=4S_{d}\), we have \(\mathbb{S}_d = S_d\), and both \(S_d\) and \(\mathbb{S}_d\) describe the same objects: the two models are then said to be self-dual. For example, for the case of spin-$1/2$ MCK model, two-channel model is self-dual.

One important consequence of the duality relationship between the two over-screened models is that the RG equations are also dual; while the strong coupling model has an irrelevant coupling \(\mathcal{J}\) that flows down to the intermediate fixed point \({\mathcal{J}^*}\), the weak coupling model has a relevant coupling \(\mathbb{J}\) that flows up to the same fixed point \({\mathbb{J}}^* = {\mathcal{J}^*}\). From the RG equation for the general spin-\(S\) MCK model, we know that \({\mathbb{J}}^* = \frac{2}{K \rho^\prime}\), where \(\rho^\prime\) is the DOS for the bath of the weak coupling Hamiltonian. This constrains the form of the scaling factor \(\gamma\):
\begin{eqnarray}
	{\mathbb{J}}^* = \frac{\gamma 4t^2}{{\mathcal{J}^*}} = \frac{2}{K \rho^\prime} \implies \gamma = \frac{1}{4t^2} {{\mathcal{J}^*}}^2 = \frac{1}{K^2 t^2 \rho \rho^\prime}~.
\end{eqnarray}

Apart from the strong-weak duality, there exists another set of dual pairs in the MCK model. This was hinted at when we looked at the degree of compensation in eq.\eqref{gamma}. Since \(\Gamma\) depends only on the magnitude of \(\delta\), both \(\pm \delta\) will give the same degree of compensation, same ground state energy and same ground state degeneracy \(\left(g^{S_d}_K = |\delta|+1\right)\). The definition of \(\delta\) implies the following duality transformation: \(K \to 2S_d, S_d \to \frac{K}{2}\). That is, we transform from a \(K-\)channel MCK model with spin-\(S_d\) impurity, to a \(2S_d\)-channel MCK with a spin-\(\frac{K}{2}\) impurity. The exactly-screened model \(K=2S_d\) maps on to itself and is therefore self-dual under this transformation.

For \(K \neq 2S_d\), we transform an over-screened model into an under-screened model and vice versa. This duality relationship allows us to infer the RG scaling behaviour of one of the models if we know that of the other. Importantly, we see that the existence of an intermediate coupling fixed point in the MCK model for a given \(K\) and \(S_d\) leads to the conclusion that the \(2S_d\)-channel spin-\(\frac{K}{2}\) model has a strong coupling fixed point.

\section{Quantum Phase transition in the MCK model under channel anisotropy}
\label{anisotropic_rg}

For a channel-anisotropic MCK model with \(K\) Kondo couplings \(\left\{\mathcal{J}_i\right\}\) for each of the \(K\) conduction channels, the zero bandwidth model is 
\begin{eqnarray}
H_K (\vec{{\mathcal{J}}}) = \sum_{i=1}^{K} {\mathcal{J}}_i\vec{S}_d.\vec{S}_i~.
\label{eq:anisotropy}
\end{eqnarray}
For the special case $\mathcal{J}_i=\mathcal{J} \forall i$, we get the usual isotropic star graph model. For the case of spin-1/2 impurity, we find that the Hamiltonian with any value of $\mathcal{J}_i>0$ has $K$ fold ground state degeneracy. This shows that the ground state degeneracy of the star graph model is extremely robust against the channel anisotropy; the ground state degeneracy does not change until at least one \(\mathcal{J}_i\) vanishes.
\begin{figure}
\centering
\includegraphics[width=0.48\textwidth]{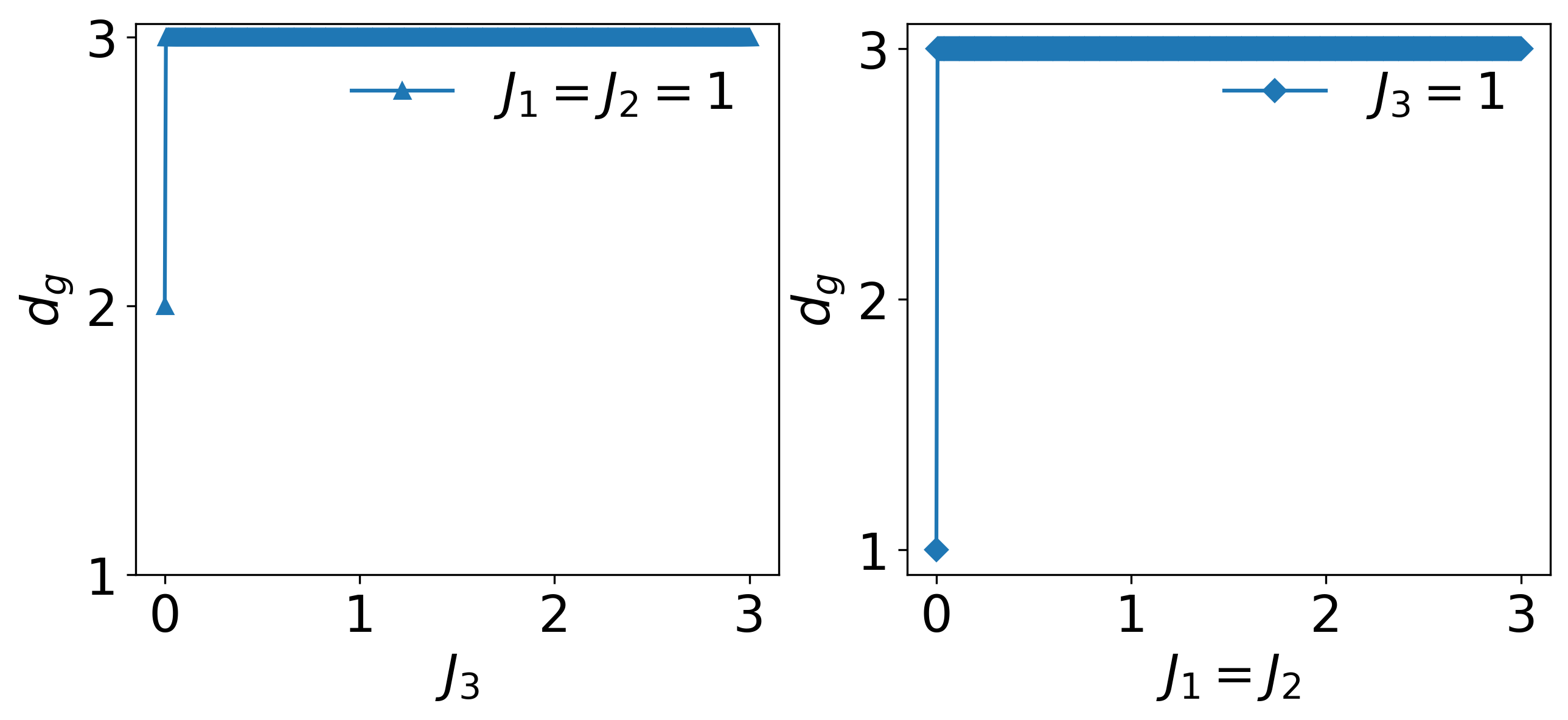}
\caption{\textit{Left:} Variation of the ground state degeneracy against the $J_3$ keeping $J_1=J_2=1$ fixed. \textit{Right:} Variation of ground state degeneracy against $J_1=J_2$ keeping $J_3$ fixed. Both plots show the robustness of the ground state degeneracy against channel anisotropy.}
\label{fig:channel-anisotropy-}
\end{figure}

We have demonstrated this numerically for a three channel anisotropic star graph model.
In Fig.~\ref{fig:channel-anisotropy-}(a), we show that if two couplings are kept constant and equal to one another (${\mathcal{J}}_1={\mathcal{J}}_2=1$) and the third coupling ${\mathcal{J}}_3$ is tuned from some finite value to zero, the degeneracy does not change from 3 to 2 until \(\mathcal{J}_3\) becomes zero. At that point, the model becomes a two channel Kondo problem. We also show in Fig.~\ref{fig:channel-anisotropy-}(b) that taking the coupling ${\mathcal{J}}_3$ fixed to $1$ and varying the common coupling ${\mathcal{J}}_1={\mathcal{J}}_2$ from non-zero to zero is equivalent to keeping ${\mathcal{J}}_1={\mathcal{J}}_2=1$ fixed and taking ${\mathcal{J}}_3$ to infinity. In this case, when the coupling ${\mathcal{J}}_1={\mathcal{J}}_2=0$, the degeneracy becomes one and reveals the single channel nature of the problem.

The above demonstration makes it clear that the ground state degeneracy can change only when at least one of the Kondo couplings vanish. This can be realised under RG flow if one considers the anisotropic MCK model.
\begin{eqnarray}
	H = \sum_{k,\alpha,l}\epsilon_{k,l} \hat n_{k\alpha,l} + \sum_{kk^\prime,\atop{\alpha,\beta,l}}{\mathcal{J}}_l \vec{S_d}\cdot\frac{1}{2}\vec{\sigma}_{\alpha\beta}c_{k\alpha,l}^\dagger c_{k^\prime\beta, l}~.
\end{eqnarray}
We consider the specific case where \(K-1\) channels have the same coupling \({\mathcal{J}}_1 = {\mathcal{J}}_2 = ... = {\mathcal{J}}_{K-1} = {\mathcal{J}}_+\) and the remaining channel has a different coupling \({\mathcal{J}}_K = {\mathcal{J}}_-\). The RG equations for such a model are
\begin{eqnarray}
	\frac{\Delta {\mathcal{J}}_\pm}{|\Delta D|} = -\frac{{\mathcal{J}}_\pm^2 \rho}{\mathcal{D}_\pm} + \frac{\rho^2 {\mathcal{J}}_\pm}{2}\left[\frac{(K-1){\mathcal{J}}_+^2}{\mathcal{D}_+} + \frac{{\mathcal{J}}_-^2}{\mathcal{D}_-}\right]~,
\end{eqnarray}
where \(\mathcal{D}_\pm = \omega - \frac{D}{2} - \frac{{\mathcal{J}}_\pm}{4}\) are the denominators of the URG equations.
Setting \({\mathcal{J}}_+ = {\mathcal{J}}_-\) leads to the critical fixed point at \({\mathcal{J}}_+^* = {\mathcal{J}}_-^* = {\mathcal{J}}_* = \frac{2}{K \rho}\). We now perturb around this fixed point by defining new variables \(j_\pm = {\mathcal{J}}_\pm - {\mathcal{J}}_*\). We also assume that the bandwidth is large enough so that \(\mathcal{D}_\pm \simeq \omega - \frac{D}{2} - \frac{{\mathcal{J}}_*}{4} = -|\mathcal{D}_*|\). Performing a linear stability analysis about \(j_\pm=0\) then reveals the following two possibilities:
\begin{itemize}
	\item[(a)] If \(j_-<0\) and \(j_+>0\), the deviation \(j_-\) is relevant and \(\mathcal{J}_-\) flows to zero. The flow of \(j_+\) are constrained such that the remaining couplings \(\mathcal{J}_+\) flow to the stable intermediate fixed point of the \(K-1\) channel MCK model: \(J_{+,*} = \frac{2}{(K-1)\rho}\).
	\item[(b)] If \(j_- > 0\) and \(j_+<0\), the couplings \(\mathcal{J}_+\) are irrelevant, and this leads to a single channel Kondo model described by the coupling \(\mathcal{J}_-\) which flows to strong coupling.
\end{itemize}

These conclusions show that the \(K\) channel intermediate fixed point is unstable under channel anisotropy~\cite{Noz_blandin_1980,andrei_jerez_1995,affleck_pang_cox_1992,zarand_2000}. If one of the couplings becomes smaller than the rest, then that coupling flows to zero while the other \(K-1\) couplings flow to the \(K-1\) channel fixed point. On the other hand, if \(K-1\) couplings are smaller than a single coupling, then the smaller couplings vanish while the remaining coupling flows to strong-coupling.

\section{Conclusions}\label{conclusions}
\par\noindent
In summary, we have explored the low-energy behaviour of the MCK models using the unitary renormalization group (URG) method, and obtained insight into the role of ground state degeneracy and quantum-mechanical frustration in shaping the non-Fermi liquid physics and criticality. We have also obtained the zero-temperature phase diagram of the MCK problem and an effective Hamiltonian with a simple zero bandwidth limit (i.e., the star graph), and shown that it determines the ground state energy, wavefunction and degeneracy of the MCK models. The star graph is found to explain much of the physics of over-screening and criticality of the MCK: this includes the singularities in the susceptibility and magnetisation, as well as a reduction in the degree to which the impurity spin is compensated. This demonstrates the ground state degeneracy of the star graph (which is unity for single-channel, and greater than $1$ for the over-screened models) as the key ingredient in leading to the qualitative and quantitative differences of the MCK from the single-channel Kondo model. The presence of ground state degeneracy also allows the construction of non-local twist operators and fractional excitations that span the degenerate ground state manifold.
Integrating out the quantum fluctuations of the impurity leads to an all-to-all effective Hamiltonian for the local conduction bath spins of the star graph. This Hamiltonian is found to contain inter-channel quantum fluctuations involving the scattering of electron-hole pairs, thereby creating a local Mott-liquid phase proximate to the impurity.
\par 
The low energy effective Hamiltonian (LEH) for the excitations of the MCK problem is then obtained by considering fluctuations of the conduction bath lying above the ground state. The LEH for the 2-channel Kondo confirms the absence of any local Fermi liquid physics, as well as the presence of a local marginal Fermi liquid arising from inter-channel off-diagonal scattering processes. This reinforces the idea that the ground state degeneracy is crucial to the appearance of non-Fermi liquid physics at the MCK fixed point. Further, we confirm that this arises from an orthogonality catastrophe in the form of singularities that are absent at exact screening. Studies of various thermodynamic properties (presented in the Supplementary Material~\cite{SM}) from this 2-channel Kondo LEH (e.g., specific heat and susceptibility) show logarithmic dependence at low temperature, in agreement with the literature~\cite{affleck_1991_overscreen,affleck_ludwig_1991,affleck_pang_cox_1992,affleck1993exact,parcollet_olivier_large_N,affleck_2005,emery_kivelson,andrei_destri_1984,Tsvelick1984,Tsvelick_1985,andrei_jerez_1995,zarand_costi_2002}.
\par 
The non-diagonal nature of the low energy theory is further investigated through several measures of entanglement. The entanglement entropy calculated from (i) the minimum \(J^z\) states, (ii)between the impurity and the outer spins as well as (iii) that between an outer spin and the rest, is observed to saturate to \(\ln 2\) in the large channel limit. The opposite behaviour is observed in the maximum $J^z$ state, where the entanglement entropy decreases with the increase of channel number. Moreover, the large values of inter-channel mutual information indicate high inter-channel correlation in the MCK ground state.
Indeed, the power-law dependence of the multi-partite information among the outer spins of the star graph model is similar to the power-law behavior of the all-to-all local Mott-liquid state. Importantly, the discontinuous behaviour in the entanglement entropy computed for the MCK ground state in the limit of vanishing bath dispersion, as compared to the smooth behaviour observed in its single-channel counterpart, points to the orthogonality catastrophe in the low-energy phase of the multi-channel models. We confirm this by noting a similar discontinuous behaviour in the Bures distance computed between the density matrices computed from the ground states of the star graph (i.e., zero bath dispersion bandwidth) and the MCK (i.e, a non-zero the bath dispersion bandwidth). This discontinuity in the Bures distance is observed to grow as the number of channels is increased. Finally, the URG study of the channel anisotropic MCK shows the expected critical nature of the channel isotropic fixed point: under any anisotropic perturbations, the RG flows go towards either the single-channel model or the symmetric MCK with one less channel coupled to the impurity. This is complemented by the study of the star graph ground state degeneracy which shows that the degeneracy changes only if one or more couplings vanishes.
Taken together, we conclude that in the presence of channel anisotropy, the degeneracy of the MCK ground state does not change until the RG reaches the stable fixed point with one (or more channels) completely disconnected.
\par 
We end by discussing some open directions. 
Our work presents a template for the study of how a degenerate ground state manifold arising from symmetry and duality properties in a multichannel quantum impurity model can lead to novel multicritical phases at intermediate coupling. Employed within an auxiliary model framework (such as dynamical mean-field theory), such quantum impurity models will likely unveil novel quantum critical points (or phases) in bulk lattice models of strongly correlated electrons and spins that arise from the competition of various quantum (or symmetry broken) orders. Since the degeneracy of the star graph is observed to be robust under anisotropy, 
it appears relevant to study the case of the MCK with superconducting leads, i.e., conduction baths with a superconducting gap. Specifically, it will be interesting to see whether the gap can separate the ground state manifold of the star graph from any low-energy excitations, providing thereby a protection for the topological quantum numbers associated with the degenerate ground state manifold. The duality transformations then allow us to view the protected ground state as that belonging to an isolated larger quantum spin. This holds potential for applications in quantum information and quantum computation (see \cite{lopes2020} for an alternate proposal based on an MCK model). Interesting behaviour may also be obtained by studying the MCK model with an easy-axis anisotropy term \(\left(J^i\right)^2\) along one of the axes ($i$).
Such a term would, if it is able to survive the RG flows towards low energy, lift the degeneracy of the ground state and make it possible to achieve perfect screening even with \(K > 2S_d\). Further, understanding the nature of the multicriticality in the MCK model displayed upon introducing channel anisotropy is in itself an important problem (see \cite{zheng2021} for a recent work).
One can also study the multi-channel lattice models~\cite{Piguet1997,shaw_1998} to better understand the role played by singlet frustration and ground state degeneracy in the competition between the local moment versus the heavy fermion sea. Finally, we note that experimental realisations of the star graph have been studied recently with regards to aspects of NMR-based quantum information processing such as algorithmic cooling and information scrambling (see \cite{mahesh2021} for a review). It will be interesting to see whether some of the results obtained by us for the spin-1/2 Heisenberg star graph can be tested experimentally in a similar manner. Finally, it is tempting to speculate whether a similar study can be conducted for a MCK system of 1D Tomonaga-Luttinger liquid conduction leads~\cite{lalbosonMCK2010}.

\ack

S. Patra and Anirban Mukherjee thank the CSIR, Govt. of India and IISER Kolkata for funding through a research fellowship. Abhirup Mukherjee thanks IISER Kolkata for funding through a junior and a senior research fellowship. S. Lal thanks the SERB, Govt. of India for funding through MATRICS grant MTR/2021/000141 and Core Research Grant CRG/2021/000852. We thank three anonymous referees for their valuable questions and comments that greatly helped improve the presentation of the work.

\section*{References}

\bibliography{MCK-manuscript}

\end{document}


\title{Supplementary Materials for ``Frustration shapes multi-channel Kondo physics: a star graph perspective"}

\author{Siddhartha Patra$^1$, Abhirup Mukherjee$^1$, Anirban Mukherjee$^1$, N. S. Vidhyadhiraja$^4$, A. Taraphder$^5$ and Siddhartha Lal$^1$}
\eads{\mailto{sp14ip022@iiserkol.ac.in}, \mailto{am18ip014@iiserkol.ac.in}, \mailto{mukherjee.anirban.anirban@gmail.com}, \mailto{raja@jncasr.ac.in}, \mailto{arghya@phy.iitkgp.ernet.in}, \mailto{slal@iiserkol.ac.in}}

\address{$^1$Department of Physical Sciences, Indian Institute of Science Education and Research-Kolkata, W.B. 741246, India}

\address{$^4$Theoretical Sciences Unit, Jawaharlal Nehru Center for Advanced Scientific Research, Jakkur, Bengaluru 560064, India}

\address{$^5$Department of Physics, Indian Institute of Technology Kharagpur, Kharagpur 721302, India}

\date{\today}

\section{Additional Properties of the Star Graph}

\subsection{Partition function of the zero-mode fixed-point Hamiltonian, in the presence of a magnetic field}
The zero-mode approximation of the fixed-point Hamiltonian is a star graph Hamiltonian:
\begin{eqnarray}
	H = J^* \vec{S_d}\cdot\vec{s}_\text{tot}~,
\end{eqnarray}
where \(\vec s_\text{tot} = \sum_l \vec s_l\) is the total spin operator for all the channels. We insert a magnetic field that acts only on the impurity and then attempt to diagonalize the Hamiltonian.
\begin{eqnarray}
	\label{stargraph_field_hamiltonian}
	H(h) = J^* \vec{S_d}\cdot\vec{s}_\text{tot} + h S_d^z~.
\end{eqnarray}
The Hamiltonian commutes with \(s_\text{tot}^2\):
\begin{eqnarray}
	\left[s_\text{tot}^2, H(h)\right] &= \left[\sum_{i=x,y,z}{s^i_\text{tot}}^2, J^* \sum_{i=x,y,z} S_d^i s^i_\text{tot}\right] = \sum_{i,j}J^* S_d^i \left\{s_\text{tot}^i, \left[s_\text{tot}^i,s_\text{tot}^j\right]\right\}\nonumber\\
							 & = \sum_{i,j}J^* S_d^i \left\{s_\text{tot}^i, i \epsilon^{ijk}s^k_\text{tot}\right\} = 0~.
\end{eqnarray}
This means the Hamiltonian is already block-diagonal in the quantum number \(s_\text{tot}\). Let us represent the quantum number of \(s_\text{tot}^z\) by \(m\). For a particular \(s_\text{tot}\), \(m\) can take values from the set \(\left[-s_\text{tot}, s_\text{tot}\right] \). The spin \(S_d^z\) can also take values \(\pm \frac{1}{2}\). From now on, we will assume we are in the subspace of a particular \(s_\text{tot} = M\), so we will ignore that quantum number and write the kets simply as \(\ket{S_d^z, m}\). So, the notation \(\ket{\uparrow,-1}\) means the state with \(S_d^z = \frac{1}{2}\) and \(m = -1\). We will now show that even inside the block of \(2\times s_\text{tot}\) (or \(2\times s_\text{tot} + 1\), depending on where it is odd or even) defined by a particular value of \(s_\text{tot}\), the Hamiltonian actually separates into decoupled \(2\times 2\) blocks. To see why, first note that the terminal states \(\ket{\downarrow, -M}\) and \(\ket{\uparrow, M}\) are already eigenstates, because they cannot scatter (the impurity can only flip down, and this would require the bath to flip up, but \(s^z_\text{tot}\) is already at its maximum value \(M\)). The other \(2M - 2\) states can be organized into \(2\times 2\) blocks formed by the states \(\ket{\uparrow, m}\) and \(\ket{\downarrow, m+1}\) for \(m \in \left[-M, M-1\right] \). The fact that this block does not interact with the other blocks can be observation: if there was some other state which when acted upon by the Hamiltonian gave a non-zero projection on \(\ket{\uparrow, m}\), it would have to come from \(S_d^z = \downarrow\), and this would mean the bath spin would have had to flip down. This means the bath spin in that state would have to be \(m+1\), and that is precisely the other state in the block. 

Defining \(\epsilon^h_m = \frac{1}{2}\left(Jm + h\right) \) and \(x^M_m = M(M+1) - m(m+1)\), the \(2\times 2\) blocks can be written as
\begin{eqnarray}
	H_m = \begin{pmatrix} \epsilon^h_m & \frac{J}{2}\sqrt{x^M_m} \\ \frac{J}{2}\sqrt{x^M_m} & -\left( \epsilon^h_m + J/2 \right)   \end{pmatrix} ~.
\end{eqnarray}
The eigenvalues are 
\begin{eqnarray}
	\label{eigenvalue}
	\lambda_{m, \pm}^{M, h} = \frac{1}{2}\left[-J/2 \pm \sqrt{J^2/4 + J^2 x_m^M + 4\epsilon^h_m\left(\epsilon^h_m + J/2\right) }\right] = -J/4 \pm \sqrt{J^2x^M_m/4 + \alpha^2}~,
\end{eqnarray}
where \(\alpha = \epsilon^h_m + J/4\).
The eigenvalues of the terminal states are \(\pm\epsilon^h_{\pm M}\). For \(h = 0\), the ground state subspace is \(K-\)fold degenerate and is formed by the negative solutions of eq.~\eqref{eigenvalue}. This common \(K-\)fold degenerate eigenvalue is \(-J(M+1)/2\).
The eigenstates for each value of \(M,m\) in the \(J=M-\frac{1}{2}\) sector are given by
\begin{eqnarray}
	\ket{M-\frac{1}{2},m+\frac{1}{2},M} =\mathcal{C}_\pm \ket{\uparrow, M, m} \pm \sqrt{1 - \mathcal{C}_\pm^2} \ket{\downarrow, M, m+1}, \\
	\mathcal{C}_\pm = \frac{J\sqrt{x_m^M}/2}{\sqrt{J^2 x_m^M/4 + \left(\alpha \mp \sqrt{\alpha^2 + J^2 x_m^M/4}\right)^2 }}~.
\end{eqnarray}
The partition function is given by
\begin{eqnarray}
	Z(h) = \sum_{M=M_\text{min}}^{M_\text{max}}\left[\sum_{m=-M, \atop{m\in \mathbb{Z}}}^{M-1}2e^{\beta J/4}\cosh \beta\sqrt{J^2x^M_m/4 + \alpha^2} + 2e^{-\beta JM/2}\cosh \beta h/2\right]~,
\end{eqnarray}
where \(M_\text{max} = K/2\) for a \(K-\)channel Kondo model, and \(M_\text{min} = 0\)  if \(K\) is even, otherwise \(1/2\). This is yet not the complete partition function, because we have not accounted for the possibility that there multiple subspaces of \(M\). For example, the \(K=3\) case states can be obtained by adding the third spin-half onto the states \(S=0,1\). \(S=0\) gives \(s_\text{tot}=1/2\) and \(S=1\) gives \(s_\text{tot} = 1/2, 3/2\). So, \(s_\text{tot} = 1/2\) appears twice. These two subspaces are actually orthogonal, because the quantum numbers for the individual channels are different. We need to count the number of instances of a particular subspace \(s_\text{tot}=M\). It turns out that this number is given by \(r^K_M = {}^{K-1}C_{K/2 - M}\), which means the correct partition function is
\begin{eqnarray}
	Z(h) &=\sum_{M=M_\text{min}}^{M_\text{max}}r^K_M\left[\sum_{m=-M, \atop{m\in \mathbb{Z}}}^{M-1}2e^{\beta J/4}\cosh \beta\sqrt{J^2x^M_m/4 + \alpha^2} + 2e^{-\beta JM/2}\cosh \beta h/2\right]~.
\end{eqnarray}

\subsection{Features of the polarised ground-states}
Of particular importance within the ground-state subspace are the zero-field maximally-polarised states, obtained by choosing \(M = \frac{K}{2}, m + \frac{1}{2}= \pm \left( M - \frac{1}{2} \right) =\pm \frac{1}{2}\left(K-1\right)\):
\begin{eqnarray}
	\ket{J=\frac{K-1}{2},J^z= \frac{\sigma}{2}\left(K-1\right),s_\text{tot}=\frac{K}{2}} \nonumber\\
	= \frac{1}{\sqrt{1 + K}} \ket{\sigma}_d\otimes\ket{s_\text{tot}^z=\sigma(\frac{K}{2}-1)} - \frac{\sqrt K}{\sqrt{1 + K}}\ket{\bar\sigma}_d\otimes\ket{s_\text{tot}^z=\sigma\frac{K}{2}},
\end{eqnarray}
where \(\sigma\) can take values \(\pm 1\) and \(\ket{\sigma}_d\) represents the up and down configurations of the impurity spin. The state \(\ket{s_\text{tot}^z=\frac{\sigma K}{2}}\) has all the bath spins pointing in the \(\sigma-\)direction: \(\ket{s_\text{tot}^z=\frac{\sigma K}{2}} = \ket{\sigma,\sigma,\ldots,\sigma} (\sigma=\uparrow \text{ or }\downarrow)\), where the notation is such that a particular symbol (say \(\uparrow\)) at the \(i^\text{th}\) position indicates the configuration of the \(i^\text{th}\) channel. The other state \(\ket{s_\text{tot}^z=\sigma(\frac{K}{2} - 1)}\) can be obtained by flipping any one out of the \(K\) spins in the previous state \(\ket{s_\text{tot}^z=\frac{\sigma K}{2}}\), leading to the normalised state \(\ket{s_\text{tot}^z=\sigma(\frac{K}{2} - 1)} = \frac{1}{\sqrt K}\ket{\bar\sigma,\sigma,\ldots,\sigma} + \frac{1}{\sqrt K}\ket{\sigma,\bar\sigma,\ldots,\sigma} + \ldots + \frac{1}{\sqrt K}\ket{\sigma,\sigma\ldots,\bar\sigma}\), a linear combination of all possible states with the spin of one channel in \(\bar\sigma(=-\sigma)\) direction and all other channels remaining in the \(\sigma\) direction. With these in mind, the polarised state can be cast in the more explicit form
\begin{eqnarray}
	\ket{J^z= \frac{\sigma}{2}\left(K-1\right)} = \frac{1}{\sqrt{K(1 + K)}} &\left[\ket{\sigma}_d\otimes\left(\ket{\bar\sigma,\sigma,\ldots,\sigma} + \ket{\sigma,\bar\sigma,\ldots,\sigma} + \ldots + \ket{\sigma,\sigma\ldots,\bar\sigma}\right) \right.\nonumber\\
										&\left.- K\ket{\bar\sigma}_d\otimes \ket{\sigma,\sigma,\ldots,\sigma}\right].\label{polarised-state}\qquad
\end{eqnarray}

In order to make the frustration faced by the impurity manifest, we now rewrite these polarised states in the form of superpositions of zero magnetisation singlet states. Let us define the singlet state \(\ket{\text{SS}_{d,l}} = \frac{1}{\sqrt 2}\left(\ket{\uparrow}_d\ket{\downarrow}_l - \ket{\downarrow}_d\ket{\uparrow}_l\right),l\in[1,K]\), between the impurity spin and the \(l^\text{th}\) channel, where \(\ket{\cdot}_l\) represents the configuration of the spin in the \(l^\text{th}\) channel. The polarised states can then be written as
\begin{eqnarray}
	\ket{J^z= \frac{\sigma}{2}\left(K-1\right)} &= \frac{1}{\sqrt{K(1 + K)}} \sum_{l=1}^K\ket{\sigma}_d\otimes\left[\ket{\bar\sigma}_l - \ket{\sigma}_l\right] \otimes\ket{\sigma,\ldots,\sigma}_{l^\prime}\nonumber \\
						    &= \frac{\sqrt 2}{\sqrt{K(1 + K)}} \sum_{l=1}^K \ket{\text{SS}_{d,l}}\otimes\ket{\sigma,\ldots,\sigma}_{l^\prime}.\label{singlet-chain}
\end{eqnarray}
This form of the state shows that the impurity is trying to participate, simultaneously, in forming \(K\) number of singlets. Put differently, the presence of \(K\) singlets shows that all \(K\) channels are trying to screen the impurity simultaneously. However, the presence of an overall non-zero magnetisation in the states means that the screening is only partial.

From these polarised states, one can now calculate the ``excess charge" supplied by the impurity site to the conduction bath in the form of gapless excitations. For this, we first note that the form eq.~\eqref{singlet-chain} of the polarised state shows that all the individual terms in the more explicit form of eq.~\eqref{polarised-state} involve the impurity hybridising with the bath. In eq.~\eqref{polarised-state}, each term of the form \(\ket{\sigma}_d\otimes\ket{\bar\sigma}_l\otimes\ket{\sigma,\ldots,\sigma}_{l^\prime}\) participates in exactly one singlet \(\ket{\text{SS}_{d,l}}\) and therefore involves hybridisation between the impurity and only the \(l^\text{th}\) channel. Each such term therefore contributes an excess charge \(n_\text{exc}^{(l)}\) of unity to the \(l^\text{th}\) conduction channel and none to the other channels: \(n_\text{exc}^{(l)} = 1 = 1 - n_\text{exc}^{(l^\prime)}\). The final term \(\ket{\bar\sigma}_d\otimes \ket{\sigma,\sigma,\ldots,\sigma}\) in eq.~\eqref{polarised-state} participates in all \(K\) singlets, and therefore involves hybridisation with all the channels. This term therefore contributes the entire excess charge of unity across all the channels, which results in an excess charge contribution of \(n_\text{exc}^{(l)}=1/K\) to each channel. With these in mind, the total excess charge contributed to each channel from the state in eq.~\eqref{polarised-state} is
\begin{eqnarray}
	n_\text{exc}^{(l)} = \frac{1}{K(1+K)}\left[1 + \left(K^2 \times \frac{1}{K}\right) \right] = \frac{1}{K}~.
\end{eqnarray}
This result then connects to the Friedel scattering phase shift (eq.(20) of the main manuscript) and the magnetisation of the impurity (eq.(21) of the main manuscript). In this way, the excess charge $n_\text{exc}^{(l)}$ of the star graph is also observed to guide the RG flow of the MCK problem to the intermediate coupling fixed point (eq.(55) of the main manuscript), and lead to the power-law behaviour of various thermodynamic quantities at the fixed point (eq.(56) of the main manuscript).

\subsection{Energy lowering due to quantum fluctuations}
The star graph Hamiltonian that sits at the heart of the RG fixed point Hamiltonian of the multichannel Kondo model contains a classical Ising term and a quantum fluctuation term:
\begin{eqnarray}
	H = {\mathcal{J}}\vec{S_d}\cdot\vec S = \mathcal{J}\left[\underbrace{S_d^z S^z}_\text{Ising term} + \underbrace{\frac{1}{2}\left(S_d^+ S^- + \text{h.c.}\right)}_\text{quantum fluctuation part} \right] = \mathcal{J}\left[\mathcal{H}^C + \mathcal{H}^Q\right] ~.
	\label{eq:stargraph_hamiltonian}
\end{eqnarray}
We are interested in studying how the presence of quantum fluctuations in the Hamiltonian of eq.~\eqref{eq:stargraph_hamiltonian} stabilises the ground state. This can be understood by looking at the contributions, per channel, of the classical and the quantum parts to the total ground state energy \(E_g\). In the over-screened case \(\left( K > 2S_d \right) \), the contributions are
\begin{eqnarray}
	E_{C} = \frac{1}{K}\langle \psi_g | \mathcal{H}^C | \psi_g \rangle, \quad E_{Q} = \frac{1}{K}\bra{\psi_g}\mathcal{H}^Q \ket{\psi_g} =\frac{E_g/\mathcal{J} - E_{C}}{K} =  -\frac{1}{2}S_d - \frac{1}{K}\left(S_d + E_{C}\right)~,
\end{eqnarray}  
where we have defined the contributions \(E_C\) and \(E_Q\) to the ground state energy coming from the Ising and fluctuation parts respectively, and we have substituted the ground state energy \(E_g\) for the over-screened case from eq.(10) in the main text. This contribution $E_Q$ is generated by the spin-flip fluctuations between the impurity spin and the outer spins of the star graph. Note that because of the degeneracy of the ground state subspace, there are multiple ground states characterised by different values of \(J^z = S^z + S_d^z\), and different ground states lead to different values of \(E_Q\).

\begin{figure}[!htb]
\centering
\includegraphics[width=0.7\textwidth]{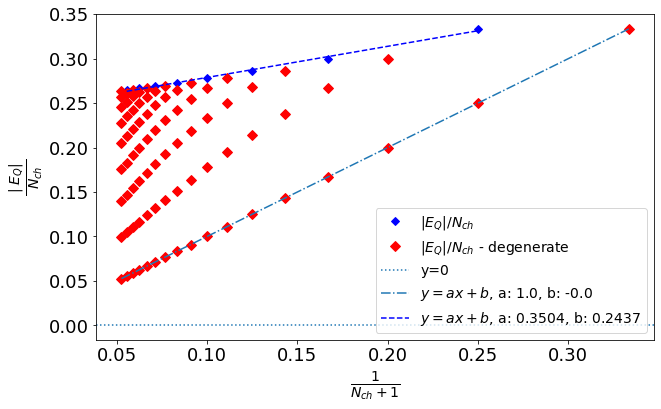}
\caption{This shows the variation of quantum energy per channel with $1/N$, where $N=N_{ch}+1$ is the total number of spins in the systems including the impurity spins.}
\label{fig:quantum_energy}
\end{figure}

The variation of \(E_Q\) is shown in Fig.\eqref{fig:quantum_energy} for the case of $S_d=1/2$, as functions of the number of channels \(K\) as well as the ground state value \(J^z\) in which \(E_Q\) is being calculated. For a particular channel $K$, we find that the maximum \(|E_Q|\) is obtained for the state with smallest magnitude of $J^z$ (\(J^z = 0 \) for odd \(K\) and \(J^z = \pm 1/2\) for even \(K\)), whereas \(|E_Q|\) is minimum in the states with largest \(J^z\) ($J^z=\pm J$). The larger values of \(|E_Q|\) indicate that the quantum fluctuations are largest in the states with minimum \(J^z\), and these states can be thought of as the counterparts to the maximally entangled singlet seen in the single-channel Kondo problem. This behaviour persists as \(K \gg 1\): the magnitude \(|Q_Q|\) associated with the $|J^z|= J$ states vanish, showing the classical nature of these states and the lack of quantum fluctuations in that limit. On the other hand, the state $|J^z| = |J^z|_{min}$ has a non-zero \(|E_Q|\) in the large \(K\) limit and hence contains some non-zero quantum fluctuations, showing the true quantum nature of this macroscopic singlet state.

\subsection{Measure of quantum fluctuations}
Here we are interested in calculating the quantum fluctuation present in the  ground state by measuring the expectation value of the spin-flip part of the Hamiltonian: ${\mathcal{Q}}\equiv \langle \psi_g | (J^x)^2+(J^y)^2 |\psi_g\rangle = \langle \psi_g | J^2 - (J^z)^2 |\psi_g\rangle$. For a general spin-$S_d$ impurity, the ground state is characterised by $J=|K/2-S_d|$, which gives $J^2=J(J+1)=(|K/2-S_d|)(|K/2-S_d|+1)$. We define $\Delta=(K/2-S_d)$ to quantify the deviation from exact screening; $\Delta=0$ represents the exactly screened problem, and $\Delta>0$ and $\Delta<0$ represents the over-screened and under-screened problems respectively. Among the degenerate ground state, the maximum quantum fluctuation will occur in the state with minimum \(|J^z|\). Since \(J\) can actually be written as \(J = |\Delta|\), \(J\) will take integer values when \(\Delta\) is integer, and the minimum value of \(|J^z|\) is then zero. Otherwise, when \(\Delta\) is half-integer, \(J\) will also take half-integer values, and the minimum value of \(|J^z|\) is \(1/2\). With these considerations, the minimum value of the quantum fluctuation, {\it per channel}, is
\begin{eqnarray}
	q_K &= \frac{1}{K}\sqrt{Q_\text{max}} = \frac{1}{K}\sqrt{\langle \psi_g | J^2 - |J^z|_{min}^2 |\psi_g\rangle}\nonumber\\
					     &= \begin{cases}
	\frac{1}{K}\sqrt{(|K/2-S_d|)(|K/2-S_d|+1)-1/4}, & \text{ where }\Delta\text{ is half-integer}\\
	\frac{1}{K}\sqrt{(|K/2-S_d|)(|K/2-S_d|+1)}, & \text{ where }\Delta\text{ is integer}
\end{cases}~.\quad
\end{eqnarray}
This expression is symmetric under the transformation \(\Delta \to -\Delta\), and represents a duality between the over-screened and under-screened models. In the limit of large channel number \(K\), \(q_K\) simplifies to $\lim_{K\rightarrow \infty} q_K= \lim_{K \to \infty}\frac{1}{2K}|\Delta(K)|$. Only the over- and under-screened models have non-vanishing (and equal) values of \(q_K\).

\subsection{Staggered magnetization of the star graph}
 Another probe to study the screening as a function of \(K\) is the staggered magnetization \(\vec{M}_s=\vec{S}-\vec{S}_d\). One can rewrite the Hamiltonian as $\vec{S}_d\cdot\vec{S}= \frac{1}{4}[J^2 - M_s^2]$. Since \(J\) and \(M_s\) commute with each other and hence also with the Hamiltonian, the eigenvalues of $\vec{J}^2$ and \(M_s^2\) act as good quantum numbers for in the ground states of the star graph model. The larger the value of \(\left<M_s^2\right>\), the stronger is the screening in that ground state. For single-channel case with $K=1$, the ground state is a unique 2-spin singlet $|\psi_g\rangle =\frac{1}{\sqrt{2}} (|\uparrow\downarrow\rangle-|\downarrow\uparrow\rangle = |J=0,J_z=0\rangle$, and the staggered magnetization per direction is \(M_s^2/3 = 1\), which shows the perfect screening.

Calculating the staggered magnetization for a general multichannel problem with \(K\) channels and a spin-\(S_d\) impurity reveals the breakdown of screening brought about by the presence of multiple channels. We find, in general, that the square of the staggered magnetization per channel is
\begin{eqnarray}
	m_s^2 &= \left< \left(\frac{1}{K}M_S\right)^2\right> = \frac{1}{K^2}\langle \psi_g | (\vec{S}_d - \vec{S})^2 |\psi_g\rangle = \frac{1}{K^2}\langle \psi_g | 2(\vec{S}_d^2 + \vec{S}^2)-\vec{J}^2 |\psi_g\rangle \nonumber\\
	      &= \frac{2S_d(S_d+1)}{K^2}+\frac{1}{4}+\frac{1}{K}+\frac{1}{4K^2}~.
\end{eqnarray}

\begin{figure}
\centering
\includegraphics[width=0.7\textwidth]{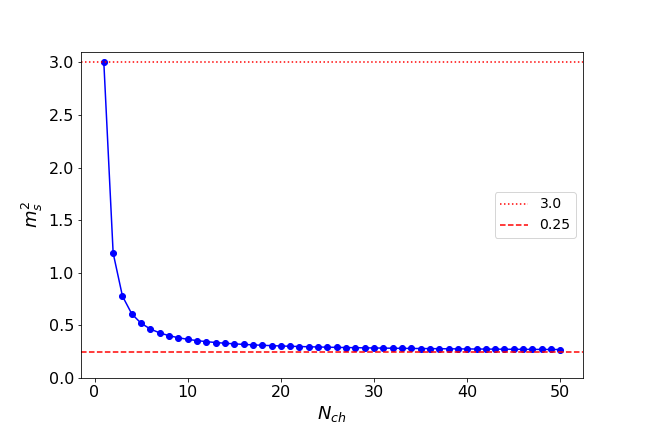}
\caption{This shows how the staggered magnetization changes with the number of channels $N_{ch}$.}
\label{fig:st_mag}
\end{figure}
This shows that as the number of channels increases, the value of \(m_s^2\) decreases from the perfectly-screened value of 3. For \(K \to \infty\) and keeping \(S_d\) finite, \(m_s^2\) approaches a lowered value of \(1/4\).

Due to the $SU(2)$ symmetry of the problem, the staggered magnetisation is the same along each of the three directions, such that the total value \(m_s^2\) is three times the value in any particular direction:
\begin{eqnarray}
\langle(m_s^x)^2\rangle=\langle(m_s^y)^2\rangle=\langle(m_s^z)^2\rangle=\frac{1}{3}m_s^2~,
\end{eqnarray}
As mentioned above, the single channel case displays a value of $\langle (m_s^z)^2 \rangle =1$, showing the perfect screening. If one takes the limit of \(K \to \infty\) keeping \(S_d = K/2\), we get the staggered magnetisation at exact-screening and large \(K\), and the value is $\langle (m_s^z)^2 \rangle =1/4$. This is reduced from the value at \(K=1\). This reduced value however is still greater than the value of \(1/12\) reached as \(K\to \infty\) away from exact-screening by keeping \(S_d\) finite. These conclusions can be easily understood from fig.\eqref{fig:st_mag} where we have shown the variation of \(\left<m_s^2 \right>\) as a function of \(K\), keeping \(S_d=1/2\).

\subsection{Thermodynamic quantities}

\subsubsection{Impurity magnetization in terms of parity operators}
Just like the complete string operator \(\pi^z\), the modified string operator \(\sigma_d^z \pi^z\) is also a Wilson loop operator that wraps around only the outer nodes of the star graph:
\begin{eqnarray}
	\pi^z_c \equiv \sigma_d^z \pi^z = \exp\left[i \frac{\pi}{2} \left(\sum_{l=1}^K \sigma^z_l - K\right)\right] ~.
\end{eqnarray}
The expectation value of the impurity magnetization along a particular direction and in specific ground states can be related to the 't Hooft operator. We will work in the state comprised of two adjacent eigenstates of \(J^z\):
\begin{eqnarray}
	\ket{g^\theta_{J^z}} \equiv \frac{1}{\sqrt 2}\left( \ket{J^z} + e^{i\theta}\ket{J^z+1}\right), && J^z < \frac{1}{2}\left( K-1 \right)~.
\end{eqnarray}
The expectation value of the impurity magnetization operator \(\sigma_d^x\) can be expressed as
\begin{eqnarray}
	\left<\sigma_d^x\right> \equiv \langle g^\theta_{J^z} \vert \sigma_d^x \vert g^\theta_{J^z}\rangle = - \langle J^z + 1 \vert \pi^x_c \vert -J^z \rangle + \text{h.c.}~.
\end{eqnarray}
This expression relates the observable impurity magnetization to the topological 't Hooft operator \cite{Maric2020}. Evaluating the matrix elements gives
\begin{eqnarray}
	\label{sigmax}
	\left<\sigma_d^x\right> = - \frac{\sqrt{K^2 - (2J^z + 1)^2}}{2(1+K)}\cos \theta~.
\end{eqnarray}
Performing a similar calculation reveals that the impurity magnetizations along \(y\) and \(z\) in the same state are given by
\begin{eqnarray}
	\label{sigmayz}
	\left<\sigma_d^y\right> = - \frac{\sqrt{K^2 - (2J^z + 1)^2}}{2(1+K)}\sin \theta, &&\left<\sigma_d^z\right> = - \frac{2J^z + 1}{(1+K)}~.
\end{eqnarray}
Combining eqs.~\eqref{sigmax} and \eqref{sigmayz}, we find
\begin{eqnarray}
	\cos^2\theta\left(\left<\sigma^x_d\right>\right)^2 + \sin^2\theta\left(\left<\sigma^y_d\right>\right)^2 + \frac{1}{4}\left(\left<\sigma^z_d\right>\right)^2 = \frac{1}{4}\left(\frac{K}{1+K}\right)^2~.
\end{eqnarray}
This relation \textit{constrains the values of the magnetization} in all the directions: the \(x\) and \(y\) magnetization values have already been shown to be related to the `t Hooft operators \(\pi^x\) and \(\pi^y\) and the magnetization along \(z\) is therefore constrained in terms of the `t Hooft operators and the quantized function on the right-hand side (the function is quantized because \(K\) can only take integer values).

\subsubsection{Thermal entropy}
The star graph model can be solved to obtain the partition function, and this then allows the computation of the Helmholtz free energy and hence the thermal entropy:
\begin{eqnarray}
\mathcal{F}= -k_B T\log Z, \nonumber\\
S = -\frac{\partial \mathcal{F}}{\partial T} = -k_B \log Z -k_B T \frac{1}{Z} \frac{dZ}{dT} = -k_B \log \sum_{\epsilon} d(\epsilon) e^{-\beta \epsilon}  -\frac{1}{\beta} \frac{k_B\sum_\epsilon \epsilon d(\epsilon) e^{-\beta \epsilon} \beta^2   }{\sum_\epsilon  d(\epsilon)e^{-\beta \epsilon}}~,
\end{eqnarray}
where \(d(\epsilon)\) is the degeneracy of the state at energy \(\epsilon\). The high and low-temperature limits take very simple forms:
\begin{eqnarray}
\lim_{\beta\rightarrow \infty} S = -k_B \log_2 d(\epsilon_{G})~,\quad\lim_{\beta\rightarrow 0} S = -k_B \log \sum_\epsilon d(\epsilon)~.
\end{eqnarray}

\begin{figure}
\centering
\includegraphics[width=0.7\textwidth]{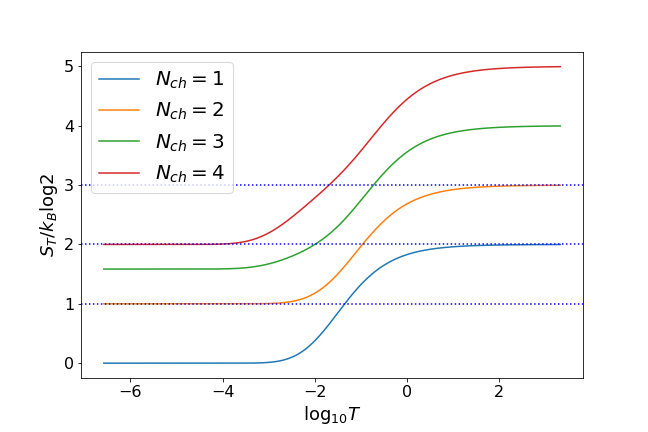}
\caption{This shows the variation of thermal entropy with the temperature.}
\label{fig:thermal_entropy}
\end{figure}

In fig.\eqref{fig:thermal_entropy}, we plot the thermal entropy (in units of $k_B \log 2$) for a range of temperatures and for different values of \(K\). At large temperatures, \(S\) saturates to integer multiples of \(k_B \log 2\), while at low-temperatures, that is not always the case. This is because the total Hilbert space dimension \(\sum _\epsilon d(\epsilon)\) is simply \(2^N\) (\(N\) being the total number of 1-particle states in the Hilbert space), and therefore \(\log \sum _\epsilon d(\epsilon) = N \log 2\) is always an integer multiple of \(N\). On the other hand, the ground state degeneracy \(d(\epsilon_G)\) is \(|K - 2 S_d|\) which is not necessarily of the form \(2^N\) and hence does not always lead to an integer multiple of \(\log 2\).

\section{Additional Topological Features of the Local Mott liquid}
\subsection{Non-local twist operators and ground state degeneracy}

The all-to-all Hamiltonian obtained by resolving the impurity-bath quantum fluctuations is written in terms of the spinor spin operators which are individually made out of two electronic degrees of freedom.
\begin{eqnarray}
H_{eff} 
 =\frac{\beta_{\uparrow}({\mathcal{J}},\omega_{\uparrow})}{4} (S^+S^-+S^-S^+)~,
\label{eq:all-to-all_1}
\end{eqnarray}
where \(\beta_{\uparrow} ({\mathcal{J}},\omega_{\uparrow})= ({\mathcal{J}}^2 \Gamma_{\uparrow})/2, ~ ~\Gamma_{\uparrow}=(\omega_{\uparrow}-{\mathcal{J}}(S_d^z-1))^{-1}\). This spin operator is defined as $\vec{S}_i=\frac{\hbar}{2} \displaystyle\sum_{\substack{ \alpha,\beta\in \{\uparrow,\downarrow\}}}  c_{0\alpha}^{(i)\dagger} \vec{\sigma}_{\alpha\beta} c_{0\beta}^{(i)}$. In the above eq.\eqref{eq:all-to-all_1} we see the $U(1)$ symmetry of the effective Hamiltonian. Using the spinor representation, we obtain the spin creation operation in terms of the electronic degree of freedom:
\begin{eqnarray}
S_i^+ = \frac{\hbar}{2} \displaystyle\sum_{\alpha,\beta\in \{\uparrow\downarrow\}} c_{0\alpha}^{(i)\dagger} {\sigma}^{+}_{\alpha\beta} c_{0\beta}^{(i)}
=\frac{\hbar}{2} c_{0\uparrow}^{(i)\dagger} c_{0\downarrow}^{(i)}, \quad S_i^z = \frac{\hbar}{2} \displaystyle\sum_{\alpha,\beta\in \{\uparrow\downarrow\}} c_{0\alpha}^{(i)\dagger} {\sigma}^{z}_{\alpha\beta} c_{0\beta}^{(i)} = \frac{\hbar}{2} (c_{0\uparrow}^{(i)\dagger} c_{0\uparrow}^{(i)} - c_{0\downarrow}^{(i)\dagger} c_{0\downarrow}^{(i)} )~.\qquad
\end{eqnarray}
This spinor is simply the Anderson pseudo-spin formulation in the spin channel. The spin-creation operator ($S_i^+$) involves the simultaneous creation of an electron-hole pair at the real space origin of the $i^{th}$ conduction channel. The condensation of such electron-hole pairs has already been shown in ~\cite{anirbanmott1,anirbanmott2}. Thus, for this effective Hamiltonian, one can define twist-translation operations to construct the gauge theory and unveil any hidden degeneracy.

Let's recall the effective Hamiltonian
\begin{eqnarray}
H_{eff} &=& \frac{\beta_{\uparrow}(\alpha,\omega_{\uparrow})}{4} \left[ \displaystyle\sum_{ij} S_i^+S_j^- ~+ \textrm{h.c.} \right]~,
\end{eqnarray}
where $i,j$ are the channel indices. Due to the all-to-all nature of the connectivity, one can draw a total of $K!$ possible unique closed paths ($\mathcal{C}_{\mu}$) such that each node (channel) is touched exactly once. These curves lead to $K!$ translation operators ($\hat{T}_{\mu}$) which keeps the Hamiltonian invariant. Let's define one such translation operator $\hat{T}_{\mu}=e^{i\hat{P}^{cm}_{{\mu}}}$ which gives a periodic shift along the closed path $\mathcal{C}_{\mu}$. The twist operator along the path $\mathcal{C}_{\mu}$ is given by
\begin{eqnarray}
\hat{\mathcal{O}}_{\mu} &=& \exp({i\frac{2\pi}{K} \displaystyle\sum_{\substack{j=1\\ \mathcal{C}_{\mu}}}^{K} j S_j^z} )~,
\end{eqnarray}
The action of the translation operator on \(S_j^z\) is to translate it by one channel index: $\hat{T}_{\mu} S_j^z \hat T^\dagger_\mu = S_{j+1}^z$ where $j+1$ and $j$ are the nearest neighbor on the closed path $\mathcal{C}_{\mu}$. Then the braiding rule between the twist and translation operators are give as
\begin{eqnarray}
\hat{T}_{\mu}\hat{\mathcal{O}}_{\mu} \hat{T}^{\dagger}_{\mu} \hat{\mathcal{O}}_{\mu}^{\dagger} = \exp\{i[2\pi S_1^z-\frac{2\pi}{K}S^z]\} = \exp\{i[\pm \pi - \frac{2\pi}{K} S^z]\} =\exp(i\frac{2\pi p}{q})~.
\end{eqnarray}
The availability of the non-trivial braiding statistics between these twist and translation operators is possible if $p\neq 0$ and $q\neq \infty$. Further simplification leads to the condition
\begin{eqnarray}
 \pm\pi -\frac{2\pi}{K} S^z = \frac{2\pi p}{q} \implies  \pm\frac{1}{2}-\frac{S^z}{K} = \frac{p}{q} , ~ ~ ~ \frac{(\pm K-2S^z)}{2K} = \frac{p}{q}~,
\end{eqnarray} 
where $p,q$ are mutual primes. We know that the $S^z$ can take values $(\mp K/2\pm m)$ where $m$ is a integer $0 \leq m \leq K$. Putting this value in the above equation leads to two possible solutions.
\begin{eqnarray}
\frac{(K-m)}{K} &=& \frac{p}{q} ~,\quad \frac{-m}{K}=\frac{p}{q}~.
\end{eqnarray}
For the first case where $K-m=p$, we can see that $m=K$ makes $p$ trivial, and hence the allowed values are $m=0,\cdots,K-1$, which represents the corresponding $S^z$ eigenvalues 
\begin{eqnarray}
&&-K/2,-K/2+1,\cdots,K/2-2,K/2-1~.
\end{eqnarray}
Similarly the second case implies, $-m=p$, but $p=0$ is not allowed as this makes the braiding statistics trivial, thus the possible $S^z$ values are 
\begin{eqnarray}
-K/2+1,-K/2+2,\cdots,K/2-1,K/2~.
\end{eqnarray}
This gives the general braiding statistics between the twist and the translation,
\begin{eqnarray}
\hat{T}_{\mu}\hat{\mathcal{O}_{\mu}} \hat{T}^{\dagger}_{\mu}\hat{\mathcal{O}_{\mu}}^{\dagger} &=& e^{i\frac{2\pi p}{K}}~,
\end{eqnarray}
where $p$ corresponds to different $S^z$ states, related as $p=\pm K/2-S^z$. Thus we can see that there are $K$ possible $S^z$ plateau states in the all-to-all model where each plateau is $K$ fold degenerate. This $p/K$ is similar to the filling factor of the fractional quantum Hall effect. There are $K!$ pairs of twist and translation operators corresponding to different closed paths $\mathcal{C}_{\mu}$ which can probe this degeneracy.

\subsection{Action on the Hamiltonian}
\begin{figure}[htpb]
	\centering
	\includegraphics[width=0.8\textwidth]{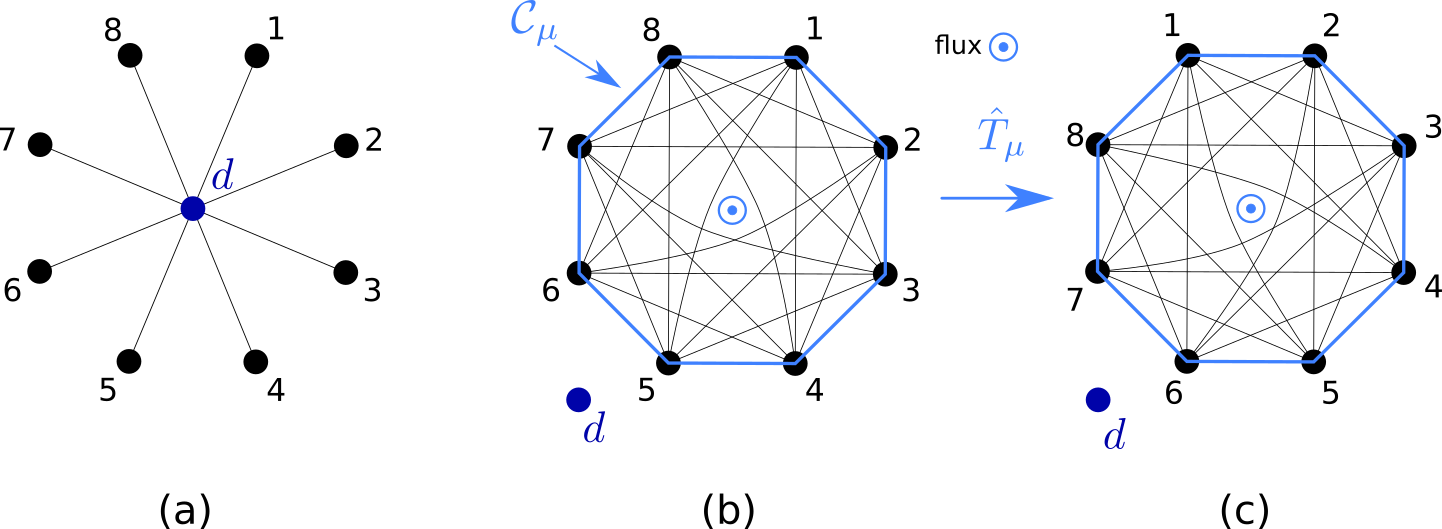}
	\caption{(a) Schematic diagram of $8-$channel multichannel Kondo model zero mode in zero bandwidth limit. (b) All-to-all model obtained by disentangling the impurity spin from the bath zero modes. $C_{\mu}$ is a particular closed curve. (c) Translated configuration along the curve $C_{\mu}$ by one unit.}
	\label{fig:stargraph-to-alltoall}
\end{figure}

We will now obtain the action of these twist operators on the Hamiltonian in eq.~\eqref{eq:all-to-all_1}. Due to the all-to-all nature of the effective Hamiltonian one can find $K!$ possible relative arrangement of those $K$ channels which keeps the Hamiltonian invariant. Here we briefly discuss the choice of the closed-loop $\mathcal{C}_{\mu}$ and the insertion of the flux. As shown in the Fig.\eqref{fig:stargraph-to-alltoall}(b) we have chosen a particular closed path which crosses all the outer spin only once. We embed that closed-loop on a plane and put the flux perpendicular to the plane through the closed loop. One can find a different closed loop where the ordering of the outer spins will be different. The action of the translation operator shifts the outer spins along this closed curve by one step.

The action of the twist operator on the $S^x$ is determined in the following calculation
\begin{eqnarray}
\hat{O}_{\mu} S^x \hat{O}^{\dagger}_{\mu} = \exp({i\frac{2\pi}{K} \displaystyle\sum_{\substack{j=1\\ \mathcal{P}_{\mu}}}^{K} j S_j^z} )S^x \exp({-i\frac{2\pi}{K} \displaystyle\sum_{\substack{j=1\\ \mathcal{P}_{\mu}}}^{K} j S_j^z} ) = e^{X} S^x e^{-X}~.
\end{eqnarray}
To simplify the calculation, we define $i\frac{2\pi}{K}=\Omega$, thus $X=\Omega\sum_j jS^z_j$. Thus we get
\begin{eqnarray}
e^X S^x e^{-X}= S^x+[X,S^x] + \frac{1}{2!}[X,[X,S^x]] +\cdots = \sum_l ( S^x_l \cos \theta_l -  S_l^y \sin \theta_l ) ~,\nonumber\\
e^X S^y e^{-X}= \sum_l  (S^y_l \cos\theta_l  + S^x_l \sin\theta_l)\nonumber~,\\
\theta_l=\frac{2\pi l}{K}~.
\end{eqnarray}
As already defined, $\theta_l=\frac{2\pi l}{K}=\frac{2\pi (K-n)}{K}$, where $n$ is an integer. We can see that in the large channel limit $\theta_l $ becomes inter multiple of $2\pi$. Thus in the large channel limit
\begin{eqnarray}
	\hspace*{-20pt}\lim_{K\rightarrow \infty}\hat{\mathcal{O}}_{\mu} H_{eff} \hat{\mathcal{O}}_{\mu}^{\dagger}\nonumber = \frac{\beta_{\uparrow}(\alpha,\omega_{\uparrow})}{2} \left[ \hat{\mathcal{O}}_{\mu} S^x \hat{\mathcal{O}}_{\mu}^{\dagger} \hat{\mathcal{O}}_{\mu} S^x \hat{\mathcal{O}}_{\mu}^{\dagger}  + \hat{\mathcal{O}}_{\mu} S^y \hat{\mathcal{O}}_{\mu}^{\dagger} \hat{\mathcal{O}}_{\mu} S^y \hat{\mathcal{O}}_{\mu}^{\dagger}  \right] = \frac{\beta_{\uparrow}(\alpha,\omega_{\uparrow})}{2} \left[  S^{x2}+S^{y2}  \right] = H_{eff},\\
\hspace*{-20pt} \lim_{K\rightarrow \infty } [\hat{\mathcal{O}},H_{eff}] = 0~.
\label{eq:hamiltonian_twisted}
\end{eqnarray}
 
Since the translation operator $\hat{T}_{\mu}=e^{i\hat{\mathcal{P}}_{\mu}}$ commutes with the Hamiltonian, so does the generator $\hat{\mathcal{P}}_{\mu}$. We can label the $j^{th}$ state with the eigenvalues of this operator $\hat{\mathcal{P}_{\mu}}$ as $|p^{j}_{\mu}\rangle$. The action of these twist and translation operators on these states are
\begin{eqnarray}
\hspace*{-20pt}\hat{T}_{\mu} |p^j_{\mu} \rangle = e^{i\hat{\mathcal{P}}_{\mu}} |p^j_{\mu} \rangle = e^{ip^j_{\mu}} |p^j_{\mu}\rangle, ~ ~ ~ ~ \hat{T}_{\mu} \hat{\mathcal{O}}_{\mu} |p^j_{\mu} \rangle =  \hat{\mathcal{O}}_{\mu} \hat{T}_{\mu} e^{i\frac{2\pi m}{K}} |p^j_{\mu}\rangle = \hat{\mathcal{O}}_{\mu}   e^{i(\frac{2\pi m}{K}+p_{\mu}^j)} |p^j_{\mu}\rangle~,\quad\frac{2\pi m}{K} \equiv p_{\mu}^{m}~, \nonumber\\
\hspace*{-20pt}\hat{T}_{\mu} \left( \hat{\mathcal{O}}_{\mu} |p^j_{\mu} \rangle  \right) = e^{i(p_{\mu}^{m}+p_{\mu}^j)} \left( \hat{\mathcal{O}}_{\mu}    |p^j_{\mu}\rangle \right)~,
\end{eqnarray}
where $m$ represents different $S^z$ plateaux. This shows that  $\hat{\mathcal{O}}_{\mu} |p^j_{\mu} \rangle $ is again an eigenstate of the translation operator, but with an eigenvalue that is different from $|p^j_{\mu} \rangle $, and is thus orthogonal to $\ket{p^j_{\mu}}$. Similarly, one can in general show that $\langle p^j_{\mu} | \hat{\mathcal{O}}^{q}_{\mu} |p^j_{\mu} \rangle=0$, where $q$ is any integer. Also in the large $K$ limit we can see from the eq.\eqref{eq:hamiltonian_twisted} that these different twisted states has same energy. Which shows that at each plateau state labeled by the $S^z$ eigenvalue has $K$ fold degenerate eigenstates labeled by the eigenvalue of the translation operators. The complete set of commuting observables is therefore formed by $H,S^z,\hat{T}$, and states can be labeled as  $|E,S^z_j,P^j_{\mu}\rangle$.

\section{Effect of conduction bath excitations on the fixed point theory}
\subsection{Non-Fermi liquid signatures in momentum space for 2-channel Kondo}
Obtaining the effective Hamiltonian involves obtaining the low energy excitations on top of the ground state of the star graph. The large-energy excitations involve spin flips. This guides the separation of the Hamiltonian into a diagonal and an off-diagonal piece:
\begin{eqnarray}
	H = H_d + V = \underbrace{H_0 + J S_d^z s_\text{tot}^z}_{H_d} + \underbrace{\frac{J}{2}S_d^+ s_\text{tot}^- + \text{h.c.}}_{V + V^\dagger}~.
\end{eqnarray}
We define \(V\) as the interaction term that decreases \(s_\text{tot}^z\) by 1: \(V \ket{s_\text{tot}^z} \to \ket{s_\text{tot}^z - 1}\). Similarly, we define \(V^\dagger \ket{s_\text{tot}^z} \to \ket{s_\text{tot}^z + 1}\). The Schrodinger equation for the ground state can be written as
\begin{eqnarray}
	E_\text{gs}\ket{\Psi_\text{gs}} = H \ket{\Psi_\text{gs}} = \left(H_d + V\right)\ket{\Psi_\text{gs}} \nonumber\\
	\implies \left(E_\text{gs} - H_d\right)\sum C_{S_d^z, s_\text{tot},s_\text{tot}^z}\ket{S_d^z, s_\text{tot}, s_\text{tot}^z} = V\sum C_{S_d^z, s_\text{tot},s_\text{tot}^z}\ket{S_d^z, s_\text{tot}, s_\text{tot}^z}~.
\end{eqnarray}
\(E_\text{gs}\) is the ground state energy, and can be replaced by the star graph ground state energy if we remove the kinetic energy cost via normal ordering: \(E_\text{gs} = -\frac{J}{2}\left(\frac{K}{2}+1\right) \). Since the interaction part \(V\) only changes \(S_d^z \to -S_d^z\) and \(s^z_\text{tot} \to s^z_\text{tot} \pm 1\), we can simplify the equation into individual smaller equations. For the state \((s_\text{tot},s^z_\text{tot}) = (1,0)\), equations are
\begin{eqnarray}
	\label{eff_ham_Sdz_10}
	\hspace*{-30pt}E_\text{gs} \ket{\frac{1}{2}, 1, 0} = \left(H_d + V \frac{1}{E_\text{gs} - H_d}V^\dagger\right) \ket{\frac{1}{2}, 1, 0}, \quad E_\text{gs} \ket{-\frac{1}{2}, 1, 0} = \left(H_d + V^\dagger \frac{1}{E_\text{gs} - H_d} V\right) \ket{-\frac{1}{2}, 1, 0}~.
\end{eqnarray}
These represent the Schrodinger equation for the states \(\ket{S_d^z, 1, 0}\), and the right hand sides therefore give the effective Hamiltonians for those states. If we combine the states into a single subspace \(\ket{1,0}= \left\{\ket{\frac{1}{2}, 1, 0}, \ket{-\frac{1}{2}, 1, 0}\right\}\), the effective Hamiltonian for this composite subspace becomes the sum of the two parts:
\begin{eqnarray}
	\label{eff_ham_10}
	H^{1,0}_\text{eff}\ket{1, 0}\bra{1, 0} = \left(H_d + V G_0 V^\dagger + V^\dagger G_0  V\right) \ket{1, 0}~,
\end{eqnarray}
where \(G_0 = \left(E_\text{gs} - H_d\right)^{-1}\). To calculate these effective Hamiltonians, we will expand the denominator in powers of in \(H_0^n/J^{n+1}, n=0,1,2,\ldots\). Expanding up to \(n=2\) and keeping at most two particle interaction terms, 
the effective Hamiltonian is
\begin{eqnarray}
	H_\text{eff}^{1, 0} = H_0 + \frac{J^2}{2\left(E_\text{gs} + \frac{J}{2}\right)}&\left[1 + \frac{ H_0 + \left(\frac{1}{2} + S_d^z\right) s^+_\text{tot}X_{1,\text{tot}} - \left(\frac{1}{2} - S_d^z\right) s^-_\text{tot}X^\dagger_{1,\text{tot}}}{2 \left(E_\text{gs} + \frac{J}{2}\right)} + \frac{H_0^2}{\left(E_\text{gs} + \frac{J}{2}\right)^2} \right.\nonumber\\
&\left.- \frac{Z_{1,\text{tot}} H_0}{\left(E_\text{gs} + \frac{J}{2}\right)^3} \right]~.
\end{eqnarray}
We employed the definitions 
\begin{eqnarray}
X_{n,\text{tot}} \equiv  \sum_l \sum_{k,k^\prime}\left(\epsilon_k - \epsilon_{k^\prime}\right)^n c^\dagger_{k \downarrow}c_{k^\prime \uparrow}, ~ ~ ~ Z_{1,\text{tot}} \equiv \sum_{k,k^\prime,l}\left( \epsilon_k - \epsilon_{k^\prime} \right) \frac{1}{2}\left(c^\dagger_{k \uparrow,l}c_{k^\prime \uparrow,l} - c^\dagger_{k \downarrow,l}c_{k^\prime \downarrow,l}\right)~.
\end{eqnarray}
There are several non-Fermi liquid terms of the form \(s^+_\text{tot}X_{1,\text{tot}}, s^-_\text{tot}X^\dagger_{1,\text{tot}},Z_{1,\text{tot}} H_0\). These arise because of the degenerate manifold and the increased availability of states in the Hilbert space for scattering, as compared to the unique singlet ground state of the single-channel Kondo model.

\subsection{Low-temperature thermodynamic behaviour}
The non-Fermi liquid (NFL) nature of the effective Hamiltonian can be demonstrated through a calculation of certain thermodynamic quantities like the impurity specific heat and the magnetic susceptibility. We begin by calculating the self-energy of this NFL hamiltonian. In real space, one can extract a diagonal piece from the effective Hamiltonian by using the fermionic anticommutation relations.
\begin{eqnarray}
H_{eff}^{off,(2)} |_{diag} &=& -(16t^2/3) [ (S_1^z)^2+ (S_2^z)^2 ] ~,
\end{eqnarray}
the corresponding momentum space Hamiltonian is obtained from the Fourier transform:
\begin{eqnarray}
H_{eff}^{off,(2)} |_{diag}= -\frac{4t^2}{3} \frac{1}{N} \left[ \displaystyle\sum_{k,\sigma} n_{k\sigma}(1-\frac{1}{N} \displaystyle\sum_{k_2}  n_{k_2,-\sigma}  ) + \displaystyle\sum_{k,\sigma} \tilde{n}_{k\sigma} ( 1-\frac{1}{N} \displaystyle\sum_{ k_2} \tilde{n}_{k_2,-\sigma}  ) \right]~.
\end{eqnarray}
The above relation leads to the self-energy correction to the kinetic energy, which is
\begin{eqnarray}
\bar{\epsilon} _k-\epsilon_k = \Sigma_k = -\frac{4t^2}{3N^2}\left(1-\frac{N}{e^{(\epsilon_k-\mu)/k_BT}+1}\right)~.
\label{eq:self-energy-NFL}
\end{eqnarray}
Using the self-energy we calculate the impurity specific heat defined as $C_{imp}=C(J^*)-C(0)$ which is defined as
\begin{eqnarray}
C_{imp} &=& \sum_{\Lambda,\sigma} \beta \left[ \frac{(\bar{\epsilon}_{\Lambda})^2 e^{\beta \bar{\epsilon}_{\Lambda}}}{( e^{\beta \bar{\epsilon}_{\Lambda}} +1)^2}  -\frac{({\epsilon}_{\Lambda})^2 e^{\beta {\epsilon}_{\Lambda}}}{( e^{\beta {\epsilon}_{\Lambda}} +1)^2} \right]~.
\end{eqnarray}
\begin{figure}[!htb]
\centering
\includegraphics[width=0.49\textwidth]{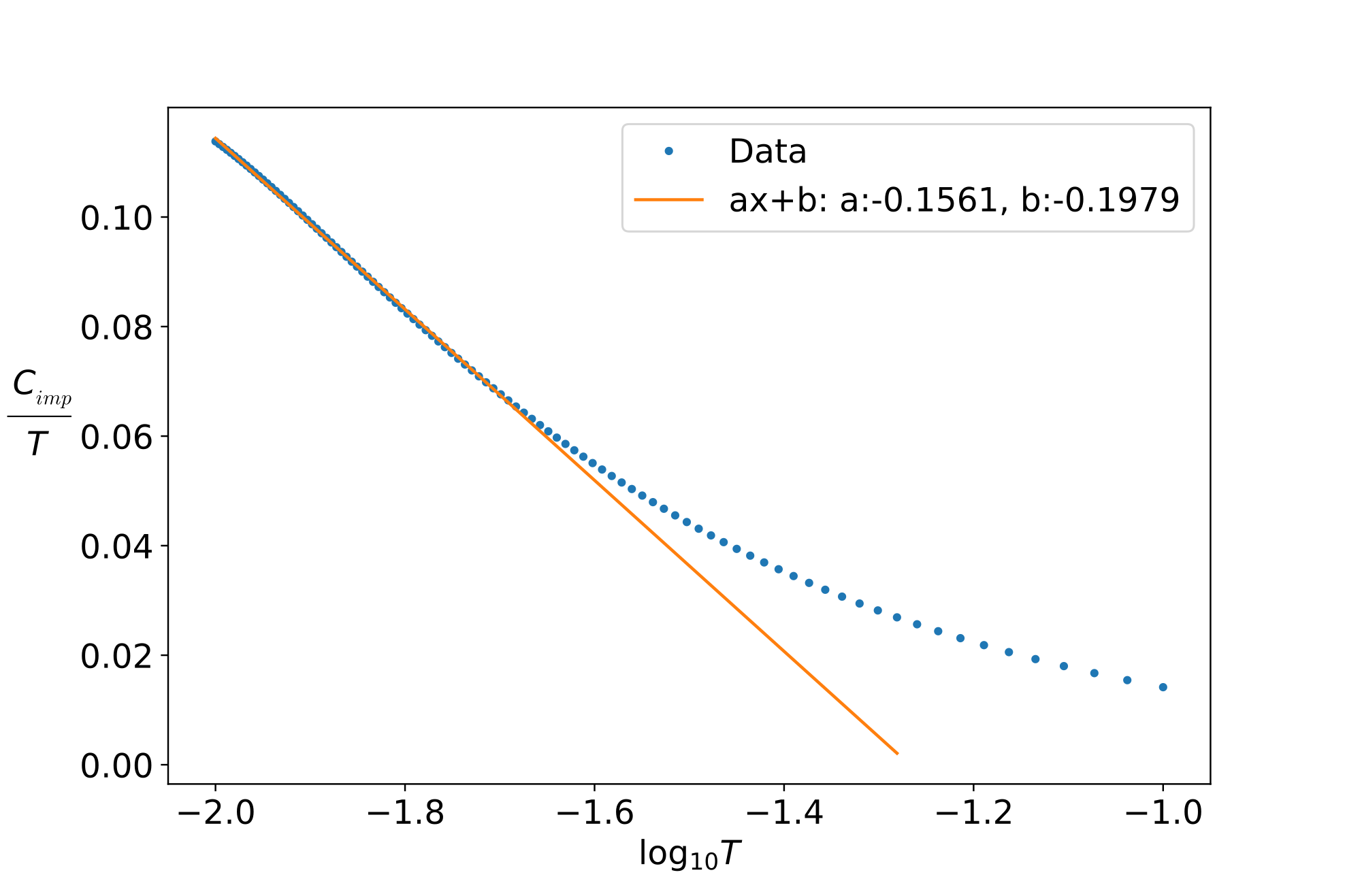}
\includegraphics[width=0.49\textwidth]{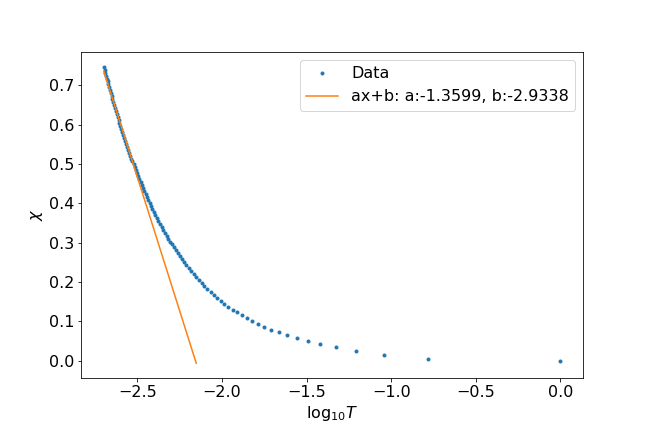}
\caption{Anomalous contribution to the impurity specfic heat (left) and the maginetic susceptibility (right) arising from the non-Fermi liquid component}
\label{fig:Cv_imp}
\end{figure}
Using the self-energy obtained above, we extract the low-temperature behaviour of the impurity specific heat in the two-channel model for $t=0.1$ and $\mathcal{J}=1$, and it is shown in the left panel of fig.~\eqref{fig:Cv_imp}. We find that at low-temperatures, the Sommerfeld coefficient \(C_\text{imp}/T\) follows a logarithmic behaviour for the two-channel case, which is in agreement with the results known in the literature~\cite{affleck_1991_overscreen,affleck_ludwig_1991,affleck_pang_cox_1992,affleck1993exact,
parcollet_olivier_large_N,affleck_2005,emery_kivelson,clarke_giamarchi_1993,zarand_2000,
vondelft_prl_1998,schofield_1997,bullaNRGreview,affleck_pang_cox_1992,pang_cox_1991,
andrei_destri_1984,Tsvelick1984,Tsvelick_1985,andrei_jerez_1995,zarand_costi_2002,
sengupta_1994,fabrizio_nozieres_1995,Coleman_tsvelik,fabrizio_gogolin_1995}.
\begin{eqnarray}
\frac{C_{imp}}{T} \propto \log T~.
\end{eqnarray}
To calculate the magnetic susceptibility, we numerically diagonalise the low-energy effective Hamiltonian, and use the following expression:
\begin{eqnarray}
\chi &=& \beta\left[\frac{\sum e^{-\beta \bar{\epsilon}_{\Lambda}} \langle \bar{S}^{z2} \rangle}{\sum e^{-\beta \bar{\epsilon}_{\Lambda}} } -\frac{\sum e^{-\beta \epsilon_{\Lambda}} \langle S^{z2 }\rangle }{\sum e^{-\beta \epsilon_{\Lambda}} } \right] ~.
\end{eqnarray}
The result is shown in the right panel of fig.~\eqref{fig:Cv_imp}, and we find that at low temperatures, the susceptibility has a logarithmic behaviour.
\begin{eqnarray}
\chi(T) &\propto& \log T~.
\end{eqnarray}

We also calculated the Wilson ratio $C_{imp}/T\chi$, and we find that taking up to three sites on each conduction channel leads to a value of $ W = 8.3$, which is greater than the known value of \(8/3\)~\cite{affleck_1991_overscreen,affleck_ludwig_1991,affleck_pang_cox_1992,affleck1993exact,
parcollet_olivier_large_N,affleck_2005,emery_kivelson,clarke_giamarchi_1993,zarand_2000,
vondelft_prl_1998,schofield_1997,bullaNRGreview,affleck_pang_cox_1992,pang_cox_1991,
andrei_destri_1984,Tsvelick1984,Tsvelick_1985,andrei_jerez_1995,zarand_costi_2002,
sengupta_1994,fabrizio_nozieres_1995,Coleman_tsvelik,fabrizio_gogolin_1995}. We find that taking less number of sites leads to a yet higher value of \(W\), which shows that taking more number of sites will lead to a value of \(W\) that is lower than \(8.3\) and closer to \(8/3\). 

\subsection{Orthogonality catastrophe in the gapless excitations of the 2CK}
By combining the ground-states of the star graph and the effect of hopping into the remaining lattice, it is possible to demonstrate an orthogonality catastrophe~\cite{anderson1967infrared,varma2002singular} in the modified ground-state (see also subsection 5.2.3 of the main manuscript for an entanglement based perspective). The full Hilbert space consists of the star graph (formed by the impurity site and the zeroth sites of all channels) and the {\it reduced} conduction bath (formed by sites 1 and beyond in all channels). All terms in the Hamiltonian conserve the total spin \(S_\text{tot}^z = S_d^z + S_0^z + s^z\) composed of the impurity spin \(S_d^z\), the total zeroth site spin \(S_0^z = \sum_l S_{0,(l)}^z\) and the spin of the reduced bath \(s^z\). We will work in the \(S_\text{tot}^z = 0\) sector. To construct the non-magnetic ground-states, we will assume the presence of gapless excitations vanishingly close to the Fermi surface in the reduced conduction bath. If \(\ket{\phi}\) represents the filled Fermi volume configuration of all the channels, these excitations can be written as \(\ket{e_\sigma^{(l)}} \equiv e^\dagger_{\sigma,(l)} \ket{\phi} = \sum_{k \in \text{FS}}c^\dagger_{k\sigma,(l)} \ket{\phi}\), where \(l \in \left\{1,2\right\} \) is the channel index. Apart from these states, we have the star graph ground-states at \(S_\text{sg}^z \equiv S_d^z + S_0^z = \pm \frac{1}{2}\): \(\ket{S_\text{sg}^z = \frac{\sigma}{2}} = \frac{1}{\sqrt 6}\left( \ket{\sigma,\sigma,\bar\sigma} + \ket{\sigma,\bar\sigma,\sigma} -2\ket{\bar\sigma,\sigma,\sigma} \right),~ \sigma=\pm 1 \). Within each ket, the three symbols indicate the configuration of the impurity spin, the zeroth site of the first channel and that of the second channel respectively. In the absence of any hopping between the star graph and the reduced conduction bath, the full non-magnetic ground states are then of the form:
\begin{eqnarray}
	\ket{l,\sigma} = \ket{S^z_\text{sg}=\frac{-\sigma}{2}}\otimes\ket{e_{\sigma}^{(l)}},~l\in\left\{ 1,2 \right\},~\sigma=\pm 1~.
\end{eqnarray}
Using degenerate perturbation theory, we now treat the effects of hopping \(H_\text{int}\) between the star graph and the reduced conduction bath
\begin{eqnarray}
	H_\text{int} = -t\frac{1}{\sqrt N}\sum_{k,\sigma,l}\left( c^\dagger_{k\sigma,(l)}c_{0\sigma,(l)} + \text{h.c.} \right)~.
\end{eqnarray}
Since all terms in the Hamiltonian conserve the number of particles in each channel, the states with different \(l\) do not mix. For each value of \(l\), the \(2\times 2\) matrix of \(H_\text{int}\) takes the form
\begin{eqnarray}
	H_\text{int} = \frac{4t^2}{J^*}\begin{pmatrix} 0 & 1 \\ 1 & 0 \end{pmatrix} ~,
\end{eqnarray}
where the non-zero off-diagonal terms arise because of second-order scattering processes. Diagonalising this gives the modified new ground-states:
\begin{eqnarray}
	\ket{\psi_\pm^{(l)}} = \frac{1}{\sqrt 2}\left(\ket{l,\uparrow} \pm \ket{l,\downarrow}\right)~,~ E_\pm = -4t^2/J^*~.
\end{eqnarray}
We can now compute the residue for scattering processes of the kind \(e^\dagger_{\sigma,(l)} e_{\bar\sigma,(l)}\) between the new ground-state and excited state, and display that they vanish identically:
\begin{eqnarray}
	Z = |\braket{\psi_+^{(l)}| e^\dagger_{\uparrow,(l)} e_{\downarrow,(l)} | \psi_-^{(l)}}|^2 = \left|\bra{\psi_+^{(l)}}\left(\ket{S_\text{sg}^z=\frac{1}{2}}\ket{e_\uparrow^{(l)}}\right)\right|^2 = 0~.
\end{eqnarray}
This shows that the single-particle excitations are no longer long-lived, signalling the orthogonality catastrophe inherent in the intermediate coupling fixed point theory of the MCK.

\subsection{Three channel LEH}
Similar to the two channel case, we here calculate the low energy effective Hamiltonian for three channel Kondo problem by introducing the real space hopping on top of the zero mode three-channel star graph model. This zero mode three channel star graph model has three fold degenerate ground states with total $2^4=16$ states in the eigenspectrum. The three degenerate ground states are given as 
\begin{eqnarray}
|\alpha_{-1}\rangle = c|1000 \rangle-b (|0100 \rangle + |0010 \rangle+ | 0001 \rangle), ~ ~ |\alpha_{+1}\rangle = b(|1110 \rangle+ | 1101 \rangle + | 1011 \rangle)-c | 0111 \rangle,\nonumber\\
|\alpha_0\rangle = -a(|1100 \rangle + |1010\rangle +|1001 \rangle) + a(|0110 \rangle+| 0101 \rangle +| 0011\rangle)~,
\end{eqnarray}
where $a=0.408$, $b=0.289$, $c=0.866$. Here the state is represented by $|n_{d},n_1,n_2,n_3\rangle$, where $n_i=1/0$ represents the spin configuration $S_i^z=\pm 1/2$ respectively. Next we use degenerate perturbation theory to get the LEH which contains diagonal and off-diagonal terms. In this case we get non-zero contribution from all the three ground states $|J^z=-1\rangle$, $|J^z=0\rangle$ and $|J^z=1\rangle$. We get the diagonal contribution to the LEH in the second order to be
\begin{eqnarray}
H^{(2)}_{eff, diag} &=& - \frac{7.2 J^2}{\alpha} \hat{I}~.
\end{eqnarray}
The contribution associated with different ground states $|J^z=-1\rangle$, $|J^z=0\rangle$ and $|J^z=1\rangle$ is given respectively as $- \frac{2.4 J^2}{\alpha} \hat{I} + \hat{\mathcal{F}},- \frac{2.4 J^2}{\alpha} \hat{I} ,- \frac{2.4 J^2}{\alpha} \hat{I} - \hat{\mathcal{F} }$,
where $\hat{\mathcal{F}}$ is a function of diagonal number operators of sites nearest neighbor to the zeroth site of each channels. The off-diagonal terms in the LEH is appearing due to the scattering between pair of degenerate states $(\alpha_0,\alpha_{+1})$ and $(\alpha_0,\alpha_{-1})$, there is no contribution in the second order coming from the scattering between $(\alpha_{-1},\alpha_{+1})$. The effective low energy Hamiltonian in the second order is given as
\begin{eqnarray}
	H^{(2)}_{eff,off-diag} = \sum_{\substack{(ijk)=\\(123),(231),(312)}}&\left[ c_{i\uparrow}c_{i\downarrow}^{\dagger} \left( -2ab \left\{ \Sigma_{jk} +c_{j\uparrow}^{\dagger}c_{j\downarrow}c_{k\uparrow}c_{k\downarrow}^{\downarrow} \right\} -ac(\Omega_{jk}+\tilde{\Omega}_{jk}) \right) + \textrm{h.c.}\right] \otimes \hat{\Xi}_l~,\qquad
\end{eqnarray}
where the different operators have the following definitions:
\begin{eqnarray}
\Sigma_{i,j} = n_{i\uparrow}(1-n_{i\downarrow}) (1-n_{j\uparrow})n_{j\downarrow} + (1-n_{i\uparrow})n_{i\downarrow} n_{j\uparrow} (1 -n_{j\downarrow}), \quad \Omega_{i,j} = 4S_i^z S_j^z n_{i\uparrow} n_{j\uparrow}\nonumber~,\\
\tilde{\Omega}_{i,j} = 4S_{i}^zS_j^z (1-n_{i\uparrow})(1-n_{j\uparrow}), \quad\hat{\Xi}_l=  ( c_{l_1\uparrow}^{\dagger} c_{l_1\downarrow} +c_{l_2\uparrow}^{\dagger} c_{l_2\downarrow} + c_{l_3\uparrow}^{\dagger} c_{l_3\downarrow} + \textrm{h.c.})~.
\end{eqnarray}

\section{Hamiltonian RG of spin-\(S\) impurity MCK Model}
\label{appendix_urg}

\subsection{Details of the URG method}

The non-trivial nature of the impurity problem arises from the fact that the many-particle correlation between the impurity spin and the conduction electrons induces a many-particle interaction among the conduction electron states. This means that within the conduction band, the states at high and low energies get entangled with each other, leading to what is called UV-IR mixing. One of the methods of obtaining the low-energy physics while taking into account the effect of the high energy degrees of freedom is the renormalisation group (RG) approach. The specific variant of RG used in this work is the unitary renormalisation group (URG) method developed by some of us in Refs.~\cite{santanukagome,anirbanmott1,anirbanmott2,1dhubjhep,siddharthacpi,mukherjeeMERG2022,kondo_urg}. The URG proceeds by applying unitary transformations on the Hamiltonian in order to decouple the high-energy \(k-\)states, leading to a reduction in the effective bandwidth and changes in the couplings for the low-energy degrees of freedom in the process. This leads to a sequence of renormalised Hamiltonians, and hence the coupling RG equations. While this is similar in spirit to the Poor Man's scaling approach of Anderson as applied to the single channel Kondo model~\cite{anderson1970}, there are several differences between the methods that will be apparent when we describe the method in more detail below.

We start by defining a UV-IR scheme for the single-particle electronic states \(\vec k = (k_F + |\vec k|) \hat k\). We label the single-particle \(\vec k-\)states in terms of their normal distance $\Lambda = |\vec k|$ from the Fermi surface (FS) and the orientation of the unit vectors $\hat{s} = \hat k$, where $\hat{s}=\frac{\nabla\epsilon_{\mathbf{k}}}{|\nabla\epsilon_{\mathbf{k}}|}|_{\epsilon_{\mathbf{k}}=E_{F}}$. Each \(\hat s\) represents a direction that is normal to the FS. Combining these two labels with the spin index \(\sigma=\uparrow,\downarrow\), each single-particle state can be uniquely labelled as $\ket{j,l} = \ket{(k_F + \Lambda_j) \hat s,\sigma}, l\equiv (\hat{s},\sigma)$. The $\Lambda$'s are arranged as follows: $\Lambda_{N}>\Lambda_{N-1}>\ldots>0$, such that \(\Lambda_N\) is farthest from the Fermi surface and is hence the most energetic (UV) while \(\Lambda_0\) is closest to the Fermi surface and is least energetic (IR). The URG proceeds by disentangling these states \(\Lambda_j\), starting from those near the UV and gradually scaling towards the IR. This leads to the Hamiltonian flow equation~\cite{anirbanurg1}
\begin{eqnarray}
\centering
\label{urg_map}
H_{(j-1)}=U_{(j)}H_{(j)}U^{\dagger}_{(j)}~,
\end{eqnarray}
where the unitary operation $U_{(j)}$ is the unitary map at RG step $j$. 
$U_{(j)}$ disentangles all the electronic states 
$|\mathbf{k}_{\Lambda_{j}\hat{s}_{m}},\sigma\rangle$
on the isogeometric curve and has the form~\cite{anirbanmott1,anirbanurg1}
\begin{eqnarray}
\centering U_{(j)}=\prod_{l}U_{j,l}, U_{j,l}=\frac{1}{\sqrt{2}}\sum_l [1+\eta_{j,l}-\eta^{\dagger}_{j,l}]~,
\end{eqnarray}
where \(l\) sums over the states on the isoenergetic shell at distance \(\Lambda_j\), and $\eta_{j,l}$ are electron-hole transition operators following the algebra
\begin{eqnarray}
	\left\{\eta_{j,l},\eta_{j,l}^{\dagger}\right\} = 1~,~\left[\eta_{j,l},\eta_{j,l}^{\dagger}\right] = 2\hat n_{j,l} - 1~.
\end{eqnarray}
The transition operator can be expressed in terms of the off-diagonal part of the Hamiltonian, \(H^X_{j,l} = Tr_{j,l}(c^{\dagger}_{j,l}H_{j})c_{j,l} + \text{h.c.}\), and the diagonal part \(H^D_{j,l}\) (kinetic energy and self-energies):
\begin{eqnarray}
	\eta_{j,l}&=&Tr_{j,l}(c^{\dagger}_{j,l}H_{j,l})c_{j,l}\frac{1}{\hat{\omega}_{j,l}-Tr_{j,l}(H^{D}_{j,l}\hat{n}_{j,l})\hat{n}_{j,l}}~.\label{e-TransOp}
\end{eqnarray}
The off-diagonal operator \(Tr_{j,l}(c^{\dagger}_{j,l}H_{j,l})c_{j,l}\) in the numerator of \(\eta_{j,l}\) contains all possible scattering vertices that  change the configuration of the Fock state \(\ket{j,l}\)~\cite{anirbanurg1}. The generic forms of $H^{D}_{j,l}$ and $H^{X}_{j,l}$ are as follows
\begin{eqnarray}
H^{D}_{j,l}=&\sum_{\Lambda\hat{s},\sigma}\epsilon^{j,l}\hat{n}_{\mathbf{k}_{\Lambda\hat{s}},\sigma}+\sum_{\alpha}\Gamma_{\alpha}^{4,(j,l)}\hat{n}_{\mathbf{k}\sigma}\hat{n}_{\mathbf{k}'\sigma'} +\sum_{\beta}\Gamma_{\beta}^{6,(j,l)}\hat{n}_{\mathbf{k}\sigma}\hat{n}_{\mathbf{k}'\sigma'}\hat{n}_{\mathbf{k}''\sigma''}+\ldots~,\nonumber\\
H^{X}_{j,l}=&\sum_{\alpha}\Gamma_{\alpha}^{2}c^{\dagger}_{\mathbf{k}\sigma}c_{\mathbf{k}'\sigma'}+\sum_{\beta}\Gamma_{\beta}^{4}c^{\dagger}_{\mathbf{k}\sigma}c^{\dagger}_{\mathbf{k}'\sigma'}c_{\mathbf{k}_{1}'\sigma_{1}'}c_{\mathbf{k}_{1}\sigma_{1}}+\ldots~.
\end{eqnarray}
The indices \(\alpha\) and \(\beta\) are strings that denote the quantum numbers of the incoming and outgoing electronic states at a particular interaction vertex \(\Gamma^n_{\alpha}\) or  \(\Gamma^m_{\beta}\). The operator $\hat{\omega}_{j,l}$ accounts for the quantum fluctuations arising from the non-commutation between different parts of the renormalised Hamiltonian and has the following form~\cite{anirbanurg1}
\begin{eqnarray}
\hat{\omega}_{j,l}&=&H^{D}_{j,l}+H^{X}_{j,l}-H^{X}_{j,l-1}~.\label{qfOp}
\end{eqnarray}
Upon disentangling electronic states $\hat{s},\sigma$ along a isogeometric curve at distance $\Lambda_{j}$, the following effective Hamiltonian $H_{j,l}$ is generated 
\begin{eqnarray}
H_{j,l}=\prod_{m=1}^{l}U_{j,m}H_{(j)}[\prod_{m=1}^{l}U_{j,m}]^{\dagger}~.
\end{eqnarray}

Accounting for only the leading tangential scattering processes, as well as other momentum transfer processes along the normal direction $\hat{s}$, the renormalised Hamiltonian \(H_{(j-1)}\) has the form~\cite{anirbanurg1}
\begin{eqnarray}
\hspace*{-0.1cm}Tr_{j,(1,\ldots,2n_{j})}(H_{(j)})+\sum_{l=1}^{2n_{j}}\lbrace c^{\dagger}_{j,l}Tr_{j,l}(H_{(j)}c_{j,l}),\eta_{j,l}\rbrace\tau_{j,l}~.\label{HRG}
\end{eqnarray}
The RG fixed point is reached when the denominator in eq.~\eqref{e-TransOp} vanishes, at a certain energy scale \(\Lambda^*\). The vanishing of the denominator can be shown to be concomitant with the vanishing of the off-diagonal component \(H^X\)~\cite{anirbanurg1}, and the fixed point value of the quantum fluctuation operator is equal to one of the eigenvalues of the Hamiltonian. The fixed point Hamiltonian \(H^*\) consists of the renormalised, still-entangled degrees of freedom \(\Lambda_j\) that lie inside the window \(\Lambda^*\): \(\Lambda_j < \Lambda^*\), as well as the integrals of motion (IOMs) that were decoupled by the URG transformations along the way. The IOMs have been stripped of any number fluctuations and are therefore diagonal in the basis of the number operators for each of the decoupled degrees of freedom.
\begin{eqnarray}
	H^* = \sum_{\Lambda_j < \Lambda^*} H^*(\Lambda_j) + \sum_{\Lambda_j > \Lambda^*}H_\text{IOMs}(\Lambda_j)~.
\end{eqnarray}
The effective Hamiltonian can be used to construct the corresponding thermal density matrix and hence the partition function at a temperature \(T\):
\begin{eqnarray}
Z^* = \mathrm{Tr}\left[ \hat{\rho}^*\right] = \mathrm{Tr}\left[ e^{-\beta \hat{H}^{*}}\right] = \mathrm{Tr}\left[ U^{\dagger} e^{-\beta \hat{H}^{*}} U\right] = \mathrm{Tr}\left[ e^{-\beta \hat{H}}\right] = Z
\label{partfunc}~,
\end{eqnarray}
where \(\beta = 1/k_B T\), \(U = \prod_{1}^{j^{*}}U_{(j)}\), $H$ is the bare Hamiltonian and $j^{*}$ is the RG step at which the IR stable fixed point is reached. The unitary transformations preserve the partition function along the RG flow.

\subsection{Derivation of RG equation for the MCK model}

The Hamiltonian for the channel-isotropic MCK model is:
\begin{eqnarray}
	H = \sum_l\left[\sum_{\substack{k\\\alpha=\uparrow,\downarrow}}\epsilon_{k,l} \hat n_{k\alpha,l} + \frac{\mathcal{J}}{2}\sum_{\substack{kk^\prime\\\alpha,\alpha^\prime= \uparrow,\downarrow}} \vec{S_d}\cdot\vec{\sigma}_{\alpha\alpha^\prime}c_{k\alpha,l}^\dagger c_{k^\prime\alpha^\prime, l}\right]~.
\end{eqnarray}

The URG equation for the single-channel Kondo model \cite{kondo_urg} shows a stable strong coupling fixed point. Ferromagnetic interactions are irrelevant. Strictly speaking, that RG equation already encodes, in principle, the multi-channel behaviour, through a modified \(\hat \omega\). To extract this information, we consider the strong coupling fixed-point \(J \gg D\) as a fixed point and analyze its stability from the star graph perspective. For the exactly-screened case, the star graph decouples from the conduction bath, leaving behind a local Fermi liquid interaction on the first site. Similarly, in the under-screened regime, the ground state is composed of states where the impurity spin is only partially screened by the conduction channels. If a particular configuration of the bath-impurity system has the total conduction bath spin down, the impurity will have a residual up spin. This induces a ferromagnetic super-exchange coupling that is irrelevant under RG, so this fixed point is stable as well. 

We now come to the over-screened case, where there is a residual spin on the conduction channel site. The neighbouring electrons will now hop in with spins opposite to that of the impurity, so an antiferromagnetic interaction will be induced, and such an interaction is relevant under the RG. This shows that the over-screened regime cannot have a stable strong coupling fixed point, and we need to search for an intermediate coupling fixed point. We therefore need the generator of the unitary transformation that incorporates third order scattering scatterings explicitly. We should take account of all possible processes that render the set of states \(\left\{\ket{\hat n_{q\beta}=1},\ket{\hat n_{q\beta}=0}\right\}\) diagonal. The higher order generator itself has two scattering processes, such that the entire renormalisation term \(c^\dagger_{q\beta} T \eta\) has in total three coherent processes. The complete generator up to third order can be written as
\begin{eqnarray}
	\eta = \frac{1}{\hat \omega - H_D}T^\dagger c \simeq \frac{1}{\omega^\prime - H_D}T^\dagger c + \frac{1}{\omega^\prime - H_D}H_X \frac{1}{\omega^\prime - H_D} T^\dagger c + \frac{1}{\omega^\prime - H_D} T^\dagger c \frac{1}{\omega^\prime - H_D} H_X~,
\end{eqnarray}
where \(H_X = J \sum_{k,k^\prime < \Lambda_j, \alpha,\alpha^\prime}\vec{S_d}\cdot\vec{s}_{\alpha \alpha^\prime}c^\dagger_{k\alpha}c_{k^\prime\alpha^\prime}\) is scattering between the entangled electrons. There are two third order terms in the above equation corresponding to the two possible sequences in which the processes can occur while keeping the total renormalisation \(c^\dagger_{q\beta}T \eta\) diagonal in \(q\beta\). The second order processes remain unchanged. The total renormalisation takes the form:
\begin{eqnarray}
	\label{full_ren}
	\Delta H_{(j)} = &\underbrace{c^\dagger T \frac{1}{\omega^\prime - H_D} T^\dagger c  + \left(c^\dagger \leftrightarrow c\right)}_{\Delta H^{(2)}_{(j)}} \nonumber\\
			 &+ \underbrace{c^\dagger T \frac{1}{\omega^\prime - H_D} H_X \frac{1}{\omega^\prime - H_D} T^\dagger c + c^\dagger T \frac{1}{\omega^\prime - H_D} T^\dagger c \frac{1}{\omega^\prime - H_D} H_X + \left(c^\dagger \leftrightarrow c\right)}_{\Delta H^{(3)}_{(j)}}.\qquad
\end{eqnarray}
\(\Delta H^{(2)}_{(j)}\) and \(\Delta H^{(3)}_{(j)}\) are the renormalisation arising from the second and third order processes respectively.

It is easier to see the RG flow of the couplings if we write the Hamiltonian in terms of the eigenstates of \(S_d^z\). These eigenstates are defined by \(S_d^z\ket{m_d} = m_d\ket{m_d}, m_d \in \left[-S, S\right]\). In terms of these eigenstates, the Hamiltonian becomes
\begin{eqnarray}
	\label{H_spin_S}
	\mathcal{H} = \sum_{k\sigma}\epsilon_{k}\tau_{k\sigma} + \sum_{m_d=-S}^S \sum_{kl,\atop{\sigma=\uparrow,\downarrow}} J^\sigma_{m_d} \ket{m_d}\bra{m_d} c^\dagger_{k \sigma}c_{l \sigma} + \sum_{kl} \sum_{m_d=-S}^{S-1} J^t_{m_d} \left(\ket{m_d+1}\bra{m_d} s^-_{kl}  + \text{h.c.}\right)~,
\end{eqnarray}
where \(k,l\) sum over the momentum states, \(\sigma\) sums over the spin indices, \(J^\sigma_m = \frac{1}{2} \sigma m J\) in the UV Hamiltonian, and \(J^t_{m} = J\frac{1}{2}\sqrt{S(S+1) - m(m+1)}\) is the coupling that connects \(\ket{m}\) and \(\ket{m+1}\). We first calculate \(\Delta H^{(2)}_{(j)}\). There will be two types of processes - those processes that start from an occupied state (particle sector) and those that start from a vacant state (hole sector). Due to particle hole symmetry of the Hamiltonian, they will be equal to each other and we will only calculate the particle sector contribution. 

In the particle sector, we have (\(\hat n_{q\beta}=1\)), so we will work at a negative energy  shell \(\epsilon_q = -D\). The renormalisation can schematically represented as \(H^I_0 \frac{1}{\omega - H^D_{q\beta}} H^I_1\). Both \(H^I_0\) and \(H^I_1\) have all three operators \(S_d^z, S_d^\pm\). We first consider specifically the case of spin-\(\frac{1}{2}\) impurity. Those terms that have identical operators on both sides can be ignored because \({S_d^z}^2 = \text{constant}\) and \({S^\pm}^2 = 0\). All the six terms that \textit{will} renormalise the Hamiltonian have a spin flip operator on at least one side of the Greens function. This means that in the denominator of the Greens function, \(S_d^z\) and \(s^z_{qq}\) have to be anti-parallel in order to produce a non-zero result for that term. This means we can identically replace \(S_d^z s^z_{qq} = -\frac{1}{4}\). Also, in the particle sector, the Greens function always has \(c_{q\beta}\) in front of it, so \(\epsilon_q \tau_{q\beta} = \frac{D}{2}\). The upshot of all this is that the denominator of all scattering processes for the spin-\(\frac{1}{2}\) impurity Hamiltonian will be \(\omega - \frac{D}{2} + \frac{J}{4}\).

We now come to the general case of spin-\(S\) impurity. The various terms that renormalise the Hamiltonian can be described in terms of the bath spin operators that come into them. For example, the term that has \(s^z\) on both sides of the intervening Greens function can be represented as \(z|z\). There are 7 such terms: \(z|z, \pm|\mp, z|\pm, \pm|z\). Each of these terms occurs both in the particle and the hole sectors. We will demonstrate the calculation of two of these terms. The \(z|z\) \(+|-\) terms evaluate in the following manner.

\begin{eqnarray}
	\hspace*{-40pt}z|z: & \sum_{kk^\prime,m,\sigma} c^\dagger_{q \sigma} c_{k^\prime \sigma} \ket{m}\bra{m}\frac{{J^\sigma_m}^2}{\omega - \frac{D}{2} + \frac{J}{2}\sigma S_d^z}\ket{m}\bra{m} c^\dagger_{k \sigma}c_{q \sigma} = -\sum_{kk^\prime,m,\sigma}n_{q \sigma}\frac{{J^\sigma_m}^2 c^\dagger_{k \sigma}c_{k^\prime\sigma} \ket{m}\bra{m}}{\omega_{m, \sigma} - \frac{D}{2} + \frac{J}{2}\sigma m}~,\nonumber\\
	\hspace*{-40pt}+|-: & \sum_{kk^\prime,m} c^\dagger_{q \uparrow} c_{k^\prime \downarrow} \ket{m}\bra{m+1}\frac{{J^t_m}^2}{\omega - \frac{D}{2} + \frac{J}{2}S_d^z}\ket{m+1}\bra{m} c^\dagger_{k \downarrow}c_{q \uparrow} = -n_{q \uparrow}\sum_{kk^\prime,m}\frac{{J^t_m}^2 c^\dagger_{k \downarrow}c_{k^\prime\downarrow} \ket{m}\bra{m}}{\omega_{m+1, \uparrow} - \frac{D}{2} + \frac{J}{2}\left( m+1 \right) }~.
\end{eqnarray}
We similarly compute the rest of the terms. We again define \(\sum_q \hat n_{q\sigma} = n(D)\). To compare with the spin-\(\frac{1}{2}\) RG equations, we will transform the general spin-\(S\) \(\omega\) to the spin-\(\frac{1}{2}\) \( \omega\), using \(\omega_{m,\sigma} \to \omega - \frac{J}{2}\left(m\sigma - \frac{1}{2}\right)\).
The renormalisation in \(J^\sigma_m\) is
\begin{eqnarray}
	\Delta J^\sigma_{m} = - n(D) \frac{\left( J^\sigma_m \right) ^2 + \left( J^t_{m-\frac{1+\sigma}{2}} \right) ^2}{\omega - \frac{D}{2} + \frac{J}{4}}~.
\end{eqnarray}
Here, we have defined \(J^t_m = 0\) for \( |m| > S\). Two relations can be obtained from this RG equation, the RG equations for the sum and difference of the couplings: \(J^\pm_m = \frac{1}{2}\left(J^\uparrow_m \pm J^\downarrow_m\right) \). The RG equation for the sum of the couplings is
\begin{eqnarray}
	\Delta J^+_m = -n(D)\frac{\sum_\sigma \left( J^\sigma_m \right) ^2 + \sum_\sigma \left( J^t_{m-\frac{1+\sigma}{2}} \right) ^2}{2\left(\omega - \frac{D}{2} + \frac{J}{4}\right)} = -n(D)\frac{J^2}{4}\frac{S(S+1)}{\omega - \frac{D}{2} + \frac{J}{4}}~.
\end{eqnarray}
This is an \(m-\)independent piece, so it can be summed over to produce an impurity-independent potential scattering term, which we ignore. 

The second is the RG equation for the difference of the couplings:
\begin{eqnarray}
	\Delta J^-_m = -n(D)\frac{1}{2}\frac{\left( J^t_{m-1} \right) ^2 - \left(J^t_{m}\right) ^2}{\omega - \frac{D}{2} + \frac{J}{4}} = -\frac{1}{4}\frac{n(D) m J^2}{\omega - \frac{D}{2} + \frac{J}{4}}~.
\end{eqnarray}
The usual \(J\) Kondo coupling is recovered through \(J = 2J^-_m/m\). Substituting this gives 
\begin{eqnarray}
	\Delta_\text{p sector} J = -\frac{1}{2}n(D)\frac{J^2}{\omega - \frac{D}{2} + \frac{J}{4}}~.
\end{eqnarray}
We can also obtain the RG equation for \(J\) from the transverse renormalisation:
\begin{eqnarray}
	\Delta J^t_m &= - \frac{n(D)J^t_m \left( J^\downarrow_m + J^\uparrow_{m+1} \right) }{\omega - \frac{D}{2} + \frac{J}{4}} = -\frac{1}{2}\frac{n(D)J^t_m J}{\omega - \frac{D}{2} + \frac{J}{4}}~.
\end{eqnarray}
Since \(J^t_m \propto J\), we have
\begin{eqnarray}
	\Delta J_\text{p sector} &= -\frac{1}{2}\frac{n(D)J^2}{\omega - \frac{D}{2} + \frac{J}{4}}~.
\end{eqnarray}
The total renormalisation from both particle and hole sectors, at this order, is
\begin{eqnarray}
	\Delta J^{(2)} &= -\frac{n(D)J^2}{\omega - \frac{D}{2} + \frac{J}{4}}~.
\end{eqnarray}

We now come to the third-order renormalisation.
Following eq.~\eqref{full_ren}, the next order renormalisation is
\begin{eqnarray}
	\label{psector_3rd_ren}
	\Delta H^{(3)}_j = c^\dagger T \frac{1}{\omega^\prime - H_D} H_X \frac{1}{\omega^\prime - H_D} T^\dagger c + c^\dagger T \frac{1}{\omega^\prime - H_D} T^\dagger c \frac{1}{\omega^\prime - H_D} H_X~.
\end{eqnarray}
The first term will be of the form
\begin{eqnarray}
\fl\sum_{q,k,l_1} c^\dagger_{q\beta, l_1}c_{k \alpha, l_1}\ket{m_1}\bra{m_2} \frac{1}{\omega - \frac{D}{2} - \frac{\epsilon_k}{2} + \frac{\beta J}{4}S_d^z} \ket{m_2} c^\dagger_{k_1 \sigma_1, l_2}c_{k_2 \sigma_2,l_2} \bra{m_3}\frac{1}{\omega - \frac{D}{2} - \frac{\epsilon_k}{2} + \frac{\beta J}{4}S_d^z}\ket{m_3}\bra{m_4}c^\dagger_{k\alpha, l_1}c_{q\beta, l_1}.\qquad
\end{eqnarray}
We have not bothered to write all the summations and the couplings correctly, because we will only simplify the denominator here. Evaluating the inner products gives
\begin{eqnarray}
	\sum_{qk,l_1} \frac{\ket{m_1}\bra{m_4}c^\dagger_{q\beta}c_{k \alpha}c^\dagger_{k_1 \sigma_1, l_2}}{\omega_{m_2,\beta} - \frac{D}{2} - \frac{\epsilon_k}{2} + \frac{\beta J m_2}{4}}  \frac{c_{k_2 \sigma_2,l_2}c^\dagger_{k\alpha}c_{q\beta}}{\omega_{m_3,\beta} - \frac{D}{2} - \frac{\epsilon_k}{2} + \frac{\beta J m_3}{4}}~.
\end{eqnarray}
We again use \(\omega_{m,\sigma} \to \omega - \frac{J}{2}\left(m\sigma - \frac{1}{2}\right)\).
\begin{eqnarray}
	\ket{m_1}\bra{m_4}c^\dagger_{k_1 \sigma_1, l_2}c_{k_2 \sigma_2,l_2}\sum_{qk,l_1} \frac{\hat n_{q\beta}\left(1 - \hat n_{k\alpha}\right)}{\left(\omega - \frac{D}{2} - \frac{\epsilon_k}{2} + \frac{J}{4}\right)^2}~.
\end{eqnarray}
We define \(\sum_q \hat n_{q\beta} = n(D)\). Performing the sums over \(k\) and \(l_1\) gives
\begin{eqnarray}
	-\frac{1}{2}\ket{m_1}\bra{m_4}c^\dagger_{k_1 \sigma_1, l_2}c_{k_2 \sigma_2,l_2} \frac{\rho n(D) K}{\omega - \frac{D}{2} + \frac{J}{4}}~.
\end{eqnarray}
\(\rho\) is the density of states which we have taken to be constant. Reinstating the complete summation and the couplings gives
\begin{eqnarray}
	\label{first_group}
	-\frac{1}{2}\frac{\rho n(D) K}{\omega - \frac{D}{2} + \frac{J}{4}}\sum_{m_1,m_4,k_1,k_2,\atop{l_2,\sigma_1\sigma_2}}\lambda_{1}\lambda_{2}\lambda_{3}\ket{m_1}\bra{m_4}c^\dagger_{k_1 \sigma_1, l_2}c_{k_2 \sigma_2,l_2}~.
\end{eqnarray}
There is no sum over \(m_2\) and \(m_3\) because they are constrained by \(m_1\) and \(m_4\) respectively. \(\lambda_i\) represent the couplings present at the three interaction vertices. \(k_{1,2}\) sum over the momenta, \(\sigma_{1,2}\) sum over the spin indices and \(l_2\) sums over the channels.

The second term in eq.~\eqref{psector_3rd_ren} can be evaluated in an almost identical fashion. The integral here will be negative of the first term, because of an exchange in the scattering processes.
\begin{eqnarray}
	\label{second_group}
	\frac{1}{2}\frac{\rho n(D) K}{\omega - \frac{D}{2} + \frac{J}{4}}\sum_{m_1,m_4,k_1,k_2,\atop{l_2,\sigma_1\sigma_2}}\lambda_{1}\lambda_{3}\lambda_{2}\ket{m_1}\bra{m_4}c^\dagger_{k_1 \sigma_1, l_2}c_{k_2 \sigma_2,l_2}~.
\end{eqnarray}
The first group of terms (those that appear in \eqref{first_group}) in the particle sector can be represented as \(a|b^\prime|c\), where \(a,b,c \in \left\{z,+,-\right\} \) and represent the operator for the conduction electrons in the three connected processes. The \(\prime\) on \(b\) indicates that it is the state of the electrons \textit{not being decoupled}. The second group of terms (those that appear in \eqref{second_group}) are therefore represented as \(a|b|c^\prime\), because in this group, the interaction \(H_X\) between the electrons that are not being decoupled occur at the very end.
We will only calculate the terms in the particle sector, the ones in hole sector will be equal to these because of particle-hole symmetry. The full list of terms is: 

$z|z^\prime|z\quad$
$z|z|z^\prime\quad$
$-|z^\prime|+\quad$
$-|+|z^\prime\quad$
$+|z^\prime|-\quad$
$+|-|z^\prime\quad$
$ z|+^\prime|z\quad$
$z|-^\prime|z\quad$
$+|+^\prime|-\quad$
$-|+^\prime|+\quad$
$z|z|+^\prime\quad$
$z|z|-^\prime\quad$
$ +|-|+^\prime\quad$
$+|-|-^\prime\quad$
$-|+|+^\prime\quad$
$z|z^\prime|z\quad$
$z|z|z^\prime\quad$
$-|z^\prime|+\quad$
$ -|+|z^\prime\quad$
$+|z^\prime|-\quad$
$+|-|z^\prime\quad$
$z|+^\prime|z\quad$
$z|-^\prime|z\quad$
$+|+^\prime|-\quad$
$-|+^\prime|+\quad$
$z|z|+^\prime\quad$
$z|z|-^\prime\quad$
$+|-|+^\prime\quad$
$+|-|-^\prime\quad$
$-|+|+^\prime\quad$
.

The total renormalisation in  \(J^\sigma_m\) is
\begin{eqnarray}
	\hspace*{-20pt}\Delta J^\sigma_m = \frac{1}{2}\frac{\rho n(D) K}{\omega - \frac{D}{2} + \frac{J}{4}}\left[\left( J^t_{m-1} \right)^2 J^\sigma_m + \left( J^t_m \right)^2 J^\sigma_m - \left( J^t_{m-1} \right)^2 J^\sigma_{m-1} - \left(J^t_m\right)^2 J^\sigma_{m+1} \right] = \frac{1}{2}\frac{\rho n(D) K}{\omega - \frac{D}{2} + \frac{J}{4}}J^3 m \sigma~.\qquad
\end{eqnarray}
Since we had defined \(J^\sigma_m \equiv \frac{1}{2}J m \sigma\), we have \(\Delta J = \frac{2}{m\sigma}\Delta J^\sigma_m\), and we get \(\Delta_\text{p sector} J = \frac{1}{4}\frac{\rho n(D) K}{\omega - \frac{D}{2} + \frac{J}{4}}J^3\).
Combining with the hole sector renormalisation, we get
\begin{eqnarray}
	\Delta J^{(3)} = \frac{1}{2}\frac{\rho n(D) K}{\omega - \frac{D}{2} + \frac{J}{4}}J^3~.
\end{eqnarray}
The total renormalisation in \(J\) after combining all orders is
\begin{eqnarray}
	\Delta J = -\frac{n(D) J^2}{\omega - \frac{D}{2} + \frac{J}{4}} + \frac{1}{2}\frac{\rho n(D) K}{\omega - \frac{D}{2} + \frac{J}{4}}J^3~.
\end{eqnarray}

\section*{References}

\bibliography{../MCK-manuscript}